\documentclass[acmtog, authorversion]{acmart}
\usepackage{bm}
\usepackage[normalem]{ulem}
\usepackage{booktabs} 
\usepackage{colortbl}
\usepackage[ruled]{algorithm2e} 

\SetAlFnt{\small}
\SetAlCapFnt{\small}
\SetAlCapNameFnt{\small}
\SetAlCapHSkip{0pt}

\usepackage{tikz}
\usepackage{mathtools}
\usepackage{xspace}
\usepackage{subcaption}
\usepackage{microtype}
\usepackage{xspace}
\usepackage{csquotes}
\usepackage{multirow}

\DeclareMathOperator*{\argmax}{arg\,max}
\DeclareMathOperator*{\argmin}{arg\,min}

\graphicspath{{./figures/}}

\setcopyright{rightsretained}
\acmJournal{TOG}
\acmYear{2024} \acmVolume{43} \acmNumber{4} \acmArticle{64} \acmMonth{7}\acmDOI{10.1145/3658187}

\usepackage[inkscapeformat=png]{svg}
\begin{teaserfigure}
  \includegraphics[width=\linewidth]{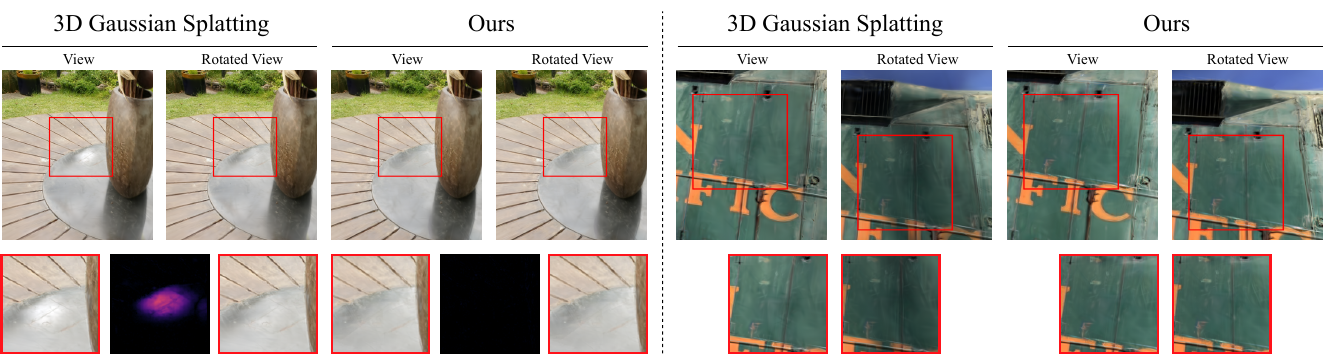}
  \caption{
  3D Gaussian Splatting~\cite{kerbl3Dgaussians} suffers from popping artifacts during view rotation due to its approximate, global sorting scheme.
  Our method is able to effectively circumvent short-range popping artifacts (left) and long-range view-inconsistencies (right) during rotation with a novel, hierarchical per-pixel sorting strategy.}
  \Description{This is the teaser figure for the article.}
  \label{fig:teaser}
\end{teaserfigure}

\begin{document}

\newcommand{\mipnerf}{Mip-NeRF 360\xspace}
\newcommand{\gs}{3DGS\xspace}
\newcommand{\ngp}{iNGP\xspace}
\newcommand{\figref}[1]{Fig.~\ref{fig:#1}\xspace}
\newcommand{\tabref}[1]{Tab.~\ref{tab:#1}\xspace}
\newcommand{\eqnref}[1]{Eqn.~\eqref{eq:#1}\xspace}
\newcommand{\appref}[1]{Appendix~\ref{app:#1}\xspace}

\newcommand{\ie}{\textit{i.e.}\xspace}
\newcommand{\eg}{\textit{e.g.}\xspace}
\newcommand{\cf}{cf.\xspace}

\newcommand{\ptd}{PTD}
\newcommand{\woptd}{w/o \ptd}

\newcommand{\FLIP}{\protect\reflectbox{F}LIP\xspace}
\newcommand{\flipview}[1]{\text{\protect\reflectbox{F}LIP}_{#1}\xspace}

\newcommand{\errmax}{\delta_{\text{max}}}
\newcommand{\erravg}{\delta_{\text{avg}}}
\newcommand{\topt}{t_{\text{opt}}}
\newcommand{\stimes}{{\times}}

\let\oldhat\hat 
\renewcommand{\vec}[1]{\mathbf{#1}} 
\renewcommand{\hat}[1]{\oldhat{\mathbf{#1}}}
\newcommand{\depth}{\zeta}

\newcommand{\ifcommentsenabled}[1]{#1}

\definecolor{mathias_color}{rgb}{.6,.4,.05}
\definecolor{michael_color}{rgb}{0,0.35,0}
\definecolor{alex_color}{rgb}{0,0,0.85}
\definecolor{markus_color}{rgb}{0,0.35,0.35}
\definecolor{bernhard_color}{rgb}{0.35,0.35,0}
\newcommand{\mathias}[1]{\ifcommentsenabled{\textcolor{mathias_color}{Mathias: #1}}}
\newcommand{\lukas}[1]{\ifcommentsenabled{\textcolor{lukas_color}{Lukas: #1}}}
\newcommand{\michael}[1]{\ifcommentsenabled{\textcolor{michael_color}{Michael: #1}}}
\newcommand{\alex}[1]{\ifcommentsenabled{\textcolor{alex_color}{Alex: #1}}}
\newcommand{\markus}[1]{\ifcommentsenabled{\textcolor{markus_color}{Markus: #1}}}
\newcommand{\bernhard}[1]{\ifcommentsenabled{\textcolor{bernhard_color}{Bernhard: #1}}}

\definecolor{edited_color}{rgb}{.7,.1,.1}
\newcommand{\new}[1]{#1} 

\definecolor{revised_color}{rgb}{.1,.1,.7}
\newcommand{\revised}[2]{#1} 

\title{StopThePop: Sorted Gaussian Splatting for View-Consistent Real-time Rendering}

\author{Lukas Radl}
\authornote{Both authors contributed equally to this work}
\email{lukas.radl@icg.tugraz.at}
\author{Michael Steiner}
\authornotemark[1]
\email{michael.steiner@tugraz.at}
\affiliation{%
  \institution{Graz University of Technology}
  \country{Austria}
}
\author{Mathias Parger}
\affiliation{%
  \institution{Huawei Technologies}
  \country{Austria}
}
\email{mathias.parger@huawei.com}

\author{Alexander Weinrauch}
\affiliation{%
  \institution{Graz University of Technology}
  \country{Austria}
}
\email{alexander.weinrauch@icg.tugraz.at}

\author{Bernhard Kerbl}
\affiliation{%
  \institution{TU Wien}
  \country{Austria}
}
\email{kerbl@cg.tuwien.ac.at}
\author{Markus Steinberger}
\affiliation{%
  \institution{Graz University of Technology}
  \country{Austria}
}
\affiliation{%
  \institution{Huawei Technologies}
  \country{Austria}
}
\email{steinberger@icg.tugraz.at}

\renewcommand\shortauthors{Radl and Steiner et al}

\begin{abstract}

Gaussian Splatting has emerged as a prominent model for constructing 3D representations from images across diverse domains. 
However, the efficiency of the 3D Gaussian Splatting rendering pipeline relies on several simplifications. 
Notably, reducing Gaussian to 2D splats with a single view-space depth introduces popping and blending artifacts during view rotation. 
Addressing this issue requires accurate per-pixel depth computation, yet a full per-pixel sort proves excessively costly compared to a global sort operation.
In this paper, we present a novel hierarchical rasterization approach that systematically resorts and culls splats with minimal processing overhead.
Our software rasterizer effectively eliminates popping artifacts and view inconsistencies, as demonstrated through both quantitative and qualitative measurements. 
Simultaneously, our method mitigates the potential for cheating view-dependent effects with popping, ensuring a more authentic representation. 
Despite the elimination of cheating, our approach achieves comparable quantitative results for test images, while increasing the consistency for novel view synthesis in motion.
Due to its design, our hierarchical approach is only $4\%$ slower on average than the original Gaussian Splatting.
Notably, enforcing consistency enables a reduction in the number of Gaussians by approximately half with nearly identical quality and view-consistency.
Consequently, rendering performance is nearly doubled, making our approach 1.6x faster than the original Gaussian Splatting, with a 50\% reduction in memory requirements. 
Our renderer is publicly available at {\color{blue}\url{https://github.com/r4dl/StopThePop}}.
\end{abstract}

\ccsdesc[500]{Computing methodologies~Rasterization}

\keywords{Parallel Computing, Point-based Rendering, Real-Time Rendering}

\maketitle

\section{Introduction}
In recent years, Neural Radiance Fields (NeRFs)~\cite{Mildenhall2020NeRF} have triggered a new surge of research around differentiable rendering of 3D representations. Leveraging the traditional volume rendering equation, NeRFs are fully differentiable, enabling continuous optimization to align the representation to diverse input views and support high-quality novel view synthesis. This differentiability also proves valuable in addressing other rendering challenges that necessitate gradient flow and optimization. 
\begin{figure*}
    \centering
    \subcaptionbox{}[0.245\linewidth]{\includegraphics[width=\linewidth,page=1]{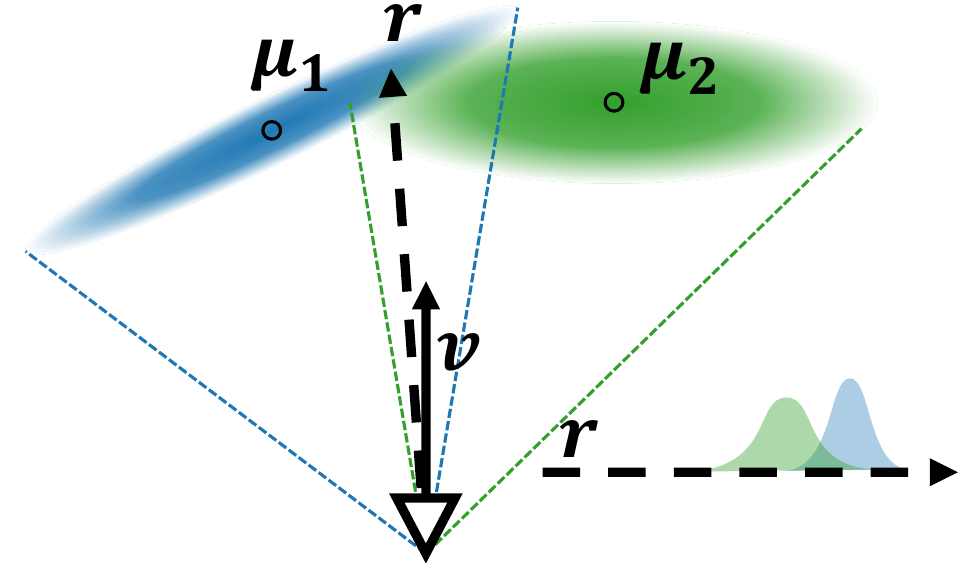}}%
    \subcaptionbox{}[0.245\linewidth]{\includegraphics[width=\linewidth,page=2]{depth2.pdf}}%
    \subcaptionbox{}[0.245\linewidth]{\includegraphics[width=\linewidth,page=3]{depth2.pdf}}%
    \subcaptionbox{}[0.245\linewidth]{\includegraphics[width=\linewidth,page=4]{depth2.pdf}}%
    \caption{Effect of collapsing 3D Gaussians into 2D splats and \gs's depth simplification: 
    (a) Integrating Gaussians along view rays $\vec r$ requires careful consideration of potentially overlapping 1D Gaussians. 
    (b) Using flattened 2D splats and view-space $z$ as depth (projection of $\vec \mu$ onto $\vec v$) puts 2D splats on spherical segments around the camera, inverting the relative positions of the two Gaussians along the example view ray.
    (c) Camera rotation inverts the order along $\vec r$, resulting in popping.
    (d) Camera translation does not alter the distance compared to (b). 
    }
    \label{fig:depth}
\end{figure*}

Various strategies have arisen to tackle challenges in NeRFs, particularly mitigating the computational costs linked to multilayer perceptron (MLP) evaluation. These approaches include adopting direct voxel representations~\cite{fridovich2022plenoxels}, employing feature hash maps~\cite{mueller2022instant}, and exploring tensor factorizations~\cite{TensorRFECCV, tang2022compressible}---departing to some extent from the original pure MLP design. 
A recent notable development in this trajectory is 3D Gaussian Splatting (\gs)~\cite{kerbl3Dgaussians}, which renders oriented 3D Gaussians with spherical harmonics (SH) as a view-dependent color representation.

Remaining faithful to the traditional volume rendering equation, \gs facilitates gradient flows from image errors to the Gaussians’ positions, shapes, densities, and colors. With an initialization based on structure-from-motion~\cite{snavelysfm}, a real-time compute-mode rasterizer, and heuristic-driven densification and sparsification, \gs converges to a high-quality representation with compact memory requirements. 
Consequently, \gs has firmly established itself as one of the most widely used methods for 3D scene reconstruction and differentiable rendering.
Colored, semi-transparent 3D Gaussians serve as a versatile representation, but their accurate rendering is challenging. 
Although the projection of a 3D Gaussian onto a view ray is straightforward, leveraging synergies between neighboring rays under perspective projection proves intricate.
Hence, \gs approximates them as flattened 2D splats~\cite{zwicker2002ewa}, necessitating depth-based sorting for rendering. 
\gs further simplifies this step by sorting based  on the view-space $z$-coordinate of each Gaussian’s mean, effectively projecting splats onto spherical shells reminiscent of \citet{broxton2020immersive}. 
While this global sorting eases the rendering algorithm, it introduces popping artifacts, \ie, sudden color changes for consistent geometry, during camera rotations due to changes in the relative depth of shells (see \figref{depth}).
\new{Such view inconsistencies due to popping can be very irritating and immersion-breaking, \eg during head rotation in a virtual reality setting.}

Fully evaluating all Gaussians in 3D along each view ray while considering their overlap would be ideal, but likely not feasible in real-time.
The next best solution involves approximating the location where each Gaussian contributes the most for each view ray, \ie, determining its depth, followed by a correct per-pixel blending.
Sorting must now happen for each view ray, rather than globally for all Gaussians; an obvious challenge as it is not uncommon to see thousands of Gaussians be considered for individual rays in \gs.
To solve this challenge, we propose a novel 3D Gaussian Splatting rendering pipeline that exploits coherence among neighboring view rays on multiple hierarchy levels, interleaving culling, depth evaluation and resorting.
We make the following contributions:
\begin{itemize}
\item A novel hierarchical 3D Gaussian Splatting renderer that leads to per-pixel sorting of Gaussian splats for both the forward and backward pass of the \gs rendering pipeline and thus removes popping artifacts.
\item An in-depth analysis of culling and depth approximation strategies, as well as pipeline optimizations and workload distribution schemes for our compute-mode \gs hierarchical renderer.
\item A discussion and evaluation of various sorting strategies of Gaussian splats and their influence on overall rendering quality and view-consistency.
\item An effective automatic method to detect popping artifacts in videos captured from trained 3D Gaussians as well as a user study confirming the results of the presented method. 
\end{itemize}
Our results indicate that a full per-pixel sorted renderer for Gaussian splats eliminates all popping artifacts but reduces rendering speed by $100\times$.
Our hierarchical renderer is virtually indistinguishable from a full per-pixel sorted renderer, but only adds an overhead of $4\%$ compared to the original \gs.
\section{Preliminaries and Related Work}
In the following, we review the renderer used in \gs.
For a complete description of the approach, \cf~\citet{kerbl3Dgaussians}.

\subsection{3D Gaussian Splatting}
NeRF-style rendering and \gs use the volume rendering equation:
\begin{equation}
C(\vec{r}) =  \int_0^t \revised{\vec{c}}{c}\left(\vec{r},t\right) \, \sigma\left(\vec{r},t\right) \, T(\vec{r},t) \,  dt \text{, \hspace{5pt}where} 
\label{eq:volume}
\end{equation}
\[
T(\vec{r},t)=e^{-\int_0^t \sigma\left(\vec{r},s\right) \, ds },
\]
$C(\vec{r})$ is the output color for a given ray $\vec{r}$, $\sigma\left(\vec{r},t\right)$ is the opacity along the ray and $c\left(\vec{r},t\right)$ is the \new{view-dependent} emitted radiance. 
\gs represents a scene as a mixture of $N$ 3D Gaussians each given by:
\revised{
\[
G(\vec x) = e^{-\frac{1}{2}(\vec{x}-\vec{\mu})^T\Sigma^{-1}(\vec{x}-\vec{\mu})}\text{, \hspace{5pt}where}
\]
\[
\Sigma = R S S^T R^T,
\]
$\vec{\mu}$ is the Gaussian's location, $R$ is a rotation matrix and $S$ is a diagonal scaling matrix, 
allowing to position, rotate and non-uniformly scale Gaussians in 3D space while ensuring that $\Sigma$ is positive semi-definite.
}{
$\vec{\mu}_i$ is the Gaussian's location, $R_i$ is a rotation matrix and $S_i$ is a scaling matrix, 
allowing to position, rotate and scale Gaussians in 3D space.
}
When evaluating a 3D Gaussian along a ray, the resulting projection is a 1D Gaussian.
It seems natural to evaluate~\eqnref{volume} considering how multiple Gaussians influence any location along the ray.
As there is no elementary indefinite integral known for Gaussians, numerical integration is likely the only option.
In practice, this would require a strict sorting of all starting and end points of all Gaussians and sampled numerical integration. 

\new{Instead}, \gs makes multiple simplifications.
First, they consider all Gaussians to be separated in space, \ie, compress their extent to a Dirac delta along the ray.
Second, the Dirac delta \new{of the $i$-th Gaussian} is located at 
\begin{equation}
    t_i = \vec \mu_i^T \vec v,
    \label{eq:viewspacedepth}
\end{equation} \ie, the projection of the mean $\vec \mu_i$ onto the view direction $\vec v$, independent of the individual ray $\vec r$.
Third, they approximate the projection of the Gaussian onto all rays, relying on an orthogonal projection approximation considering the first derivative of the 3D Gaussian to construct a 2D splat $G_2$~\cite{zwicker2002ewa}.

These approximations enable faster rendering:
\eqnref{volume} becomes
\begin{equation}
C(\vec{r}) = \sum_{i=1}^{N_{\vec{r}}} \revised{\vec{c}}{c}_i  \alpha_i \prod_{j=1}^{i-1}(1-\alpha_j),
\label{eq:volume_desc}
\end{equation}
where $i$ iterates over the $N_{\vec{r}}$ Gaussians that influence the ray in the ordering of $t_i$, and $\alpha_i$ is the opacity of the Gaussian along the ray, \ie, $G_2(x,y)$, multiplied by a learned per-Gaussian opacity value.

Because $t_i$ is independent of $\vec{r}$, a global sort of all $t$ is possible.
Na\"ively, this would lead to $N_{\vec{r}} = N$ for all rays.
To reduce the number of Gaussians considered per ray, \gs \new{splits the image into $16\stimes16$ pixel tiles, and} runs a combined depth and tile sorting pre-pass, before evaluating \eqnref{volume_desc}. 
For each \revised{tile}{$16\stimes16$ pixel tile} and each Gaussian that may potentially contribute to any pixel in this tile---considering the 2D bounding box around the $1\%$ Gaussian contribution threshold---a sorting key is generated with the tile index in the higher order bits and the depth in the lower bits.
Sorting those combined keys leads to a $t_i$-sorted list for each tile.

\subsection{Radiance Field Methods}
Contrary to \gs, NeRFs~\cite{Mildenhall2020NeRF} require sampling a continuous, implicit neural scene representation densely.
Therefore, real-time rendering as well as handling unbounded scenes proves difficult.
Many follow-up works investigated NeRF extensions to handle unbounded scenes~\cite{barron2021mipnerf, barron2022mipnerf360, barron2023zip} as well as faster rendering~\cite{mueller2022instant, fridovich2022plenoxels, TensorRFECCV}, 3D scene editing~\cite{Nguyen2022Snerf, kuang2022palettenerf, jambon2023nerfshop},
avatar generation~\cite{zielonka2023instant}, scene dynamics~\cite{pumarola2020d, park2021nerfies} and 3D \revised{object}{mesh} generation~\cite{jain2021dreamfields, poole2022dreamfusion, raj2023dreambooth3d}.

\subsection{\gs Follow-up Work}
Following the code release and subsequent publication of \gs, several extensions have popped up investigating various paradigms,
including the editing of trained Gaussians~\cite{chen2023gaussianeditor, fang2023GaussianEditor}, text-to-3D~\cite{tang2023dreamgaussian, yi2023gaussiandreamer} and 4D novel view synthesis~\cite{luiten2023dynamic, wu20234dgaussians}.
Mip-Splatting~\cite{Yu2023MipSplatting} proposes a 3D smoothing filter and 2D Mip filter to remedy aliasing in \gs.
Besides them, most approaches merely leverage Gaussians as graphics primitives, whereas our approach tackles current problems with \gs.

\subsection{Software Rasterization}
Our compute-mode rendering pipeline for \gs is related to other software-based rendering pipelines. 
Early works like Pomegranate \cite{eldridge2000pomegranate} and the Larrabee project \cite{seiler2008larrabee} showed that software pipelines on custom hardware are viable for rendering.
Special compute-mode rendering pipelines have been proposed for REYES \cite{zhou2009renderants,tzeng2010task}, triangle rasterization \cite{liu2010freepipe,laine2011high,patney2015piko,kenzel2018high,karis2021deep} and point clouds \cite{schutz2021rendering}.
Similarly to these efforts, we show that taking into account the specifics of the rendering problem, a compute-mode renderer for \emph{sorted} Gaussian splats can execute in real-time on modern GPUs.

\new{
\subsection{Order Independent Transparency}
Correctly and efficiently rendering semi-transparent primitives, such as Gaussian splats, proves intricate for rasterization-based renderers.
Methods approximating order independent transparency \cite{wyman2016exploring} investigate this paradigm.
$k$-buffers~\cite{callahan2005visibilitysorting, bavoilk2007kbuffer} operate with a fixed per-pixel memory budget, circumventing the large memory requirement of $A$-buffers~\cite{carpenter1984abuffer}.
When this budget is exceeded, new incoming fragments are either merged~\cite{salvi2011adaptive, salvi2014multilayer} or the closest fragment gets written to the color buffer~\cite{callahan2005visibilitysorting};
both cases require a nearly-sorted order for incoming fragments. 
Our work combines hierarchical levels of $k$-buffers with \gs's tile-based rasterization. 
}

\section{Real-time Sorted Gaussian Splatting}

We present a novel per-pixel sorted 3D Gaussian splatting approach, departing from the current global sorting paradigm. Utilizing fast per-pixel depth calculations and a hierarchical intra-tile cooperative sorting approach, our method enhances the accuracy of the resulting sort order. To streamline computations, we incorporate per-tile opacity culling and a fast and GPU-friendly load balancing scheme.

\begin{figure}
    \centering
    \begin{subcaptionblock}{0.49\linewidth}
    \includegraphics[width=\linewidth,page=5]{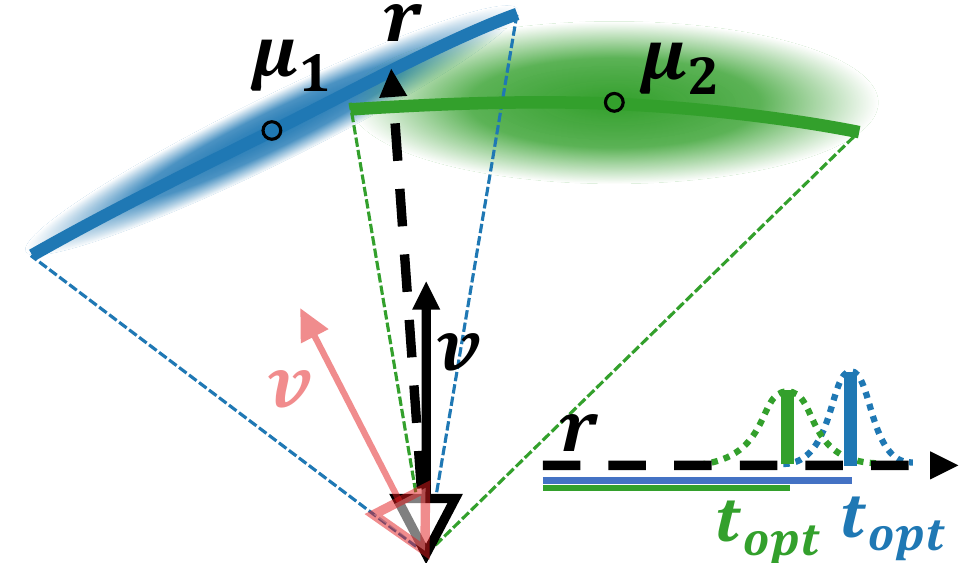}%
    \end{subcaptionblock}%
    \begin{subcaptionblock}{0.49\linewidth}
    \includegraphics[width=\linewidth,page=6]{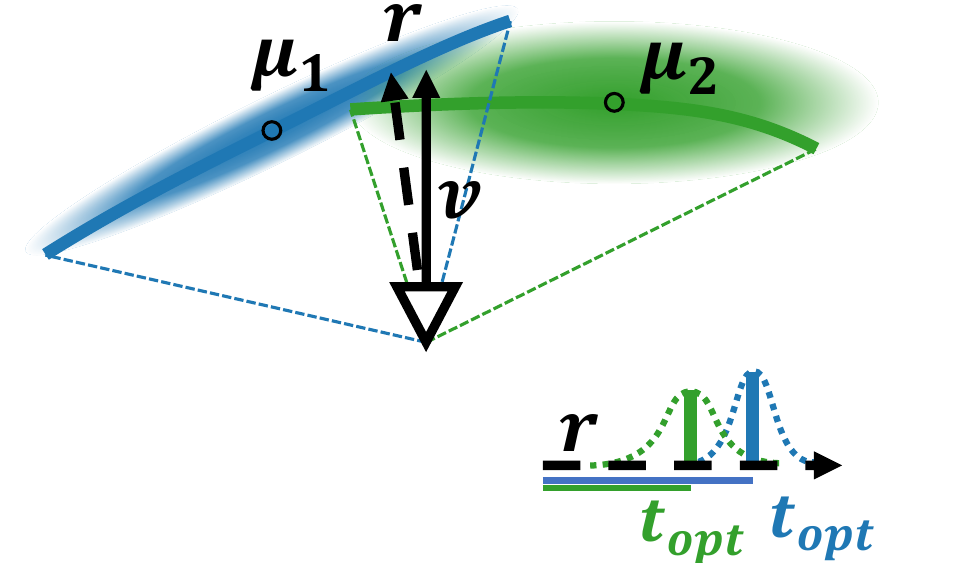}%
    \end{subcaptionblock}%
    \caption{\revised{Our approach to compute $t_{opt}$ avoids popping by placing splats at the point of maximum contribution along the view ray $\vec r$, creating sort orders independent of camera rotation (red view vector)}{Our approach to compute $t_{opt}$ places splats along the maximum point of contribution along the view ray $\vec r$, creating sort orders independent of camera rotation (red view vector) and thus avoiding popping}. 
    Note that the shape of $t_{opt}$ is a curved surface and changes with \revised{the camera position}{$\vec o$}; \cf \figref{depth}.
    }
    \label{fig:depth_ours}
\end{figure}

\subsection{Global Sorting}

\begin{figure*}[ht!]
    \centering

    \begin{subcaptionblock}[C]{.02\linewidth}
    \rotatebox[origin=c]{90}{full sorted}
    \end{subcaptionblock}
    \begin{subcaptionblock}[C]{.24\linewidth}
    \begin{tikzpicture}
     \node[anchor=south west,inner sep=0] (image) at (0,0) {\includegraphics[width=\linewidth]{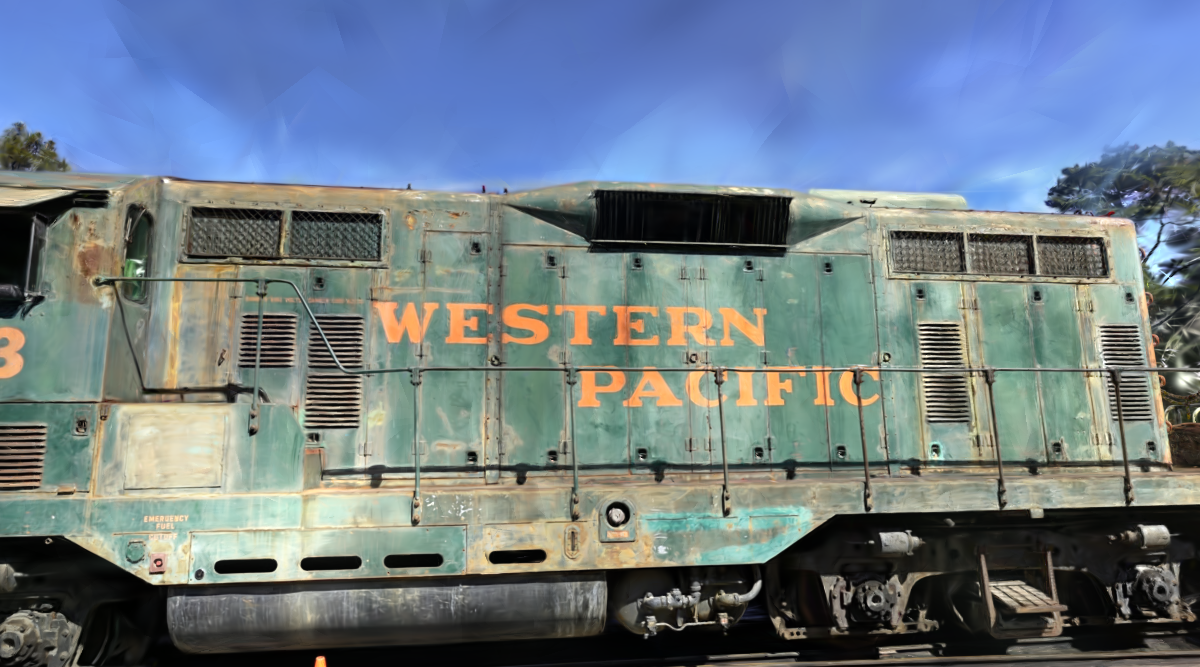}};
      \begin{scope}[x={(image.south east)},y={(image.north west)}]
       \draw[red,thick] (0.8333,0.700) rectangle (0.90667,0.7736);
       \end{scope}
    \end{tikzpicture}
    \caption*{\gs trained}
    \end{subcaptionblock}
    \begin{subcaptionblock}[C]{.24\linewidth}
    \includegraphics[width=\linewidth]{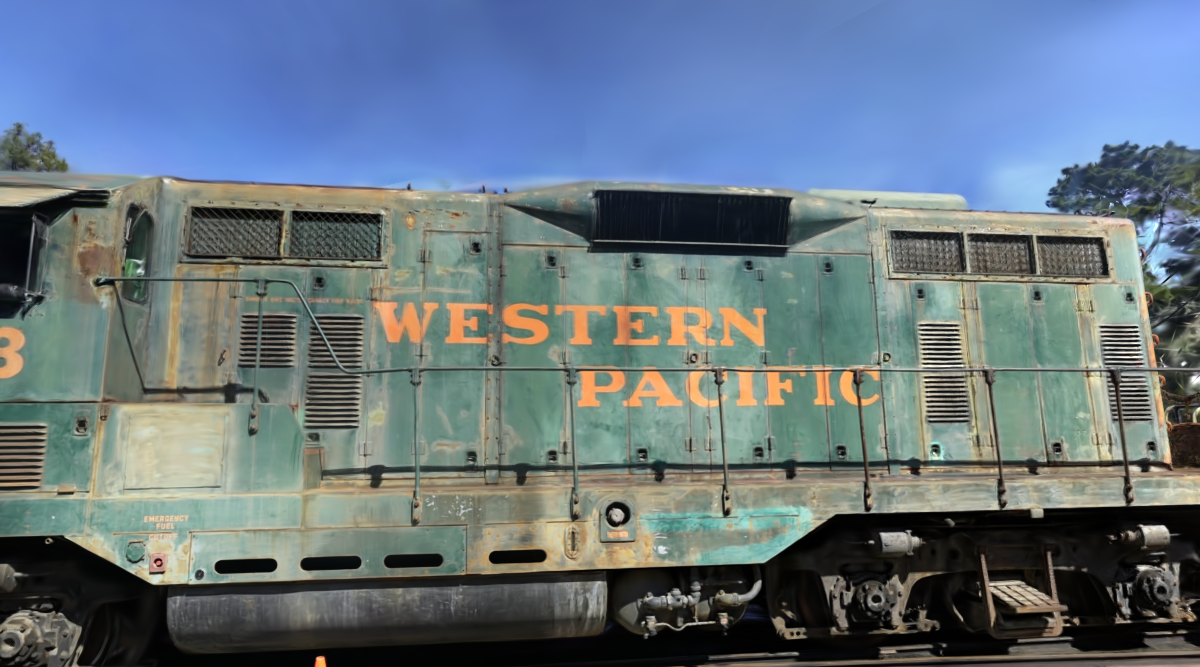}
    \caption*{Ours trained}
    \end{subcaptionblock}
    \begin{subcaptionblock}[C]{.24\linewidth}
    \begin{tikzpicture}
     \node[anchor=south west,inner sep=0] (image) at (0,0) {\includegraphics[width=\linewidth]{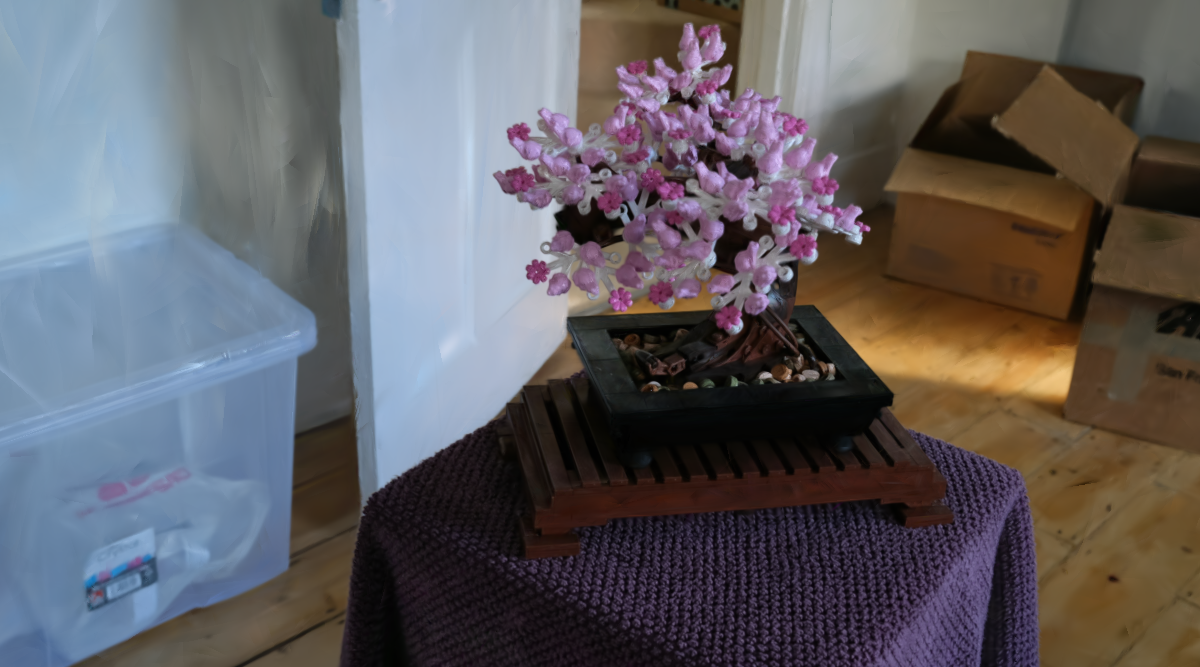}};
      \begin{scope}[x={(image.south east)},y={(image.north west)}]
       \draw[red,thick] (0.225,0.6102) rectangle (0.31167,0.6986);
       \end{scope}
    \end{tikzpicture}
    \caption*{\gs trained}
    \end{subcaptionblock}
    \begin{subcaptionblock}[C]{.24\linewidth}
    \includegraphics[width=\linewidth]{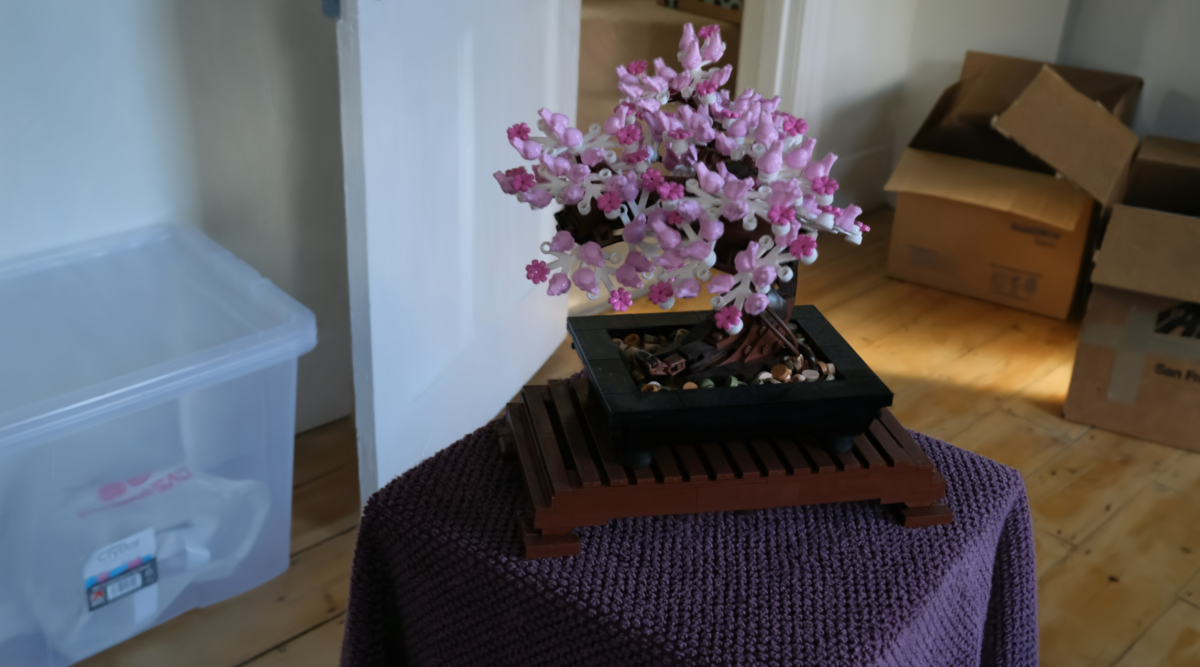}
    \caption*{Ours trained}
    \end{subcaptionblock}

\medskip
\hspace{.025\linewidth}\hrulefill\par
\vspace{6pt}

    \begin{subcaptionblock}[C]{.02\linewidth}
    \rotatebox[origin=c]{90}{\small{\gs rendered}}
    \end{subcaptionblock}
    \begin{subcaptionblock}[C]{.19\linewidth}
    \pdfpxdimen=\dimexpr 1 in/72\relax
    \includegraphics[clip, width=\linewidth, viewport=1000px 467px 1088px 516px]{train_sorted.png}
    \end{subcaptionblock}
    \begin{subcaptionblock}[C]{.19\linewidth}
    \pdfpxdimen=\dimexpr 1 in/72\relax
    \includegraphics[clip, width=\linewidth, viewport=1000px 467px 1088px 516px]{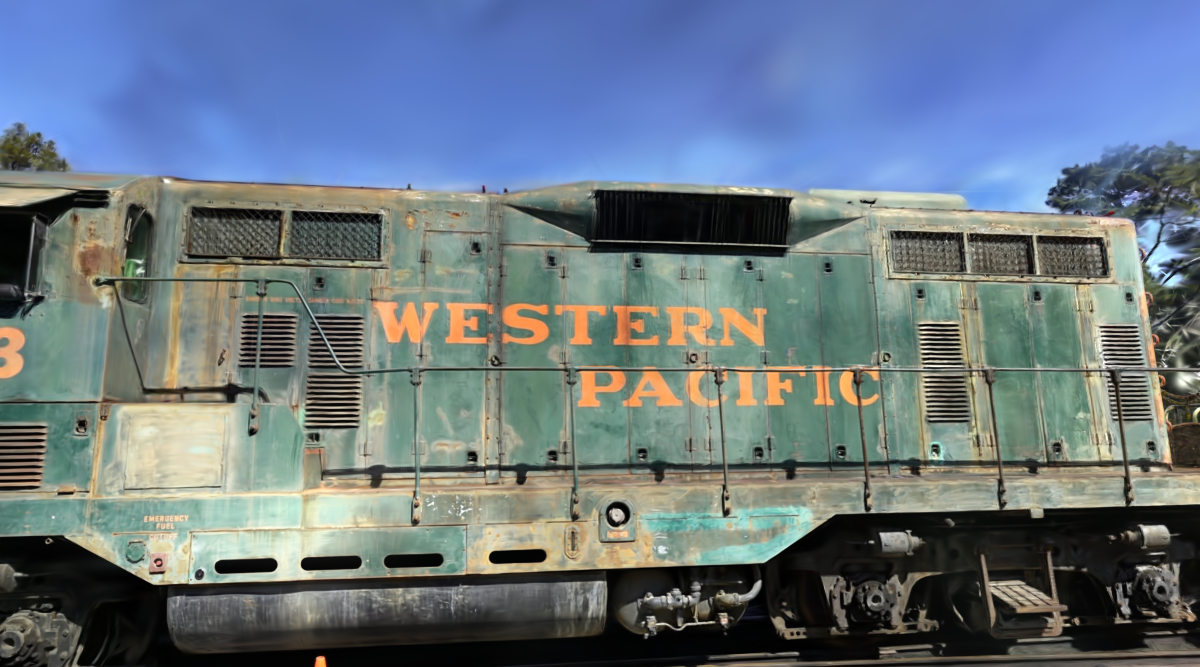}
    \end{subcaptionblock}
    \begin{subcaptionblock}[C]{.19\linewidth}
    \pdfpxdimen=\dimexpr 1 in/72\relax
    \includegraphics[clip, width=\linewidth, viewport=1000px 467px 1088px 516px]{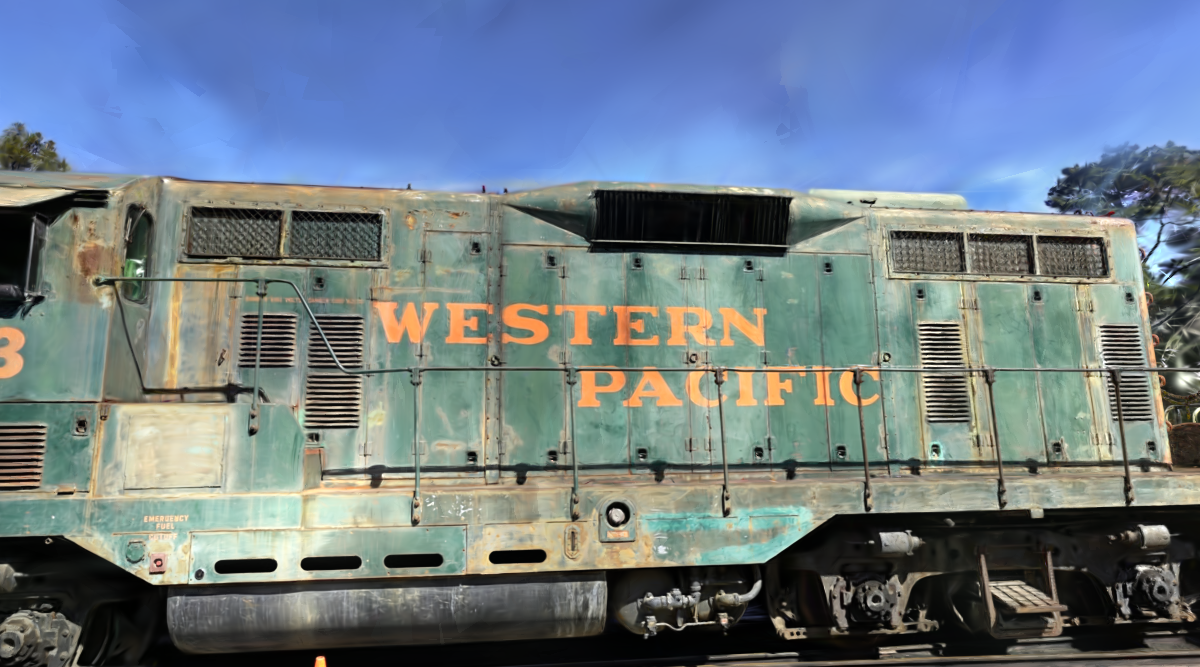}
    \end{subcaptionblock}
    \begin{subcaptionblock}[C]{.19\linewidth}
    \pdfpxdimen=\dimexpr 1 in/72\relax
    \includegraphics[clip, width=\linewidth, viewport=1000px 467px 1088px 516px]{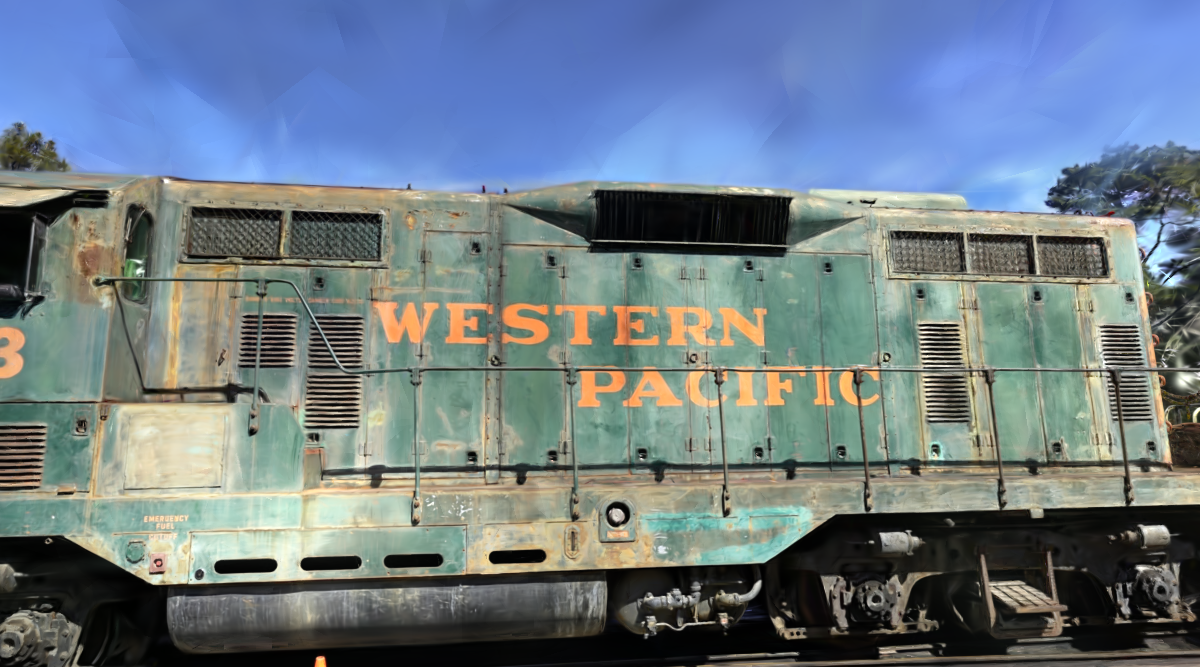}
    \end{subcaptionblock}
    \begin{subcaptionblock}[C]{.19\linewidth}
    \pdfpxdimen=\dimexpr 1 in/72\relax
    \includegraphics[clip, width=\linewidth, viewport=1000px 467px 1088px 516px]{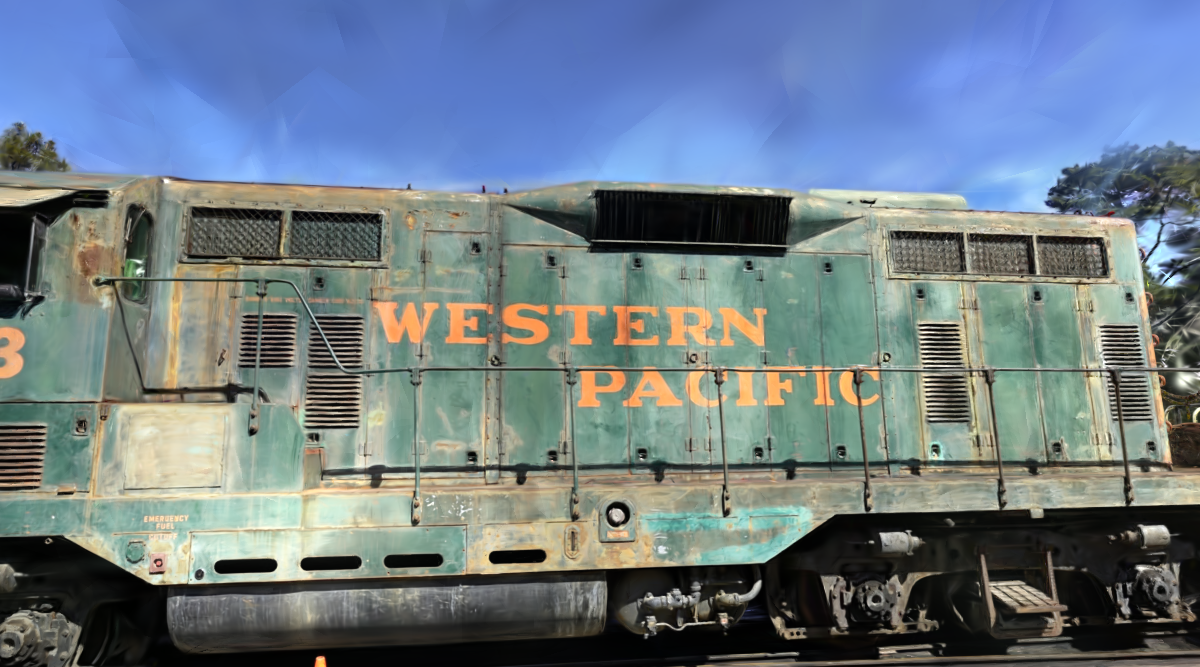}
    \end{subcaptionblock} \\

    \begin{subcaptionblock}[C]{.02\linewidth}
    \rotatebox[origin=c]{90}{\small{Sort Error}}
    \end{subcaptionblock}
    \begin{subcaptionblock}[C]{.19\linewidth}
    \includegraphics[width=\linewidth]{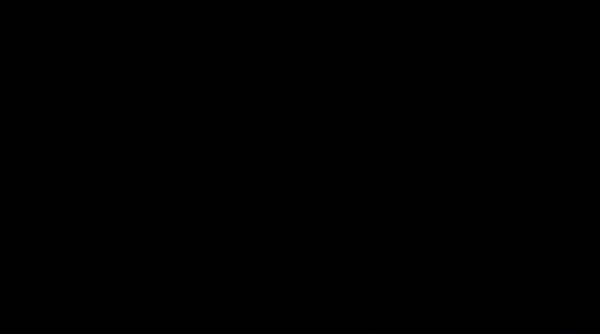}
    \end{subcaptionblock}
    \begin{subcaptionblock}[C]{.19\linewidth}
    \includegraphics[width=\linewidth]{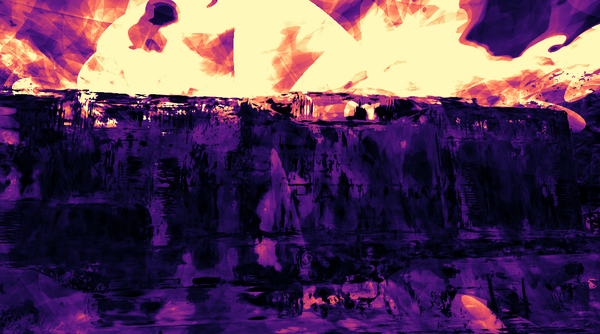}
    \end{subcaptionblock}
    \begin{subcaptionblock}[C]{.19\linewidth}
    \includegraphics[width=\linewidth]{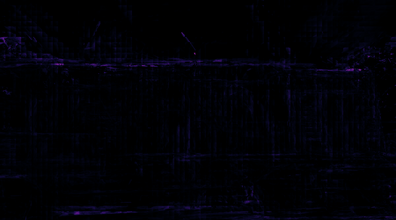}
    \end{subcaptionblock}
    \begin{subcaptionblock}[C]{.19\linewidth}
    \includegraphics[width=\linewidth]{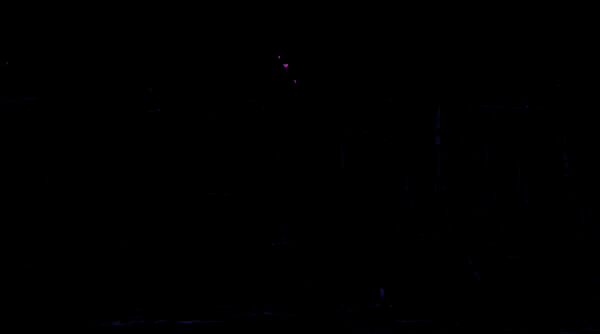}
    \end{subcaptionblock}
    \begin{subcaptionblock}[C]{.19\linewidth}
    \includegraphics[width=\linewidth]{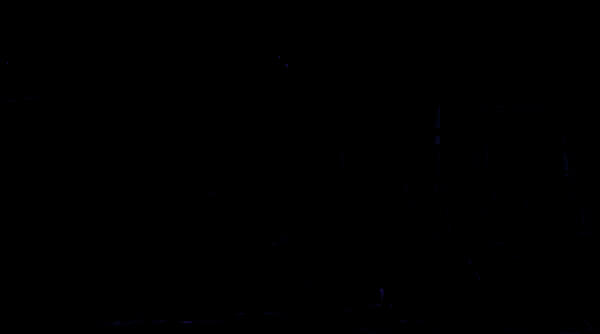}
    \end{subcaptionblock} \\

    \begin{subcaptionblock}[C]{.02\linewidth}
    \rotatebox[origin=c]{90}{\small{\gs rendered}}
    \end{subcaptionblock}
    \begin{subcaptionblock}[C]{.19\linewidth}
    \pdfpxdimen=\dimexpr 1 in/96\relax
    \includegraphics[clip,width=\linewidth, viewport=270px 407px 374px 466px]{bonsai_sorted.png}
    \end{subcaptionblock}
    \begin{subcaptionblock}[C]{.19\linewidth}
    \pdfpxdimen=\dimexpr 1 in/96\relax
    \includegraphics[clip,width=\linewidth, viewport=270px 407px 374px 466px]{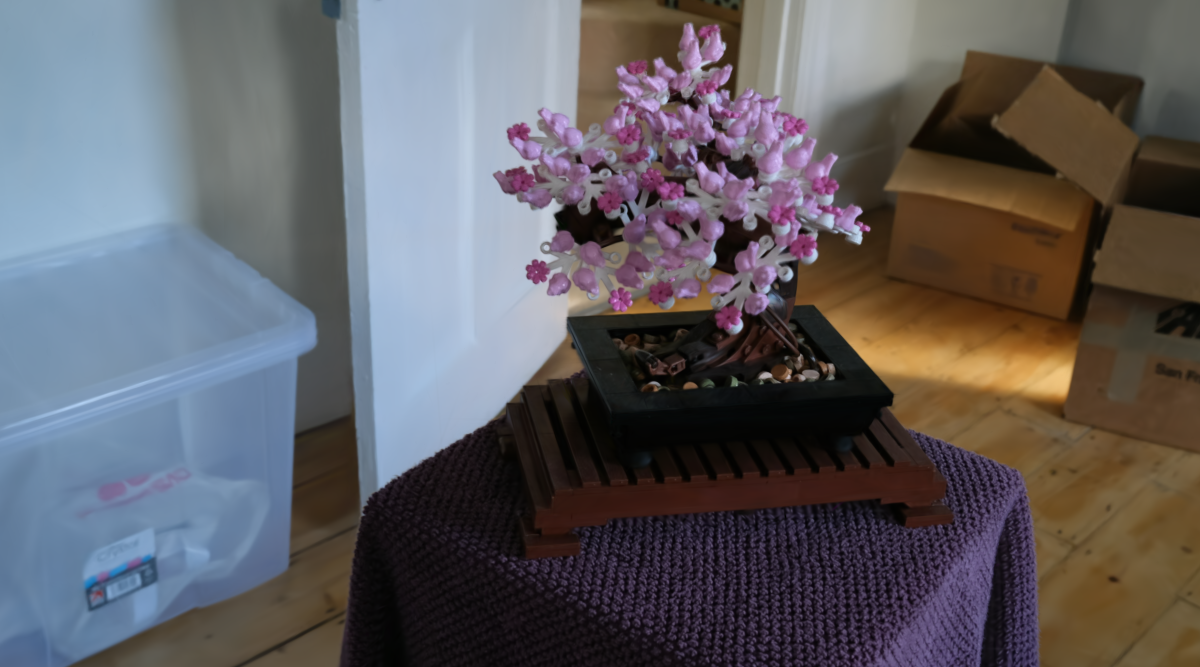}
    \end{subcaptionblock}
    \begin{subcaptionblock}[C]{.19\linewidth}
    \pdfpxdimen=\dimexpr 1 in/96\relax
    \includegraphics[clip,width=\linewidth, viewport=270px 407px 374px 466px]{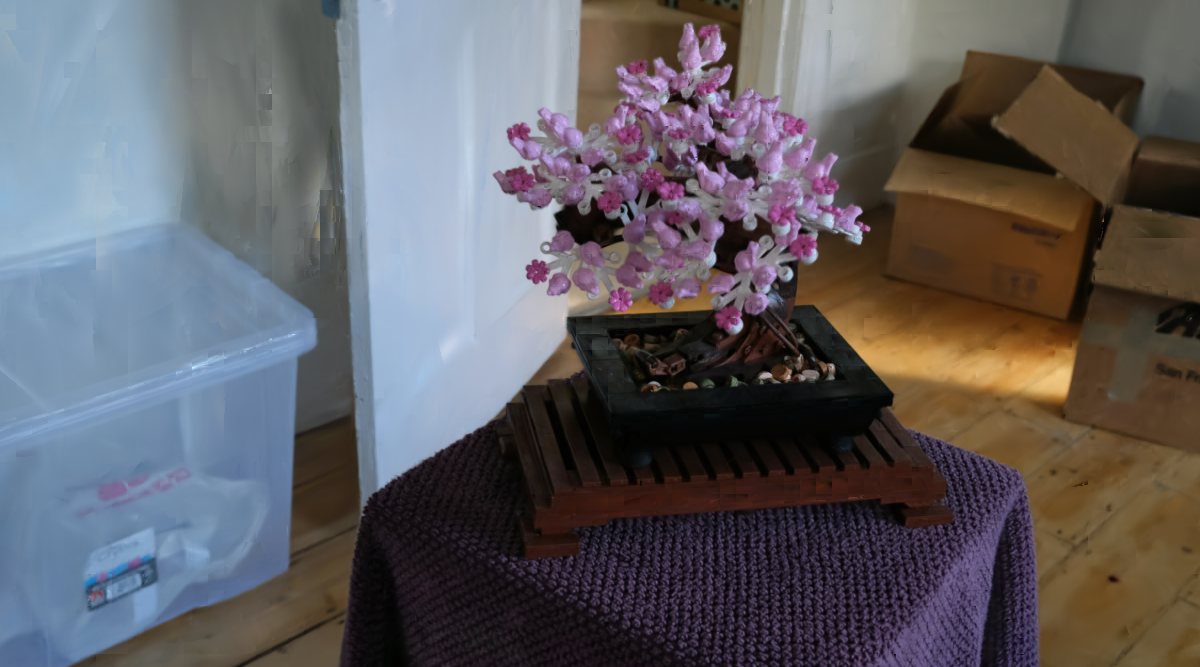}
    \end{subcaptionblock}
    \begin{subcaptionblock}[C]{.19\linewidth}
    \pdfpxdimen=\dimexpr 1 in/96\relax
    \includegraphics[clip,width=\linewidth, viewport=270px 407px 374px 466px]{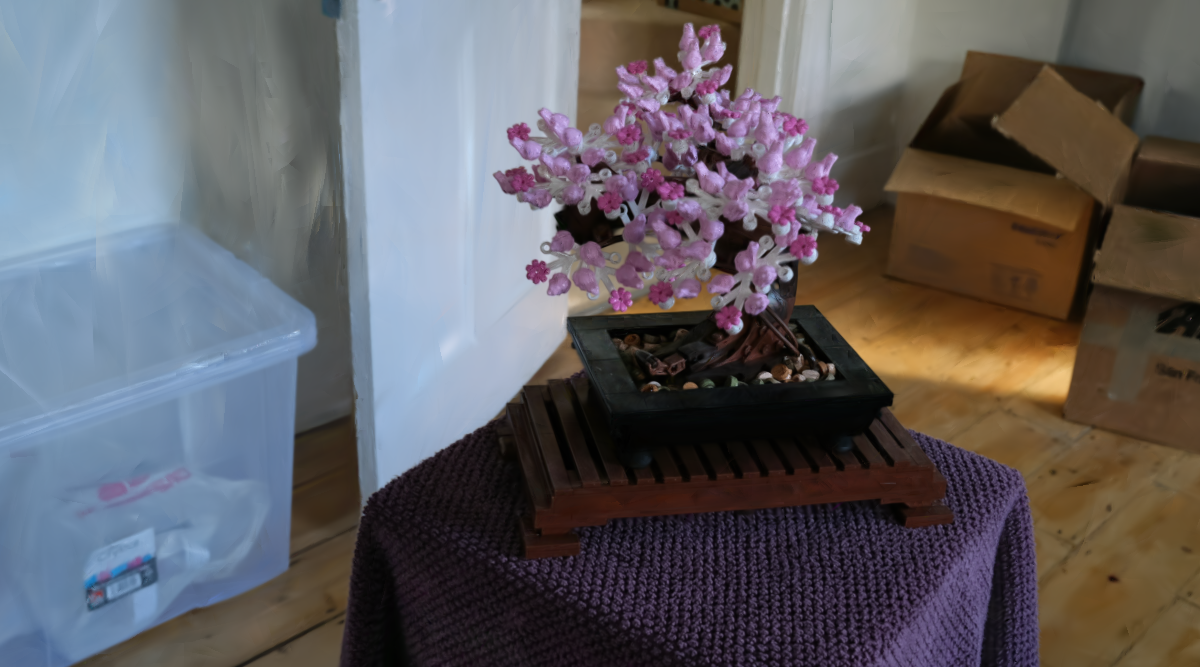}
    \end{subcaptionblock}
    \begin{subcaptionblock}[C]{.19\linewidth}
    \pdfpxdimen=\dimexpr 1 in/96\relax
    \includegraphics[clip,width=\linewidth, viewport=270px 407px 374px 466px]{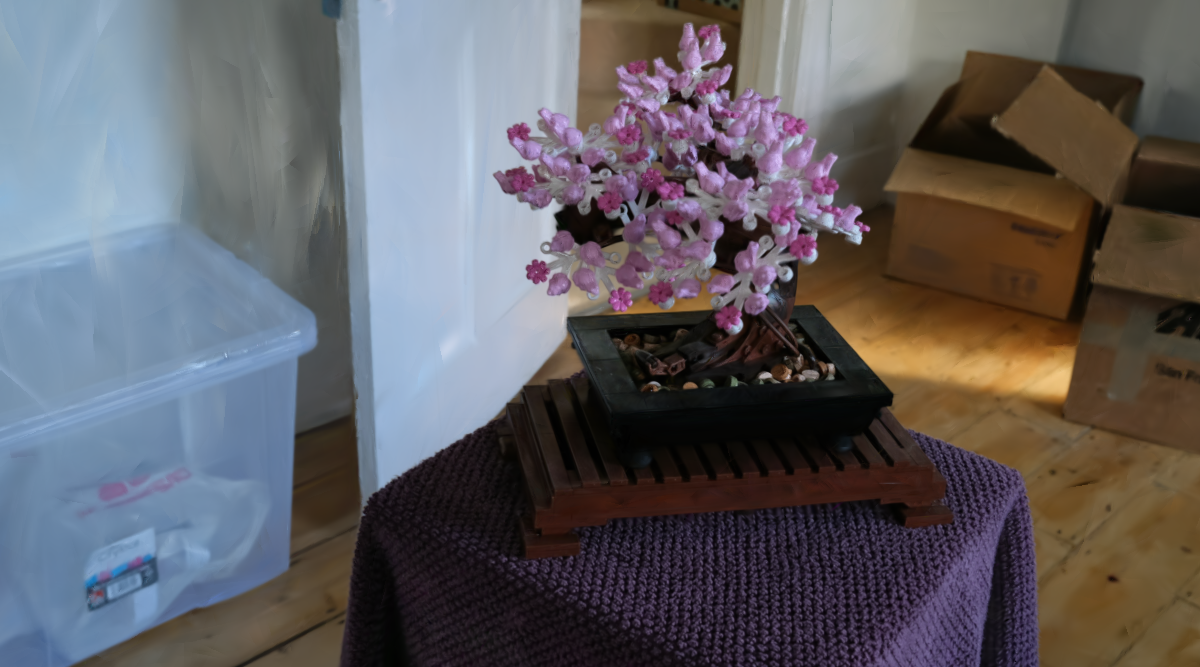}
    \end{subcaptionblock} \\

    \begin{subcaptionblock}[C]{.02\linewidth}
    \rotatebox[origin=c]{90}{\small{Sort Error}}
    \end{subcaptionblock}
    \begin{subcaptionblock}[C]{.19\linewidth}
    \includegraphics[width=\linewidth]{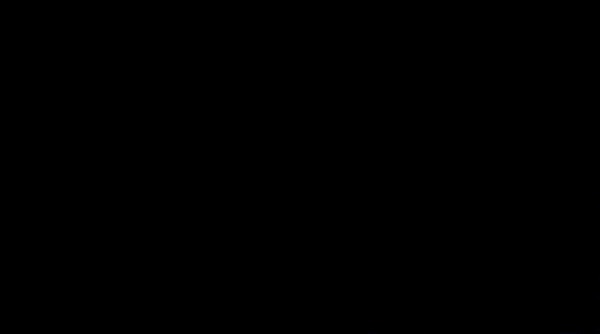}
    \caption*{full sort}
    \end{subcaptionblock}
    \begin{subcaptionblock}[C]{.19\linewidth}
    \includegraphics[width=\linewidth]{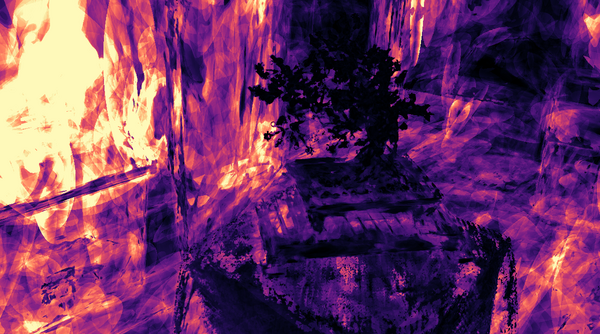}
    \caption*{\gs}
    \end{subcaptionblock}
    \begin{subcaptionblock}[C]{.19\linewidth}
    \includegraphics[width=\linewidth]{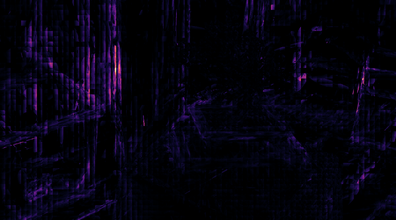}
    \caption*{resort 4}
    \end{subcaptionblock}
    \begin{subcaptionblock}[C]{.19\linewidth}
    \includegraphics[width=\linewidth]{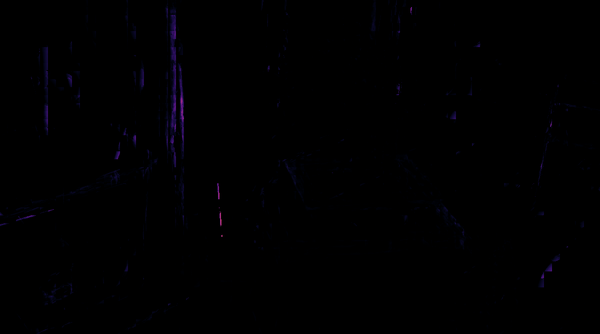}
    \caption*{resort 24}
    \end{subcaptionblock}
    \begin{subcaptionblock}[C]{.19\linewidth}
    \includegraphics[width=\linewidth]{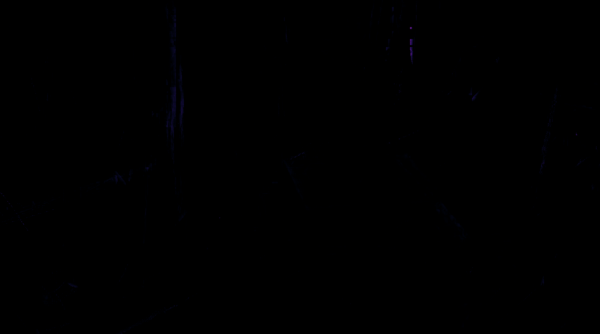}
    \caption*{Ours}
    \end{subcaptionblock} \\
    
    \vspace{2pt}
    \begin{subcaptionblock}[b]{\linewidth}
    \centering
    0 \includegraphics[height=120pt,angle=270,origin=b]{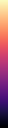} 10.0
    \end{subcaptionblock}
    
    \caption{Correct rendering of a trained \gs scene with per-pixel sorting reveals how \gs cheats with the location of Gaussians. 
    Our approach, on the other hand, considers correct sorting during training and rendering.
    Below, we show the sort error of different resorting windows and our full approach \cf \tabref{sortcost}.
    We intentionally use the trained \gs model here, as our trained version does not show these kinds of artifacts for visualization.
    The error visualization captures the sum over the depth difference of all wrongly sorted neighbors.
    For resorting with a window size of 4, tile artifacts are still visible. 
    Our approach hardly diverges from fully sorted rendering, while running $100\times$ faster; it is also about $5\times$ faster than resort 24 and on average only 4\% slower than \gs.} 
    \label{fig:sort_compare}
\end{figure*}

\gs \cite{kerbl3Dgaussians} performs a global sort based on the view-space $z$-coordinate of each Gaussian's mean $\mu$, see \eqnref{viewspacedepth}.
This leads to a consistent sort order during translation, but not during rotation, as illustrated in \figref{depth}.
While \gs may use this fact during training to introduce differences between views (and thus reduce the loss), it is in general undesirable, as camera rotations can lead to popping artifacts, which are particularly disturbing when inspecting the optimized 3D scene.
Our objective is to stabilize color computations under rotation by splatting Gaussians based on the point of highest contribution along each view ray.
Note that, although we improve rendering consistency, we still approximate true 3D Gaussians, neglecting any overlap between them.

\subsection{Per-pixel Depth and Na\"ive Sorting}
When replacing a 1D Gaussian along the view ray with a Dirac impulse, the mean/maximum of this 1D Gaussian is arguably the best discrete blend location.
This maximum, $t_{opt}$, can be computed from the derivative of the 3D Gaussian along the view ray $\vec{r}(t)=\vec{o}+t\vec{d}$:
\begin{align}
t_{opt} & =\frac{\vec{d}^T\Sigma^{-1}(\vec{\mu}-\vec{o})}{\vec{d}^T\Sigma^{-1}\vec{d}}. \label{eq:dopt}
\end{align}
Please see \appref{depth} for the step-by-step derivation.

Consider a simple 2D case with an isotropic Gaussian $\Sigma^{-1}=\vec I$, the camera at $(0,0)$ and the Gaussian at $\vec \mu = (0,\mu_y)$. 
It is easy to see that the depth function follows a cosine as $\vec d$ is normalized:
\[
t = \frac{\vec d^T \vec I \vec{\mu}}{\vec d^T \vec I \vec d} = d_y \cdot \mu_y = \cos(\theta) \mu_y,
\]
where $\theta$ is the angle of the view ray. Thus, we conclude that there is no simple primitive, like, \eg, a plane to represent the $t_{opt}$ which could be rasterized traditionally, see \figref{depth_ours}.
Therefore, we compute $t_{opt}$ on a per-ray basis.

When reconstructing surfaces, Gaussians often turn very flat, as such, $\Sigma^{-1}$ may become large and lead to instabilities in the computation.
Bounding the entries of $S^{-1}$ to $10^{3}$ removes those instabilities in our experiments, by effectively thickening very thin Gaussians, with minimal impact on the computed depth.

With the computation of $t_{opt}$ in place, we can eliminate all popping artifacts and ensure perfect view-consistency by sorting all Gaussians per ray by their $t_{opt}$ value.
Unfortunately, even the simplest \gs reconstructions consist of tens of thousands of Gaussians, often leading to thousands of potentially contributing Gaussians per view ray.
\new{Furthermore, early ray termination cannot be performed before sorting, as it is dependent on the sort order.}
Even an optimized parallel per-ray sort on top of the original \gs tile-based rasterizer leads to slowdowns of more than $100\times$, not only making the approach impractical for real-time rendering, but also impeding optimization.

\subsection{Per-tile Sorting and Local Resorting}
\begin{figure}[t]
    \centering
    \subcaptionbox{w/o per-tile depth}[0.48\linewidth]{\includegraphics[width=\linewidth,page=3]{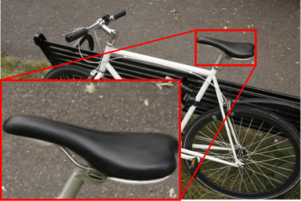}}
    \vspace{0.05cm}
    \subcaptionbox{w/ na\"ive per-tile depth}[0.48\linewidth]{\includegraphics[width=\linewidth,page=1]{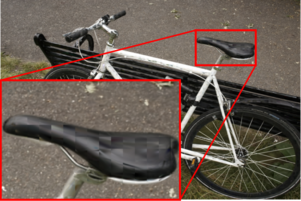}} 
    \caption{Comparison of \gs with and without per-tile depth calculation. Per-tile depth calculation lowers sorting errors ($\delta_{max}=4.01, \delta_{avg}=0.284$ compared to $\delta_{max}=5.43, \delta_{avg}=0.898$). However, doing this without additional per-pixel sorting leads to artifacts at the tile borders.}
    \label{fig:pertiledepth}
\end{figure}

Although it is not possible to describe $t_{opt}$ with a simple primitive for rasterization, we may still rely on the fact that $t_{opt}$ is smooth across neighboring rays.
As such, the sorting order of neighboring rays should also be similar.
Because sorting in \gs already happens with a combined tile/depth key, we could replace the global depth with an accurate per-tile depth value for each Gaussian, \eg, using the tile center ray for \eqnref{dopt}.
As can be seen in \figref{pertiledepth}, using per-tile depth clearly leads to artifacts along the tile borders.

With that in mind, we propose a simple per-ray resorting extension.
Instead of immediately blending the next Gaussian when walking through the tile list, we keep a small resorting window in registers.
When loading a Gaussian, we evaluate its $t_{opt}$ and use insertion sort to place it in the resorting window.
If the window overflows, we blend the sample with the smallest depth.
\new{This simple method follows the idea of $k$-buffers~\cite{callahan2005visibilitysorting, bavoilk2007kbuffer} without fragment merging, which requires the Gaussians along a ray to be nearly-sorted.}
Although this sorting strategy is easy to implement, it already achieves good results for a resorting window of about $16$ to $24$, removing the majority of visible popping artifacts in our tested scenes.
To confirm the improvement in blending order, we compute a per-ray sort error $\delta$:
If two consecutive Gaussians are out of order, we accumulate their difference in $t_{opt}$\revised{}{, detailed in \tabref{sortcost} and \figref{sort_compare}}.
\revised{We present a visual example in \figref{sort_compare}, with corresponding runtimes and $\delta$ in \tabref{sortcost} --- evidently, even though $\delta$ decreases with a larger resorting window, there is a non-negligible increase in runtime}{As can be seen, there is a non-negligible cost even for such a simple approach}.

\begin{figure}[t]
    \centering
    \subcaptionbox{\revised{w/o~tile-based culling}{w/o~culling}}[0.45\linewidth]{\includegraphics[width=\linewidth,page=3]{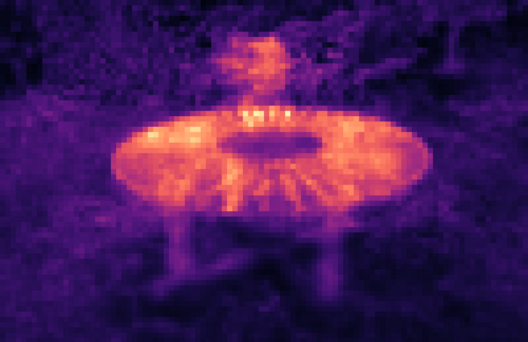}}%
    \vspace{0.1cm}
    \subcaptionbox{\revised{w/~tile-based culling}{w/~culling}}[0.45\linewidth]{\includegraphics[width=\linewidth,page=1]{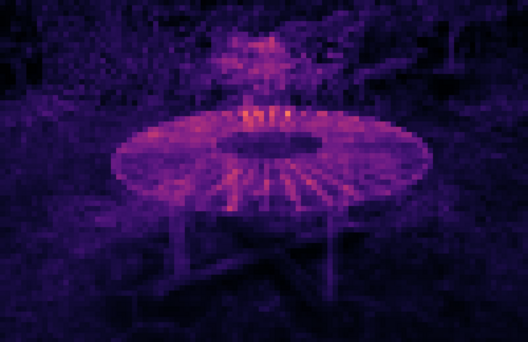}}
    \vspace{0.1cm}
    \includegraphics[height=70pt]{magma_colormap.png}
    \subfloat{\rotatebox{90}{0 \qquad \qquad \qquad 5k}}
    \caption{Number of Gaussians per tile with and without \revised{tile-based culling}{per-tile culling enabled} for the \mipnerf \revised{G}{g}arden scene. 
    The average number of Gaussians per tile is reduced by $\sim44\%$.}
    \label{fig:culling}
\end{figure}

\begin{table}[t]
\small
    \setlength{\tabcolsep}{4pt}
    \centering
    \caption{Maximum sort error over all pixels and average sort error for two representative example views from \figref{sort_compare}.
    A \emph{full} sort per ray increases rendering times (relative to \gs) by more than $100\times$.
    Local resorting with a sort window of $16$ to $24$ removes the majority of visible popping artifacts, yet increases rendering time $2$ to $6\times$.
    Our hierarchical approach improves sort quality further and keeps processing time low.
    Note that a larger sorting window may lead to more Gaussians being fetched and thus our measurement of $\errmax$ may increase with larger sort windows.
    }
    \label{tab:sortcost}
    \begin{tabular}{@{}crcccccccc@{}}
    \toprule
          & & \multirow{2}{*}{\gs}    &  \multirow{2}{*}{Full}   & \multicolumn{4}{c}{Resorting Window} &  \multirow{2}{*}{Ours} \\\cmidrule(lr){5-8}
         & & &  & 4 & 8 & 16 & 24 &  \\
         \midrule
         \parbox[t]{2mm}{\multirow{3}{*}{\rotatebox[origin=c]{90}{{Train}}}}
         & $\errmax$ & 28.445 & \cellcolor{blue!50}0.000 & 5.867 & 3.882 & \cellcolor{blue!10}3.544  & 4.580 & \cellcolor{blue!30}0.575 \\
         & $\erravg$ & 3.688 & \cellcolor{blue!50}0.000 & 0.124 & 0.045 & 0.014 & \cellcolor{blue!10} 0.007 & \cellcolor{blue!30}0.003 \\
         & time\textsubscript{[ms]}  & \cellcolor{blue!30}1.00 & 142.03 & \cellcolor{blue!10}1.21 & 1.66 & 2.70 & 4.22 & \cellcolor{blue!50} 0.92 \\
         \midrule
         \parbox[t]{2mm}{\multirow{3}{*}{\rotatebox[origin=c]{90}{{Bonsai}}}}
        & $\errmax$ & 33.543 & \cellcolor{blue!50}0.000 & 12.786 & 8.954 & 6.391 & \cellcolor{blue!10}5.595 & \cellcolor{blue!30}3.098 \\
        & $\erravg$ & 3.786 & \cellcolor{blue!50}0.000 & 0.265 & 0.110 & 0.039 & \cellcolor{blue!10}0.019 & \cellcolor{blue!30}0.006 \\
        & time\textsubscript{[ms]} & \cellcolor{blue!50}1.00 & 179.70 & \cellcolor{blue!10}1.76 & 2.58 & 4.33 & 6.88 & \cellcolor{blue!30}1.47 \\
         \bottomrule
    \end{tabular}
\end{table}

\subsection{Hierarchical Rendering}

\begin{figure*}
       \centering
       \includegraphics[width=\linewidth]{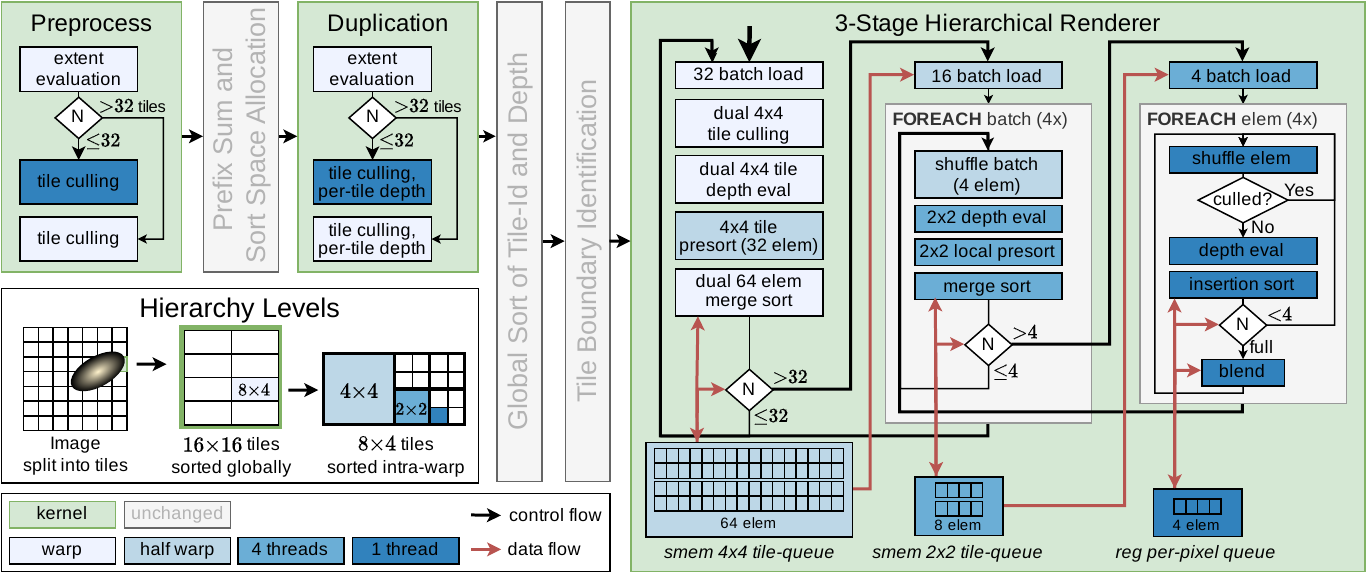}
       \caption{Overview of the detailed steps in our pipeline. We add load balancing, tile culling and per-tile depth evaluation to the first two stages of \gs.
       Our hierarchical rasterizer utilizes three sorted queues, going from \revised{$4\stimes4$}{$4\times4$} tiles over \revised{$2\stimes2$}{$2\times2$} tiles to individual rays.
       \new{The queues store only id and the tile's $t_{opt}$ per Gaussian, while additional information is re-fetched from global memory on demand, and shared between threads via shuffle operations.}
       Depending on the queue fill levels, we switch between different cooperative group sizes while ensuring the queues remain filled for \revised{effective}{appropriate} sorting.
       Our pipeline achieves an \revised{overall}{effective} sorting window of 25-72 elements\revised{}{, depending on the current queue fill levels}.}
       \label{fig:pipeline}
\end{figure*}

Local resorting is already able to significantly improve the per-pixel sort order, which greatly reduces popping artifacts.
To tackle the imposed performance overhead,
we insert additional resorting levels between tiles and individual threads, creating a sort hierarchy.
In this way, we can share sorting efforts between neighboring rays, while incrementally refining the sort order as we move towards individual rays.
By additionally culling non-contributing Gaussians at every level of the hierarchy, we can drastically reduce sorting costs.
We propose a hierarchical rendering pipeline that relies on the innate memory and execution hierarchy of the GPU to minimize the number of memory access operations, as outlined in \figref{pipeline}.
For a fair comparison, we intentionally only alter the blend order of Gaussians and leave the other parts of \gs untouched, including the 2D splatting approximation from~\citet{zwicker2002ewa}.

\paragraph{Tile-based culling} 
We propose a fast tile-based culling approach that bounds Gaussians to exactly those tiles they contribute to.
For each ray, \citet{kerbl3Dgaussians} disregard Gaussians with a contribution below $\epsilon_O = 1/255$, which forms an exact culling condition.
Like \gs, we start with an axis-aligned bounding rectangle using the largest eigenvalue of the 2D covariance matrix to determine which tiles may potentially be touched during both \emph{Preprocess} and \emph{Duplication}.
This conservative estimate gives very large bounds for highly anisotropic Gaussians.

For exact culling, we then calculate the point $\vec{\hat{x}}$ inside each tile $X$ that maximizes the 2D Gaussian's contribution $G_2(\vec x)$:
\begin{equation}
    \vec{\hat{x}}=\argmax_{\vec{x} \in X} G_2(\vec x) = \argmin_{\vec{x} \in X} (\vec{x}-\vec{\mu_2})^T\Sigma_2^{-1}(\vec{x}-\vec{\mu_2}).
\end{equation}
If $\mu_2 \in X \Rightarrow \vec{\hat{x}}=\vec{\mu_2}$.
If $\mu_2 \notin X$, then $\vec{\hat{x}}$ must lie on one of the two tile edges closest to $\mu_2$, due to Gaussians being monotonic along rays pointing away from $\vec{\mu_2}$. 
We can then \revised{compute the maximum along those two edges (similar to \eqnref{dopt}, but in 2D)}{perform line search along those edges} and clamp the resulting values to obtain ${\hat{x}}$ (see the \appref{culling} for the full algorithm).
Finally, we evaluate $G_2(\vec{\hat{x}})$ 
to perform the comparison with $\epsilon_O$, which significantly reduces the number of Gaussians per tile (\cf \figref{culling}).

\paragraph{Tile-depth Adjustment}
For pre-sorting we require a representative $t_{opt}$ per tile.
Intuitively, the center ray of the tile should be a valid compromise for all rays in the tile.
However, this completely ignores the fact that a Gaussian in general does not uniformly contribute to all rays in a tile.
Especially for small Gaussians whose main extent is approximately parallel to the view rays, the center ray may result in depth estimates far away from any contribution made by the Gaussian.

Arguably, the weighted integral $\int_X G_2(\vec x) t_{opt}(\vec x) d\vec{x}$ is a better estimate.
Yet, even a numerical approximation considering all rays in the tile $X$ is too compute-intensive.
Thus, we approximate it with a single sample: the one with the highest weight within a tile, $\ie$, $\vec{\hat{x}}$.
Since $\vec{\hat{x}}$ was already calculated during culling, we only need to construct the corresponding ray to evaluate $t_{opt}$.
The optimized depth location reduces $(\errmax, \erravg)$ from $(1.553$, $0.006)$ to $(0.575$, $0.003)$ and $(3.917$, $0.014)$ to $(3.098$, $0.006)$ for the views in \tabref{sortcost}.

\paragraph{Load Balancing} 
Similar to other compute-mode rasterization methods, primitives that cover a large portion of the screen may become an issue if a single thread evaluates their coverage. 
For \gs, this is the case in the first two stages of the rendering pipeline, which operate on a per-Gaussian basis.
For our method, tile-culling and per-tile depth calculations increase the workload of these stages, which further amplifies this problem.

To remedy this issue, we propose a two-stage load balancing scheme:
In the first phase, each thread responsible for a Gaussian which covers fewer than a predetermined maximum number of tiles, performs its own processing.
We empirically determined that a maximum of \revised{$32$ tiles}{$32$} results in good performance.
Most threads are typically idle after this initial phase. 
In the second phase, we distribute the remaining workload within each warp using warp voting and shuffle instructions.
\revised{For close-ups and high-resolution rendering, where single Gaussians often cover a large portion of the screen}{Especially for close-ups}, our approach can speed up \emph{Preprocess} and \emph{Duplication} by up to $10\times$.

\paragraph{Hierarchically Sorted Rendering}
With the goal of establishing a hierarchical rendering \revised{pipeline, a na\"ive approach is to design one kernel per hierarchy level}{approach, relying on one kernel per hierarchy level may seem attractive}.
However, such an approach would require communication via slow global memory between the levels and would prohibit early ray termination after reaching the opacity threshold.
Thus, we opt for combining the final three levels of our rendering hierarchy in a single kernel,
where multiple threads cooperatively sort and manage shared queues, as detailed in \figref{pipeline}.
We use a large \revised{$4\stimes4$ tile-queue of 64 elements}{tail-queue of 64 elements for $4\times4$ tiles} (managed by 16 threads), feeding into four eight-element \revised{$2\stimes2$ tile-queues}{mid-queues, each representing a $2\times2$ tile}.
Finally, each \revised{$2\stimes2$ tile-queue}{$2\times2$ tile} feeds into four {\revised{per-pixel queues}{head-queues}} with four elements, \revised{managed by one thread each}{each being kept for a single ray and managed by one thread}.
For one \revised{$16\stimes16$}{$16\times16$} tile, we thus start 256 threads, allocate 16 \revised{$4\stimes4$ tile-queues}{tail-queues} and 64 \revised{$2\stimes2$ tile-queues}{mid-queues} in shared memory as well as one \revised{per-pixel queue}{head-queue} per thread in registers. \new{Each queue stores only the Gaussian's \emph{id} and the current level's depth $t_{opt}$. 
Additional information is loaded on demand from global memory and shared between threads of the hierarchical level via shuffle operations, \eg $\mu, \Sigma^{-1}$ during depth calculation, or $\mu_2, \Sigma_2^{-1}$ during culling and blending. 
}

The queues follow a \emph{push} methodology to keep queue fill rates as high as possible, ensuring that resorting remains effective.
While 16 threads (a \emph{halfwarp}) are assigned to each \revised{$4\stimes4$ tile-queue}{tail-queue}, we load and feed batches of 32 into \revised{two $4\stimes4$ tile-queues}{tail-queues} at once, allowing all threads within a warp to load data together.
\revised{}{Thus, each thread is responsible for inserting Gaussians into two \revised{$4\stimes4$ tile-queues}{tail-queues}.} 
\revised{After loading}{At first}, each thread performs tile\new{-based} culling (as described before, \revised{but}{just} for a $4\stimes4$ tile), followed by computing $t_{opt}$. 
For culled Gaussians, we set $t_{opt} = \infty$.
Then, each halfwarp sorts the 32 newly loaded elements using Batcher Merge Sort~\cite{batcher1968sorting} before writing them to the back of the {\revised{$4\stimes4$ tile-queue}{tail-queue}}. 
Typically, there are now two individually sorted parts in the \revised{$4\stimes4$ tile-queue}{tail-queue}: the already present elements (up to 32) and the newly added (up to 32).
As both are sorted, we use efficient merge sort to combine them.
Culled Gaussians are now at the back of the queue and \revised{can be discarded}{we can discard them}.

While there are more than 32 elements in the {\revised{$4\stimes4$ tile-queue}{tail-queue}}, we push batches of size 16 into the \revised{$2\stimes2$ tile-queue}{mid-queue}.
Each thread in the halfwarp re-fetches the data needed for computing $t_{opt}$ for a single Gaussian.
Each group of four threads then pushes sub-batches of size four into their \revised{$2\stimes2$ tile-queue}{mid-queue}, relying on shuffle instructions to update $t_{opt}$ for \revised{each $2\stimes2$}{the $4\stimes4$} tile.
We follow the same approach as before: we sort the four new entries according to depth, for which we use a simple coordination using shuffle instructions.
We then use merge sort to combine the new elements with the existing ones.

After the \revised{$2\stimes2$ tile-queue}{mid-queue} is filled, we draw four elements from it and insert them into the \revised{per-pixel queue}{head-queue}.
Again, we batch-load the needed data using the four threads assigned to the respective \revised{$2\stimes2$ tile-queue}{mid-queue}, and again use shuffle instructions to communicate all relevant information for each Gaussian to all other threads in the $2\stimes2$ tile.
We evaluate $t_{opt}$ and $\alpha$ for the respective rays and insert the newly computed data into the \revised{per-pixel queue}{head-queue}.
If the Gaussian's $\alpha$ is below $\epsilon_O$, we simply discard it.
As we add elements one by one into the \revised{per-pixel queue}{head-queue}, we rely on simple insertion sort.
Only if the \revised{per-pixel queue}{head-queue} is full, we take one element from it and perform blending, freeing up space for the next element from the \revised{$2\stimes2$ tile-queue}{mid-queue}.

Due to the hierarchical structure, we effectively construct an overall sort window varying between \revised{$25$}{$24$} and $72$, where the minimum is hit if the \revised{$4\stimes4$ tile-queue}{tail-queue} is drained down to $17$ elements, \revised{with $4$ elements remaining in the other queues}{$4$ elements each remain in mid and head}.
$72$ elements are sorted if we fill the \revised{$4{\stimes}4$ tile-queue with}{tail-queue} $64$ elements and then move $4$ elements through the half-filled \revised{$2\stimes2$ tile-queue}{mid-queue} and the filled \revised{per-pixel queue}{head-queue}.
While our sort setup typically achieves better sorting than a simple per-thread sort window of \revised{$25$}{$24$}, we may occasionally achieve worse sorting, as the \revised{higher-level}{lower} queues are shared between threads.

\new{
The sizes of the three queues are variable, with some restrictions.
The $4\stimes4$ tile-queue size is constrained to $32n+32$, with $n\in\mathbb{Z}^+$, as this enables efficient warp-wide merge sort.
Similarly, the $2\stimes2$ tile-queue must be of size $4m+4$, with $m\in\mathbb{Z}^+$, as it is managed by four threads.
The per-pixel queue size can be chosen arbitrarily.
We heuristically decided on $(64/8/4)$ for the three queue sizes, as this achieves a large enough sort window, while limiting shared memory requirements and register pressure, ultimately leading to better performance.
We provide ablations for our chosen queue sizes and load balancing thresholds in \appref{performance}.
}

\subsection{Backward Pass}
Contrary to \gs, we perform gradient computations in front-to-back blending order, avoiding the large memory overhead required for storing per-pixel sorted Gaussians---which would be needed to restore the correct blending order\revised{}{(\cf supplemental for details). In our experiments, this did not impact the stability of the gradient computations}.

\new{
Gradient computation in \gs, independent of direction, requires the final accumulated transmittance $T_{N_{\vec{r}}} = \prod_{j=1}^{N_r} (1 - \alpha_j)$ and the final per-pixel color $C(\vec{r})$.
To compute gradients for the $i$-th Gaussian along a view ray, we require the contribution of all subsequently blended Gaussians.
Crucially, rather than accumulating the contribution of subsequent Gaussians back-to-front, we use subtraction and division, \ie
\begin{align}
    \sum_{j=i+1}^{N_{\vec{r}}}\vec{c}_j \alpha_j \prod_{k=1}^{j-1}(1-\alpha_k) &= C(\vec{r}) - C_i, \\
    \prod_{k=i}^{N_{\vec{r}}}(1-\alpha_k) &= \frac{T_{N_{\vec{r}}}}{T_i},
\end{align}
where $C_i$ is the accumulated output color including the $i$-th Gaussian in front-to-back order. 
As we perform the same rendering routine as in the forward pass, including early stopping, the backward pass is equally efficient.
Note that this does not change the stability of the gradient computations; \gs also relies on a division.
Arguably, our approach may even lead to more accurate gradients as the Gaussians blended first along a ray have a higher contribution to the final color and computing those first, will accumulate less floating point errors compared to reversing the computation.
}

\new{
It is imperative that the same exact sort order is used between forward and backward pass to ensure correct gradients.
Like \gs, we keep the global sort order in memory, which ensures that potentially equal depth values do not lead to different sorting results. 
In our implementation, we use stable sorting routines throughout: 
Batcher Merge Sort~\cite{batcher1968sorting} is stable by design, our merge sort routines rely on each thread's rank to establish sort orders among equal depths, and our insertion sort is trivially stable.
}

\begin{table*}[ht!]
    \centering
    \footnotesize
    \setlength{\tabcolsep}{4pt}
    \caption{Image metrics for our method, \gs and related work.
    Results with dagger ($\dagger$) are reproduced from~\citet{kerbl3Dgaussians} to facilitate cross-method comparisons. 
    Our quality is comparable to \gs. 
    With Opacity Decay, our approach loses slightly less quality than \gs.
    }
    \label{tab:image_metrics}
    \begin{tabular}{lcccccccccccccccc}
    \toprule
    Dataset & \multicolumn{4}{c}{Deep Blending}  & \multicolumn{4}{c}{\mipnerf Indoor}   & \multicolumn{4}{c}{\mipnerf Outdoor}  & \multicolumn{4}{c}{Tanks \& Temples} \\
    Metric & PSNR\textsuperscript{$\uparrow$} & SSIM\textsuperscript{$\uparrow$} & LPIPS\textsuperscript{$\downarrow$} &  \FLIP\textsuperscript{$\downarrow$}         &  PSNR\textsuperscript{$\uparrow$} & SSIM\textsuperscript{$\uparrow$} & LPIPS\textsuperscript{$\downarrow$} &  \FLIP\textsuperscript{$\downarrow$}              & PSNR\textsuperscript{$\uparrow$} & SSIM\textsuperscript{$\uparrow$} & LPIPS\textsuperscript{$\downarrow$} & \FLIP\textsuperscript{$\downarrow$}               & PSNR\textsuperscript{$\uparrow$} & SSIM\textsuperscript{$\uparrow$} & LPIPS\textsuperscript{$\downarrow$} &  \FLIP\textsuperscript{$\downarrow$}  \\
    \midrule
    \revised{Mip-NeRF 360}{}\textsuperscript{$\dagger$} & 29.40 & \cellcolor{blue!15} 0.900 & \cellcolor{blue!15} 0.245 & 0.138 & \cellcolor{blue!45} 31.57 & 0.914 & \cellcolor{blue!45} 0.182 & \cellcolor{blue!45} 0.088 & 24.42 & 0.691 & 0.286 & 0.170 & 22.22 & 0.758 & 0.256 & 0.232 \\
    \revised{Instant-NGP (base)}{}\textsuperscript{$\dagger$} & 23.62 & 0.797 & 0.423 & 0.258 & 28.65 & 0.840 & 0.281 & 0.120 & 22.63 & 0.536 & 0.444 & 0.203 & 21.72 & 0.723 & 0.330 & 0.245 \\
    \revised{Instant-NGP (big)}{}\textsuperscript{$\dagger$} & 24.96 & 0.817 & 0.390 & 0.222 & 29.14 & 0.863 & 0.241 & 0.114 & 22.75 & 0.567 & 0.403 & 0.200 & 21.92 & 0.745 & 0.304 & 0.241 \\
    \revised{Plenoxels}{}\textsuperscript{$\dagger$} & 23.09 & 0.794 & 0.425 & 0.244 & 24.84 & 0.765 & 0.366 & 0.182 & 21.69 & 0.513 & 0.467 & 0.229 & 21.09 & 0.719 & 0.344 & 0.262 \\
    3DGS & \cellcolor{blue!15} 29.46 & 0.900 & 0.247 & \cellcolor{blue!15} 0.131 & \cellcolor{blue!30} 30.98 & \cellcolor{blue!45} 0.922 & \cellcolor{blue!15} 0.189 & \cellcolor{blue!30} 0.094 & \cellcolor{blue!30} 24.59 & \cellcolor{blue!30} 0.727 & \cellcolor{blue!30} 0.240 & \cellcolor{blue!30} 0.167 & \cellcolor{blue!45} 23.71 & \cellcolor{blue!45} 0.845 & \cellcolor{blue!30} 0.178 & \cellcolor{blue!45} 0.199 \\
    Ours & \cellcolor{blue!45} 29.86 & \cellcolor{blue!30} 0.904 & \cellcolor{blue!45} 0.234 & \cellcolor{blue!30} 0.127 & \cellcolor{blue!15} 30.62 & \cellcolor{blue!30} 0.921 & \cellcolor{blue!30} 0.186 & 0.099 & \cellcolor{blue!45} 24.60 & \cellcolor{blue!45} 0.728 & \cellcolor{blue!45} 0.235 & \cellcolor{blue!45} 0.167 & \cellcolor{blue!15} 23.21 & \cellcolor{blue!30} 0.843 & \cellcolor{blue!45} 0.173 & 0.216 \\
    \midrule
    3DGS (Opacity Decay) & 28.94 & 0.894 & 0.262 & 0.134 & 30.57 & \cellcolor{blue!15} 0.918 & 0.198 & \cellcolor{blue!15} 0.097 & 24.45 & 0.718 & 0.261 & \cellcolor{blue!15} 0.169 & \cellcolor{blue!30} 23.52 & 0.839 & 0.194 & \cellcolor{blue!30} 0.205 \\
    Ours (Opacity Decay)& \cellcolor{blue!30} 29.84 & \cellcolor{blue!45} 0.905 & \cellcolor{blue!30} 0.241 & \cellcolor{blue!45} 0.126 & 30.03 & 0.917 & 0.194 & 0.103 & \cellcolor{blue!15} 24.46 & \cellcolor{blue!15} 0.722 & \cellcolor{blue!15} 0.254 & 0.169 & 23.18 & \cellcolor{blue!15} 0.839 & \cellcolor{blue!15} 0.184 & \cellcolor{blue!15} 0.214 \\
    \bottomrule
    \end{tabular}
\end{table*}

\section{Evaluation}
For evaluation, we follow \citet{kerbl3Dgaussians} and use 13 real-world scenes from \mipnerf~\cite{barron2022mipnerf360}, Deep Blending~\cite{DeepBlending2018} and Tanks \& Temples~\cite{Knapitsch2017}.

\paragraph{Opacity Decay}
A viable approach to reduce the total number of Gaussians after optimization is replacing \gs's opacity reset with a standard Opacity Decay during training.
Every $50$ iterations, we multiply each Gaussian's opacity with a constant $\epsilon_{\text{decay}} = 0.9995$.
We find that this modification results in significantly fewer, but larger Gaussians, potentially causing exacerbated popping.

\subsection{Quantitative Evaluation}
\paragraph{Image Metrics}
For our quantitative evaluation, we report PSNR, SSIM, LPIPS~\cite{zhang2018unreasonable} and \FLIP~\cite{Andersson2020Flip} in \tabref{image_metrics}.
\new{To facilitate cross-method comparisons, we reproduce the results from~\citet{kerbl3Dgaussians} for Mip-NeRF 360~\cite{barron2022mipnerf360}, Instant-NGP~\cite{mueller2022instant} and Plenoxels~\cite{fridovich2022plenoxels}.} 
For Deep Blending and \mipnerf Outdoor, we outperform \gs.
For Tanks \& Temples and \mipnerf Indoor, our model performs slightly worse, which we attribute to \gs's ability to fake view-dependent effects with popping.
When enabling Opacity Decay, which results in ${\new{\sim}}50\%$ fewer Gaussians, our method retains more quality than \gs.
In general, our approach performs comparably to \gs in terms of standard image quality metrics.

\paragraph{Popping}

View inconsistencies between subsequent frames, such as popping, cannot be detected with standard image quality metrics. 
To detect such artifacts, we follow recent best practice in 3D style transfer~\cite{Nguyen2022Snerf} and measure \new{the} consistency between novel views and warped novel views with optical flow~\cite{Lai_2018_ECCV}.
\new{While ground-truth images or videos may seem attractive, they vary significantly in location and thus view-dependent effects or only exist for a small subset of our used datasets.}
For our method and \gs, we capture videos from three separate camera paths per scene, exhibiting both rotation and translation.
We then \new{directly} warp each frame $\vec{F}_{i}$ to a subsequent frame $\vec{F}_{i+t}$ \new{with offset $t$} using optical flow predictions from state-of-the-art RAFT~\cite{teed2020raft}.

\begin{figure*}[h!]
\centering
     \begin{minipage}[t]{0.24\linewidth}
        \centering
        \text{Non-warped view $\vec{F}_{i+1}$}
    \end{minipage}
     \begin{minipage}[t]{0.24\linewidth}
        \centering
        \text{Warped view $\hat{\vec{F}}_{i+1}$}
    \end{minipage}
     \begin{minipage}[t]{0.24\linewidth}
        \centering
        \text{$\flipview{1}$}
    \end{minipage}
     \begin{minipage}[t]{0.24\linewidth}
        \centering
        \text{$\text{MSE}(\vec{F}_{i+1}, \hat{\vec{F}}_{i+1})$}
    \end{minipage} \par
\vspace{4pt}
\centering
    \includegraphics[width=0.24\linewidth]{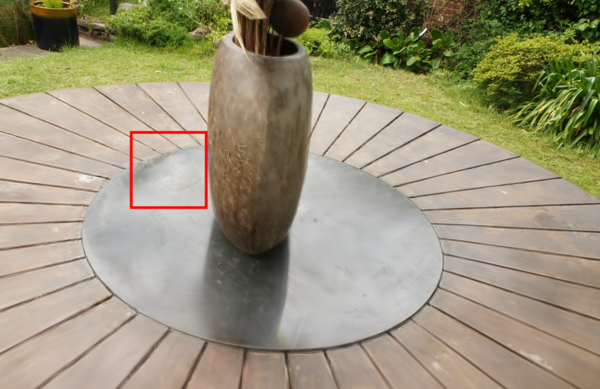}
    \includegraphics[width=0.24\linewidth]{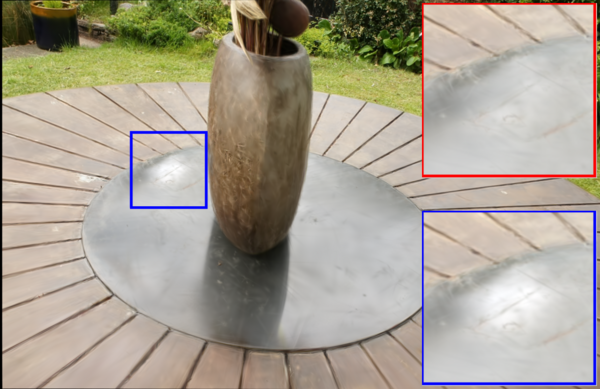}
    \includegraphics[width=0.24\linewidth]{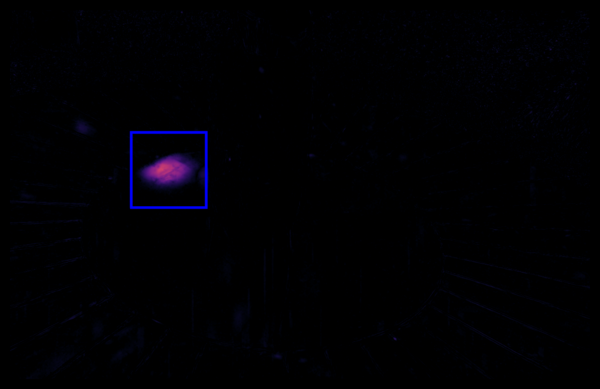}
    \includegraphics[width=0.24\linewidth]{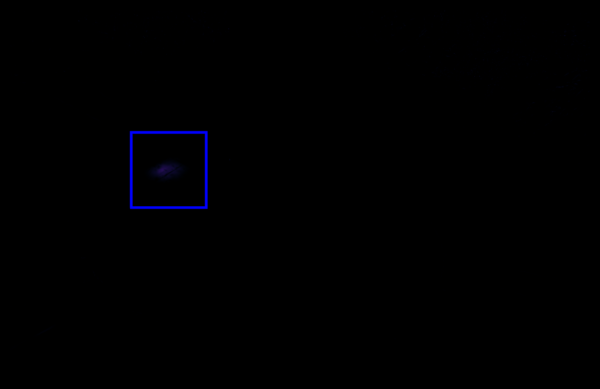}
    \caption{Comparison between \FLIP and MSE to measure differences between \revised{rendered}{non-warped} frames $\vec{F}_{i+1}$ and warped frames $\hat{F}_{i+1}$ for \gs.
    \revised{Notably, using MSE does not yield large errors even when disturbing popping artefacts are encountered --- \FLIP, on the other hand, weighs such artifacts accordingly.
    }{Notable, even egregious popping artefacts do not result in a large MSE for this specific frame.
    On the other hand, \FLIP yields a large error.}}
    \label{fig:flip_consistency_mse}
\end{figure*}

Measuring the error between \new{the} warped frame $\hat{F}_{i+t}$ and the \revised{rendered}{actual} frame $\vec{F}_{i+t}$ with MSE does not prove effective to detect popping artifacts \new{(see~\figref{flip_consistency_mse})}.
MSE tends to weigh small inaccuracies that originate from warping higher than popping artifacts.
\FLIP~\cite{Andersson2020Flip} proves significantly more reliable in our experiments, as it approximates the difference perceived by humans when flipping between images \new{--- a scenario in which popping artifacts are particularly disturbing}.
For each frame, we calculate a consistency error $E_{i+t} = \text{\FLIP}(\hat{F}_{i+t}, \vec{F}_{i+t})$. 
For each video, consisting of $\revised{N_F}{N}$ frames, we then compute the mean \FLIP error as
\begin{equation}
    \flipview{t} = \frac{1}{\revised{N_F}{N}-t}\sum_{i=0}^{\revised{N_F}{N}-t} E_{i+t}.
\end{equation}
Note that the error metric includes a base error floor due to dis-occlusions under translation and correct view-dependent shading.
\revised{
To mitigate these issues, we use an occlusion detection method from~\citet{ruder2016artistic}, do not consider the outermost $20$ pixels, and subtract the per-pixel minimum $\flipview{t}$ score --- clearly, this does not perturb the inter-method differences.
}{
Thus, a zero score is not possible, and averages over all pixels and frames are considered.
Therefore, low score differences may already point towards significant popping.}

We use $t=1$ and $t=7$ to measure short-range and long-range consistency, \revised{following~\citet{Nguyen2022Snerf}}{respectively}.
\tabref{view-consistency} shows our obtained results.
\revised{The large margins, particularly for $\flipview{7}$, highlight that our method is more view-consistent than \gs}{Our method is more view-consistent than \gs, especially when looking at $\flipview{7}$ results}.
We argue that $\flipview{7}$ is a more reliable metric, allowing errors due to popping to accumulate over multiple frames, as can be seen in~\figref{flip_consistency}.
Please see the supplementary video for further evidence. 
With Opacity Decay, our approach achieves virtually identical results, indicating that our method can handle large Gaussians.
For \gs, popping is significantly increased, indicating that \gs may increase the number of Gaussians to hide imperfections in the renderer, while our approach achieves comparable view-consistency scores.

\begin{table}[ht!]
    \centering
    \footnotesize
    \setlength{\tabcolsep}{3.8pt}
    \caption{View-consistency metrics for videos. 
    We measure $\flipview{t}$ for timesteps \revised{$t \in \{1,7\}$}{$t=1, t=7$} (lower is better).
    Our method outperforms \gs with and without Opacity Decay.
    }
    \label{tab:view-consistency}
    \begin{tabular}{lcccccccc}
    \toprule
    Dataset& \multicolumn{2}{c}{DB}  & \multicolumn{2}{c}{M360 Indoor}   & \multicolumn{2}{c}{M360 Outdoor}  & \multicolumn{2}{c}{T\&T} \\
    Metric & $\flipview{1}$ & $\flipview{7}$ & $\flipview{1}$ & $\flipview{7}$ & $\flipview{1}$ & $\flipview{7}$ & $\flipview{1}$ & $\flipview{7}$ \\
    \midrule
    &\multicolumn{8}{c}{{Without Opacity Decay}} \\\cmidrule{2-9}
    3DGS & 0.0061 & 0.0114 & 0.0069 & 0.0134 & \cellcolor{blue!45} 0.0083 & 0.0148 & 0.0102 & 0.0286 \\
    Ours & \cellcolor{blue!45} 0.0053 & \cellcolor{blue!45} 0.0059 & \cellcolor{blue!45} 0.0060 & \cellcolor{blue!45} 0.0077 & 0.0085 & \cellcolor{blue!45} 0.0122 & \cellcolor{blue!45} 0.0076 & \cellcolor{blue!45} 0.0113 \\
    \midrule
    &\multicolumn{8}{c}{{With Opacity Decay}} \\\cmidrule{2-9}
    3DGS & 0.0063 & 0.0122 & 0.0072 & 0.0149 & \cellcolor{blue!45} 0.0083 & 0.0154 & 0.0107 & 0.0315 \\
    Ours & \cellcolor{blue!45} 0.0052 & \cellcolor{blue!45} 0.0055 & \cellcolor{blue!45} 0.0060 & \cellcolor{blue!45} 0.0073 & 0.0083 & \cellcolor{blue!45} 0.0115 & \cellcolor{blue!45} 0.0076 & \cellcolor{blue!45} 0.0114 \\
    \bottomrule
    \end{tabular}
\end{table}

\new{
\paragraph{Depth Evaluation.}
\gs enables efficient extraction of depth values $\depth \in \mathbb{R}_+$ with volumetric rendering:
\begin{equation}
    \depth = \sum_{i=1}^{N_{\vec{r}}} \phi(\mu_i; \vec o, \vec d)\  \alpha_i \prod_{j=1}^{i-1}(1-\alpha_j),
\end{equation}
where $\phi(\cdot)$ describes the depth of a single Gaussian with location $\mu$ (in \gs's case, $\phi(\mu; \vec o, \vec d) = \|\mu - \vec o\|_2$).
Clearly, this depth estimate is dependent on the sort order, leading to problems for \gs's approximate global sort.
Our approach improves sort quality and places 2D splats at the points of maximum contribution ($\phi(\mu; \vec o, \vec d) = t_{opt}$, \cf \eqnref{dopt}).
}

\new{
We establish a metric to compare multi-view consistency in depth estimates, leveraging the sparse point cloud $\mathcal{P} = \{\vec{p}_i\}$ from COLMAP~\cite{schönberger2016sfm}, which serves as initialization for \gs.
If $\vec{p}_i$ is visible from a camera with position $\vec o$, we reconstruct the estimated location $\hat{\vec{p}}_i = \vec o + \depth \cdot \vec d$, with rendered depth $\depth$ and view direction $\vec d$ of the corresponding pixel of $\vec{p}_i$.
The black background for real-world scenes used by \gs enables cheating by not fully accumulating opacity and blending the background color.
For a fair comparison, if any of the tested methods has $T_{N_{\vec{r}}} > 1 \times 10^{-2}$ for a point $\vec{p}_i$, we do not consider this point in our set of tested points $\bar{\mathcal{P}}$.
To minimize errors due to resolution, we render at the resolution used for COLMAP when computing $\hat{\vec{p}}_i$.
Finally, we establish the depth error $E_{depth}$ as 
\begin{equation}
    E_{depth} = \frac{1}{|\bar{\mathcal{P}}|}\sum_{\vec{p}_i \in \bar{\mathcal{P}}} \|\hat{\vec{p}}_i - \vec{p}_i\|_2.
\end{equation}
}

\new{
We compute $E_{depth}$ for all test set views and report our results in \tabref{depth-consistency}.
On average, our method achieves better scores than \gs, especially for the outdoor scenes of Mip-NeRF 360~\cite{barron2022mipnerf360}.
Opacity decay leads to significantly fewer and larger Gaussians, resulting often in lower accumulated opacity and, consequently, more discarded points.
Both methods achieve better results for $E_{depth}$ in this case, as these removed points often correspond to the background, where depth estimates are generally less precise.
}

\begin{table}
    \centering
    \small
    \setlength{\tabcolsep}{4pt}
    \caption{\new{Depth-consistency metric $E_{depth}$ for 3D points $\bar{\mathcal{P}}$ from COLMAP~\cite{schönberger2016sfm} (lower is better).
    We report the mean results over all test set views.
    Our method outperforms \gs with and without Opacity Decay
    In total, we consider $|\bar{\mathcal{P}}|=17404$ points without opacity decay and $|\bar{\mathcal{P}}|=11306$ with opacity decay.}
    }
    \label{tab:depth-consistency}
    \begin{tabular}{lccccc}
    \toprule
    Dataset& {DB}  & {M360 Indoor}   & {M360 Outdoor}  & {T\&T} & Average\\\midrule
    Method &\multicolumn{5}{c}{{Without Opacity Decay}} \\\cmidrule{2-6}
    3DGS & 0.133 & \cellcolor{blue!45} 0.219 & 0.764 & 1.108 & 0.552 \\
    Ours & \cellcolor{blue!45} 0.122 & 0.242 & \cellcolor{blue!45} 0.387 & \cellcolor{blue!45} 0.947 & \cellcolor{blue!45} 0.388 \\
    \midrule
    &\multicolumn{5}{c}{{With Opacity Decay}} \\\cmidrule{2-6}
    3DGS & 0.095 & \cellcolor{blue!45} 0.127 & 0.637 & 0.967 & 0.447 \\
    Ours & \cellcolor{blue!45} 0.073 & 0.168 & \cellcolor{blue!45} 0.408 & \cellcolor{blue!45} 0.916 & \cellcolor{blue!45} 0.361 \\
    \bottomrule
    \end{tabular}
\end{table}

\begin{figure*}[ht!]
\centering

    \begin{subcaptionblock}[C]{.02\linewidth}
    \rotatebox[origin=c]{90}{ }
    \end{subcaptionblock}
     \begin{minipage}[t]{0.19\linewidth}
        \centering
        \text{View $\vec{F}_i$}
    \end{minipage}
     \begin{minipage}[t]{0.19\linewidth}
        \centering
        \text{$\flipview{1}$}
    \end{minipage}
     \begin{minipage}[t]{0.19\linewidth}
        \centering
        \text{Warped View $\hat{F}_{i+1}$}
    \end{minipage}
     \begin{minipage}[t]{0.19\linewidth}
        \centering
        \text{$\flipview{7}$}
    \end{minipage}
     \begin{minipage}[t]{0.19\linewidth}
        \centering
        \text{Warped View $\hat{F}_{i+7}$}
    \end{minipage} \par
\vspace{4pt}

    \begin{subcaptionblock}[C]{.02\linewidth}
    \rotatebox[origin=c]{90}{\gs}
    \end{subcaptionblock}
    \begin{subcaptionblock}[C]{.19\linewidth}
    \pdfpxdimen=\dimexpr 1 in/72\relax
    \includegraphics[width=\linewidth]{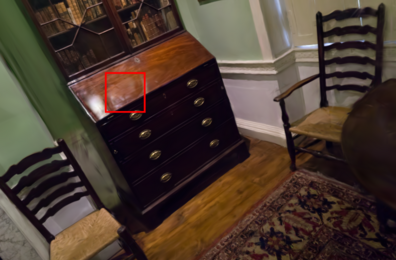}
    \end{subcaptionblock}
    \begin{subcaptionblock}[C]{.19\linewidth}
    \includegraphics[width=\linewidth]{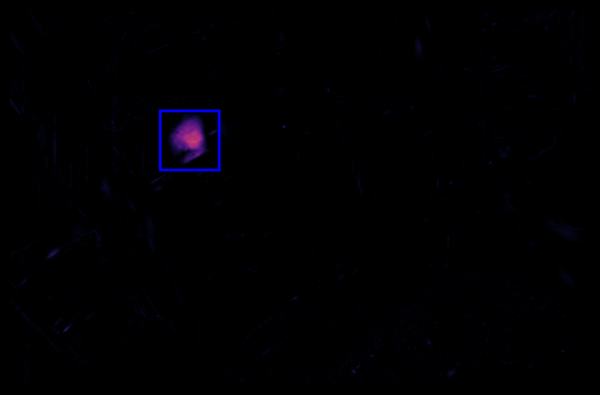}
    \end{subcaptionblock}
    \begin{subcaptionblock}[C]{.19\linewidth}
    \pdfpxdimen=\dimexpr 1 in/72\relax
    \includegraphics[width=\linewidth]{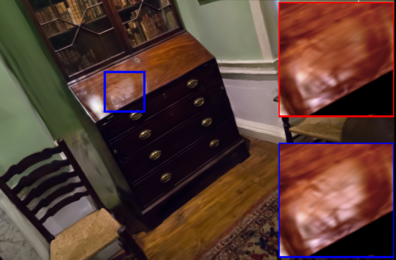}
    \end{subcaptionblock}
    \begin{subcaptionblock}[C]{.19\linewidth}
    \includegraphics[width=\linewidth]{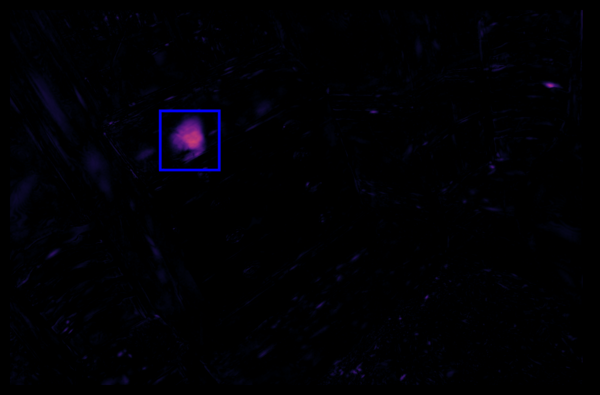}
    \end{subcaptionblock}
    \begin{subcaptionblock}[C]{.19\linewidth}
    \includegraphics[width=\linewidth]{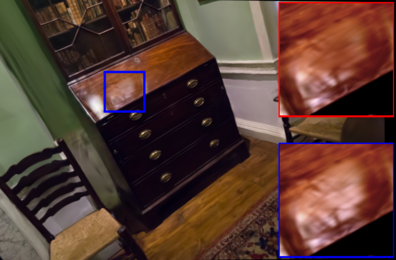}
    \end{subcaptionblock} \\

    \begin{subcaptionblock}[C]{.02\linewidth}
    \rotatebox[origin=c]{90}{Ours}
    \end{subcaptionblock}
    \begin{subcaptionblock}[C]{.19\linewidth}
    \pdfpxdimen=\dimexpr 1 in/72\relax
    \includegraphics[width=\linewidth]{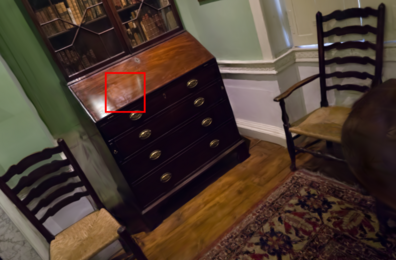}
    \end{subcaptionblock}
    \begin{subcaptionblock}[C]{.19\linewidth}
    \includegraphics[width=\linewidth]{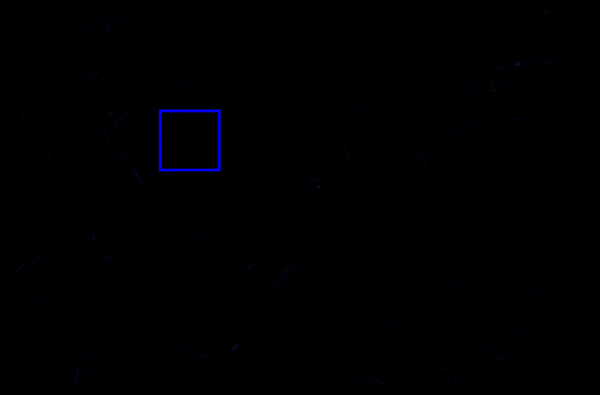}
    \end{subcaptionblock}
    \begin{subcaptionblock}[C]{.19\linewidth}
    \pdfpxdimen=\dimexpr 1 in/72\relax
    \includegraphics[width=\linewidth]{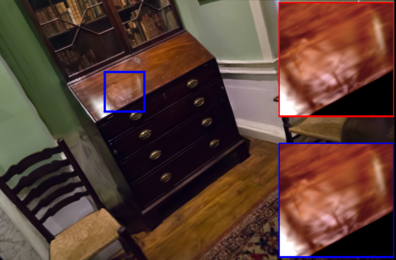}
    \end{subcaptionblock}
    \begin{subcaptionblock}[C]{.19\linewidth}
    \includegraphics[width=\linewidth]{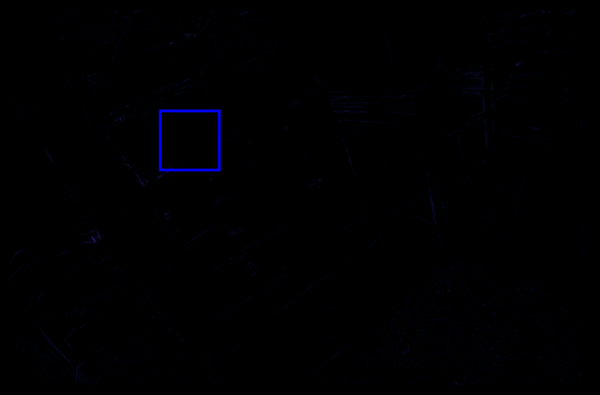}
    \end{subcaptionblock}
    \begin{subcaptionblock}[C]{.19\linewidth}
    \includegraphics[width=\linewidth]{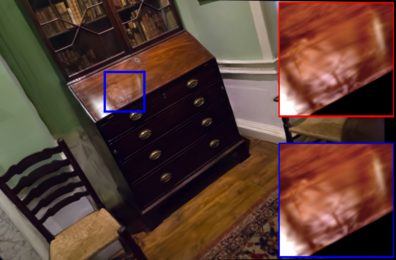}
    \end{subcaptionblock} \par
    \vspace{2pt}

    \begin{subcaptionblock}[C]{.02\linewidth}
    \rotatebox[origin=c]{90}{\gs}
    \end{subcaptionblock}
    \begin{subcaptionblock}[C]{.19\linewidth}
    \pdfpxdimen=\dimexpr 1 in/72\relax
    \includegraphics[width=\linewidth]{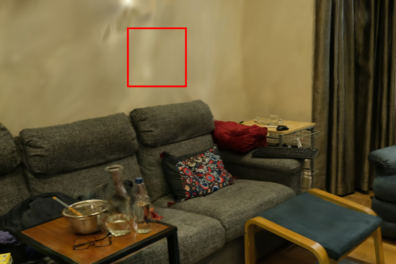}
    \end{subcaptionblock}
    \begin{subcaptionblock}[C]{.19\linewidth}
    \includegraphics[width=\linewidth]{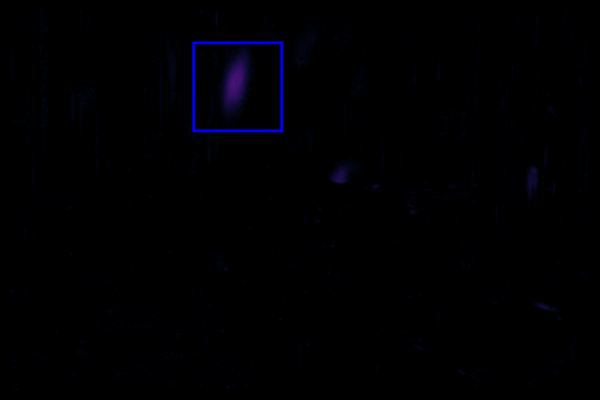}
    \end{subcaptionblock}
    \begin{subcaptionblock}[C]{.19\linewidth}
    \pdfpxdimen=\dimexpr 1 in/72\relax
    \includegraphics[width=\linewidth]{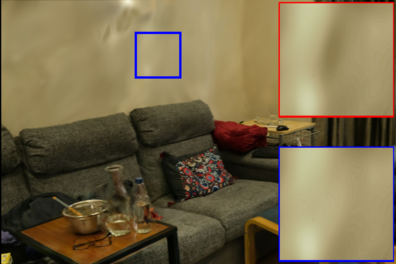}
    \end{subcaptionblock}
    \begin{subcaptionblock}[C]{.19\linewidth}
    \includegraphics[width=\linewidth]{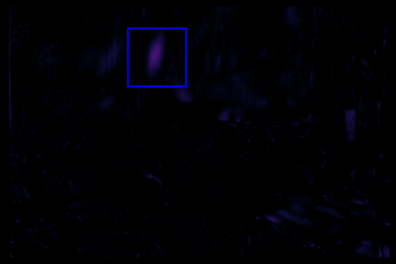}
    \end{subcaptionblock}
    \begin{subcaptionblock}[C]{.19\linewidth}
    \includegraphics[width=\linewidth]{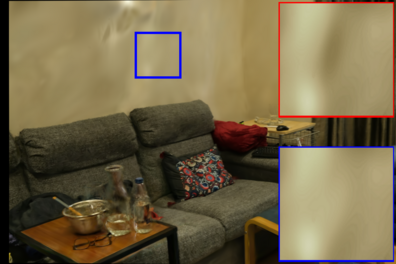}
    \end{subcaptionblock} \\
    
    \begin{subcaptionblock}[C]{.02\linewidth}
    \rotatebox[origin=c]{90}{Ours}
    \end{subcaptionblock}
    \begin{subcaptionblock}[C]{.19\linewidth}
    \pdfpxdimen=\dimexpr 1 in/72\relax
    \includegraphics[width=\linewidth]{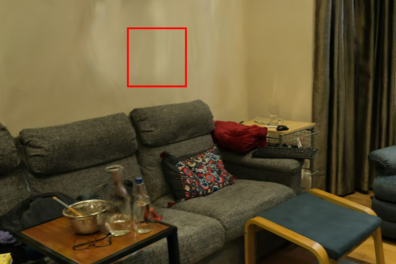}
    \end{subcaptionblock}
    \begin{subcaptionblock}[C]{.19\linewidth}
    \includegraphics[width=\linewidth]{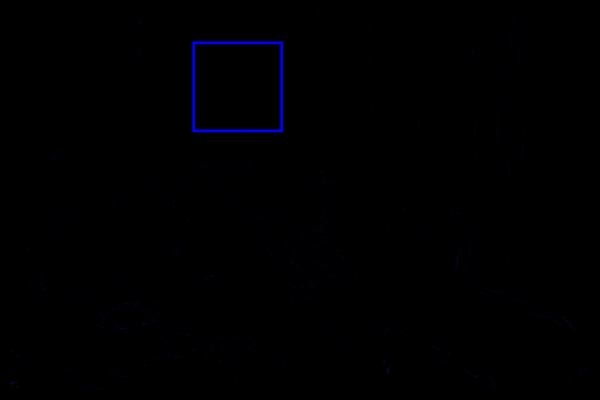}
    \end{subcaptionblock}
    \begin{subcaptionblock}[C]{.19\linewidth}
    \pdfpxdimen=\dimexpr 1 in/72\relax
    \includegraphics[width=\linewidth]{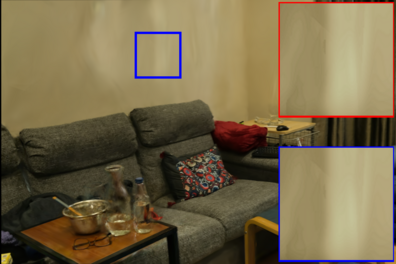}
    \end{subcaptionblock}
    \begin{subcaptionblock}[C]{.19\linewidth}
    \includegraphics[width=\linewidth]{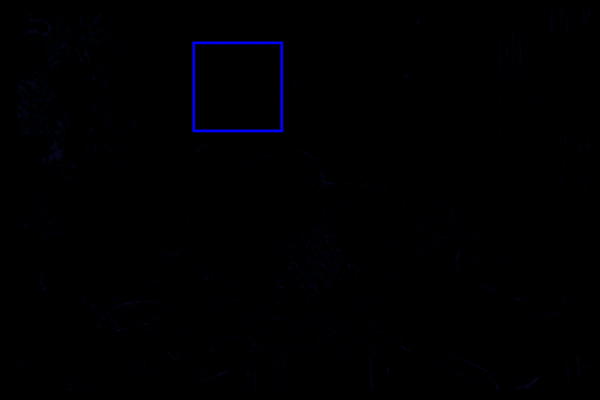}
    \end{subcaptionblock}
    \begin{subcaptionblock}[C]{.19\linewidth}
    \includegraphics[width=\linewidth]{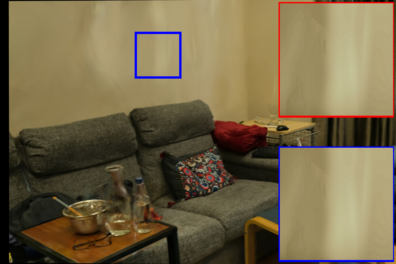}
    \end{subcaptionblock} \par
    \vspace{2pt}

    \begin{subcaptionblock}[C]{.02\linewidth}
    \rotatebox[origin=c]{90}{\gs}
    \end{subcaptionblock}
    \begin{subcaptionblock}[C]{.19\linewidth}
    \pdfpxdimen=\dimexpr 1 in/72\relax
    \includegraphics[width=\linewidth]{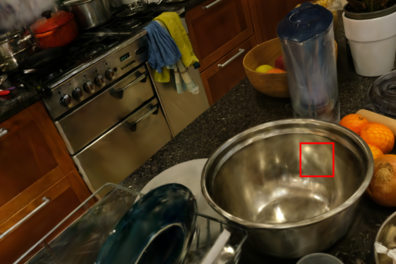}
    \end{subcaptionblock}
    \begin{subcaptionblock}[C]{.19\linewidth}
    \includegraphics[width=\linewidth]{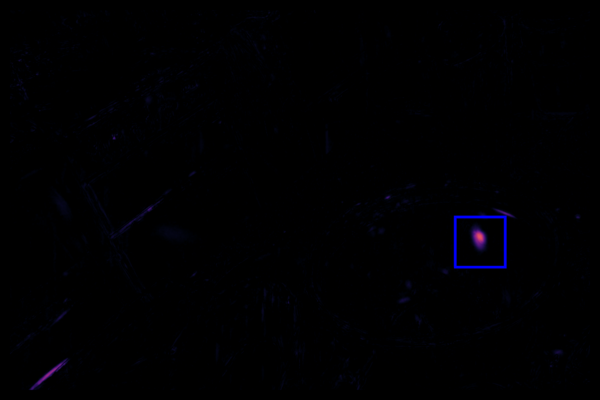}
    \end{subcaptionblock}
    \begin{subcaptionblock}[C]{.19\linewidth}
    \pdfpxdimen=\dimexpr 1 in/72\relax
    \includegraphics[width=\linewidth]{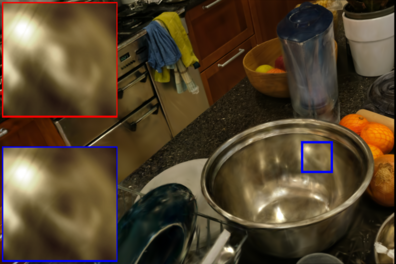}
    \end{subcaptionblock}
    \begin{subcaptionblock}[C]{.19\linewidth}
    \includegraphics[width=\linewidth]{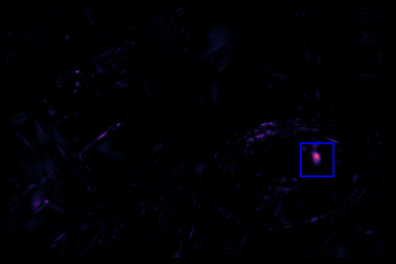}
    \end{subcaptionblock}
    \begin{subcaptionblock}[C]{.19\linewidth}
    \includegraphics[width=\linewidth]{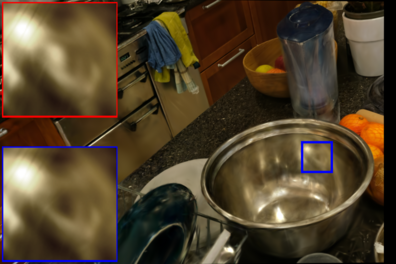}
    \end{subcaptionblock} \\

    \begin{subcaptionblock}[C]{.02\linewidth}
    \rotatebox[origin=c]{90}{Ours}
    \end{subcaptionblock}
    \begin{subcaptionblock}[C]{.19\linewidth}
    \pdfpxdimen=\dimexpr 1 in/72\relax
    \includegraphics[width=\linewidth]{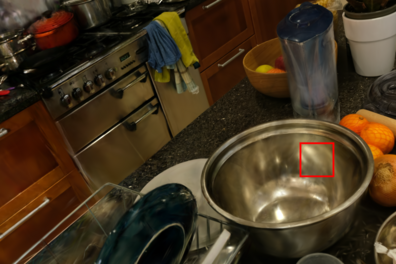}
    \end{subcaptionblock}
    \begin{subcaptionblock}[C]{.19\linewidth}
    \includegraphics[width=\linewidth]{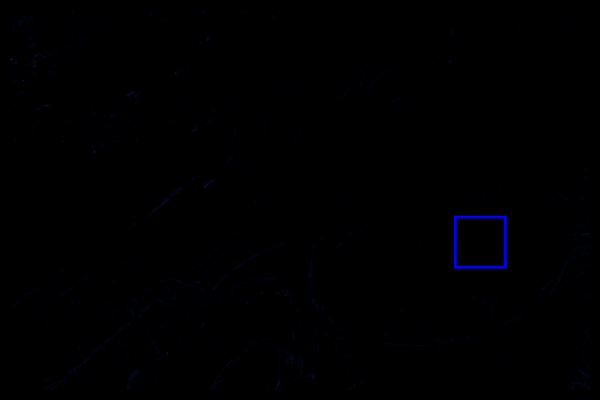}
    \end{subcaptionblock}
    \begin{subcaptionblock}[C]{.19\linewidth}
    \pdfpxdimen=\dimexpr 1 in/72\relax
    \includegraphics[width=\linewidth]{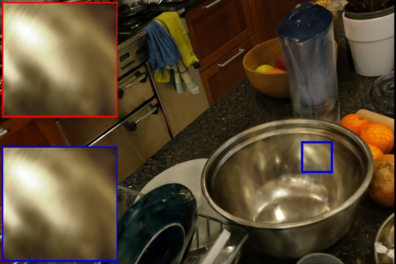}
    \end{subcaptionblock}
    \begin{subcaptionblock}[C]{.19\linewidth}
    \includegraphics[width=\linewidth]{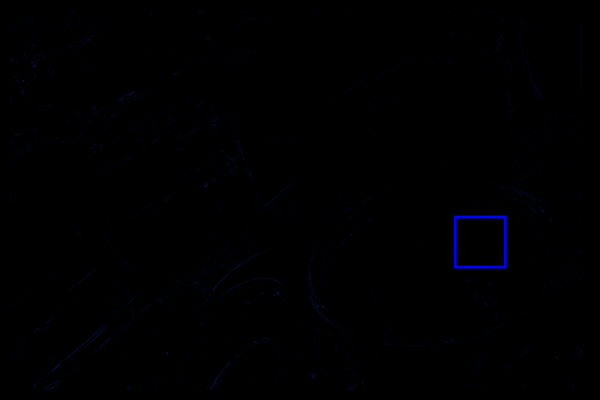}
    \end{subcaptionblock}
    \begin{subcaptionblock}[C]{.19\linewidth}
    \includegraphics[width=\linewidth]{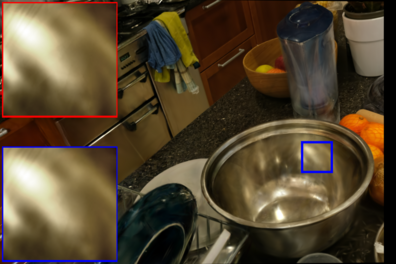}
    \end{subcaptionblock} 
        \caption{Visualization of our proposed popping detection method with detailed views inset.
    We warp view $\vec{F}_i$ to $\hat{F}_{i+1}, \hat{F}_{i+7}$ using optical flow and use \FLIP to measure errors between warped and non-warped views.
    While $\flipview{1}$ is able to effectively detect popping artifacts, the obtained errors are only accumulated over a single frame.
    On the contrary, $\flipview{7}$ is able to accumulate errors due to popping over multiple frames, making this metric more reliable.
    \new{We increased contrast for the zoomed-in views to better highlight view-inconsistencies.}
    } 
    \label{fig:flip_consistency}

\end{figure*}

\subsection{Qualitative Evaluation}
\begin{figure*}[ht!]
    \begin{subcaptionblock}[C]{.02\linewidth}
    \rotatebox[origin=c]{90}{\footnotesize{\gs}}
    \end{subcaptionblock}
    \begin{subcaptionblock}[C]{.1575\linewidth}
    \pdfpxdimen=\dimexpr 1 in/72\relax
    \includegraphics[clip,width=\linewidth, viewport=0px 100px 1559px 967px]{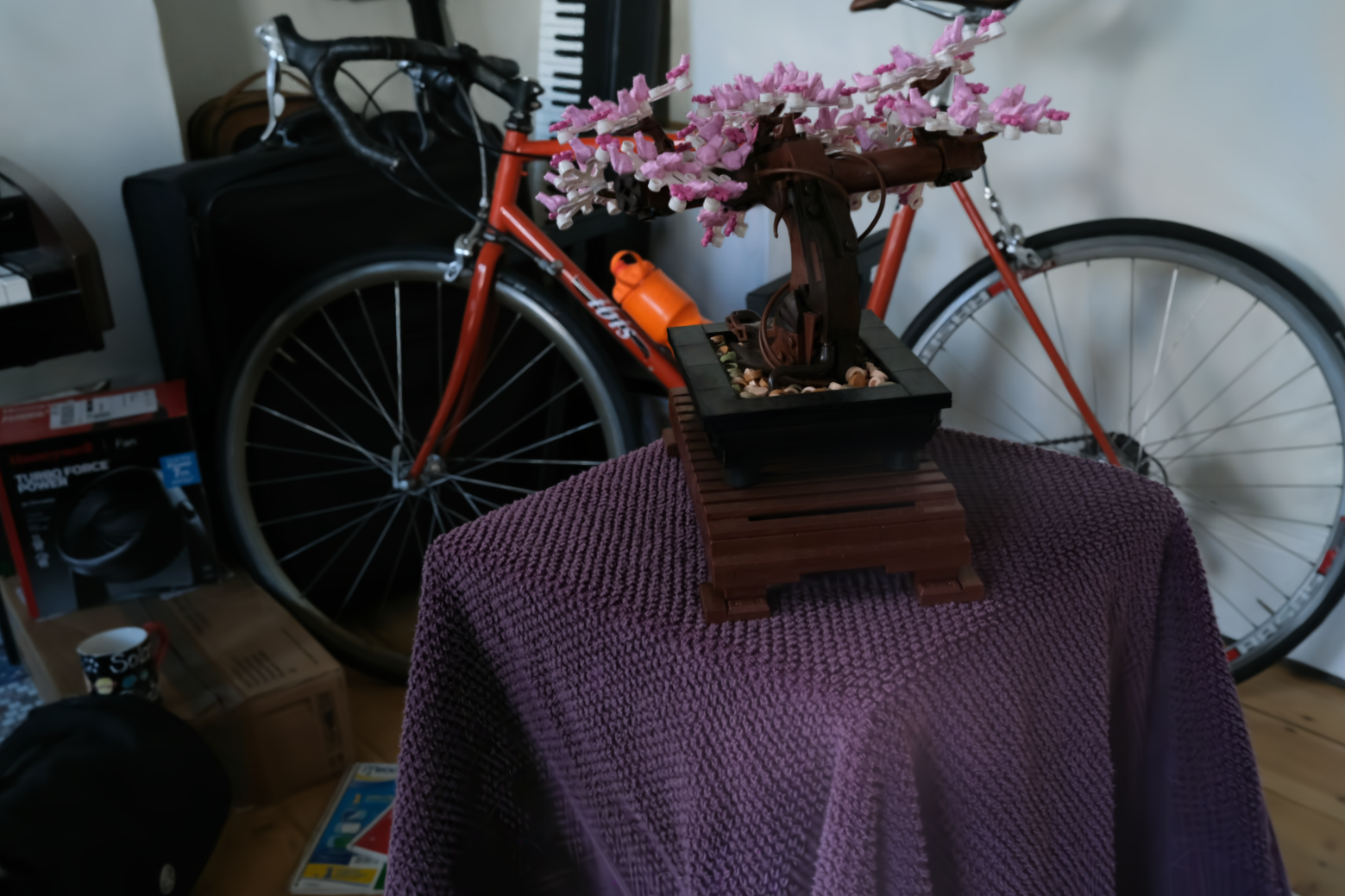}
    \end{subcaptionblock}
    \begin{subcaptionblock}[C]{.1575\linewidth}
    \pdfpxdimen=\dimexpr 1 in/72\relax
    \includegraphics[clip,width=\linewidth, viewport=0px 100px 1559px 967px]{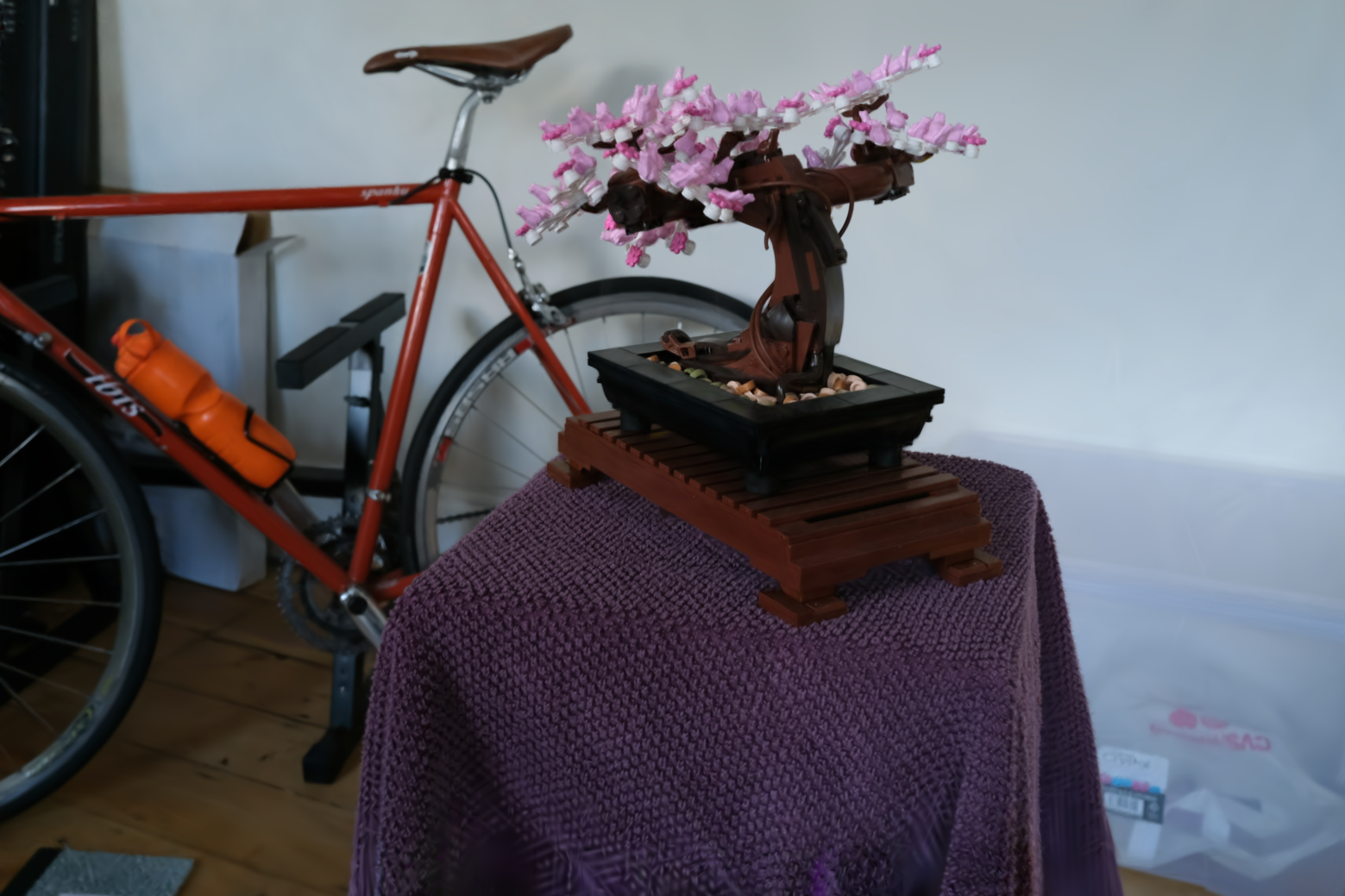}
    \end{subcaptionblock}
    \begin{subcaptionblock}[C]{.1575\linewidth}
    \includegraphics[width=\linewidth]{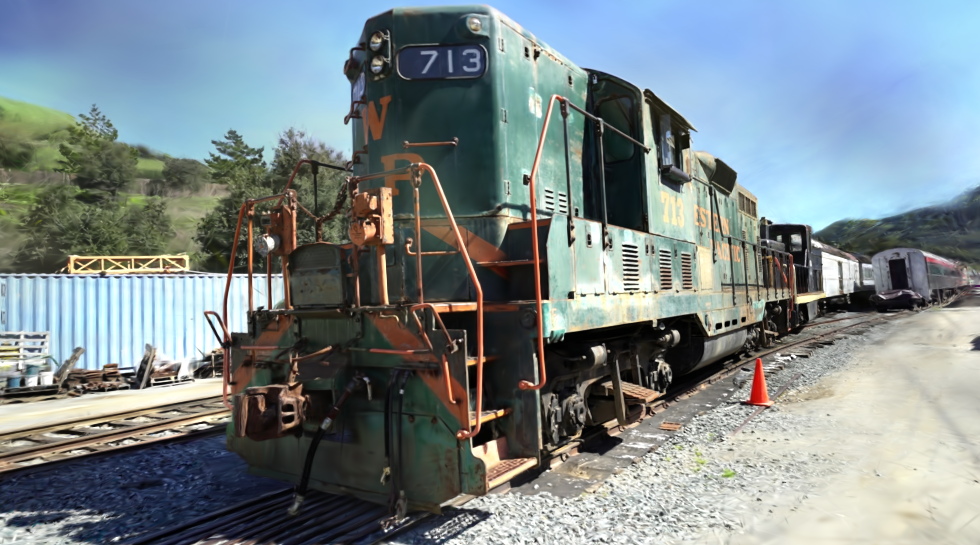}
    \end{subcaptionblock}
    \begin{subcaptionblock}[C]{.1575\linewidth}
    \pdfpxdimen=\dimexpr 1 in/72\relax
    \includegraphics[width=\linewidth]{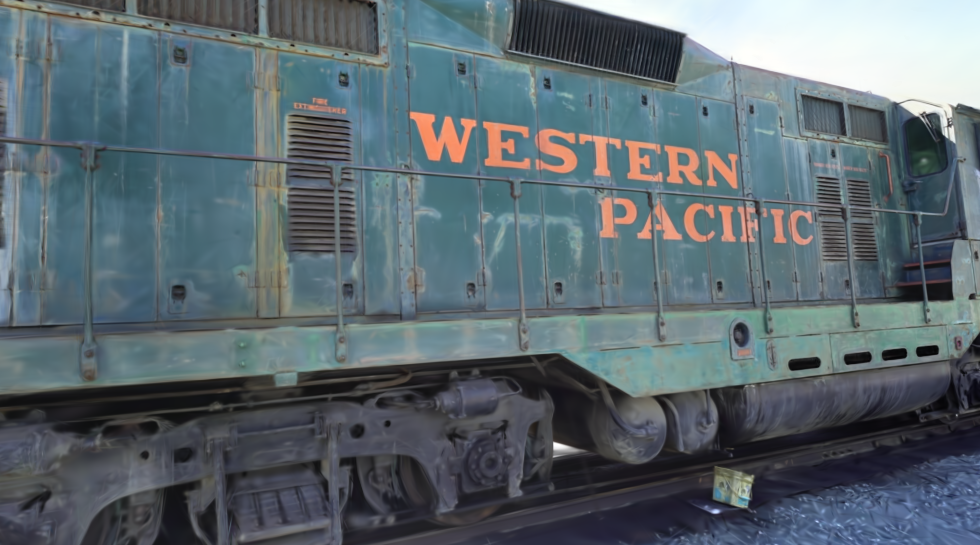}
    \end{subcaptionblock}
    \begin{subcaptionblock}[C]{.1575\linewidth}
    \includegraphics[clip,width=\linewidth, viewport=0px 50px 1264px 753px]{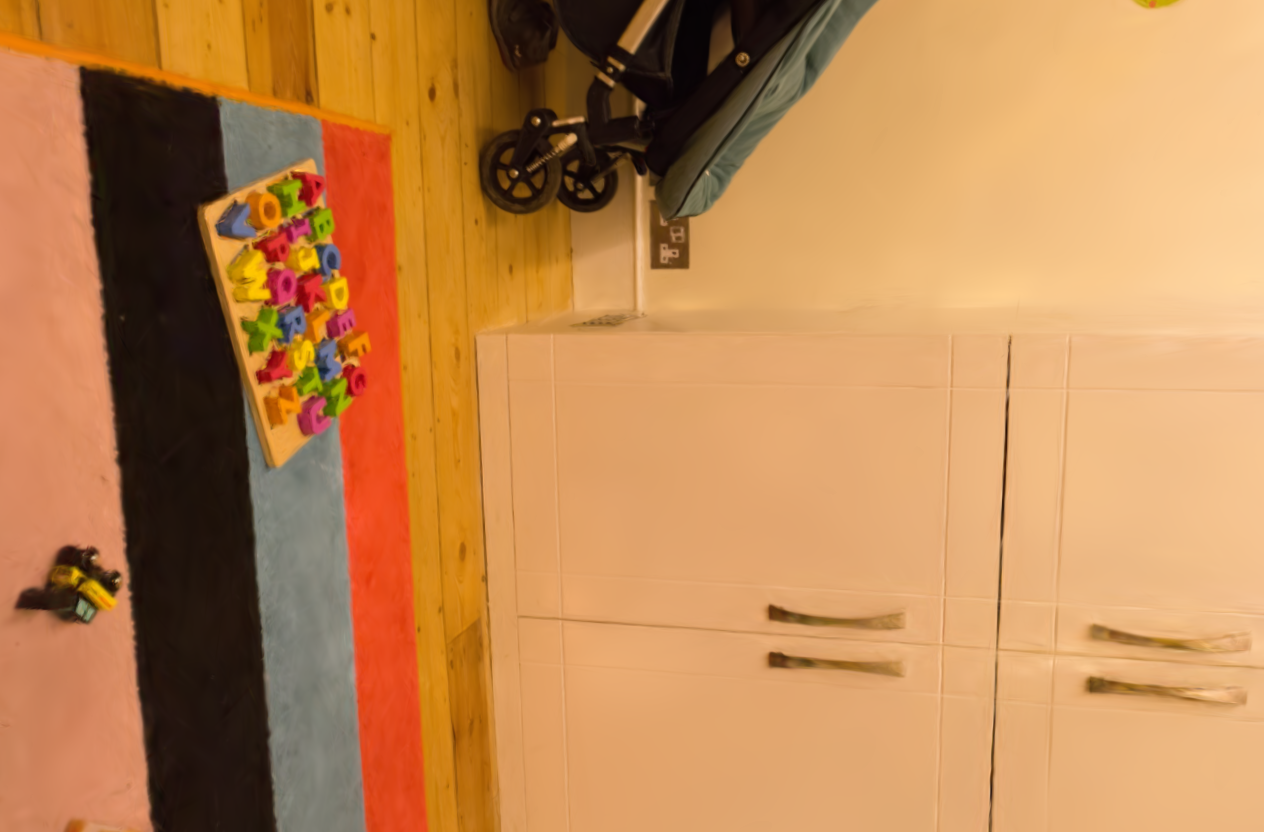}
    \end{subcaptionblock}
    \begin{subcaptionblock}[C]{.1575\linewidth}
    \includegraphics[clip,width=\linewidth, viewport=0px 50px 1264px 753px]{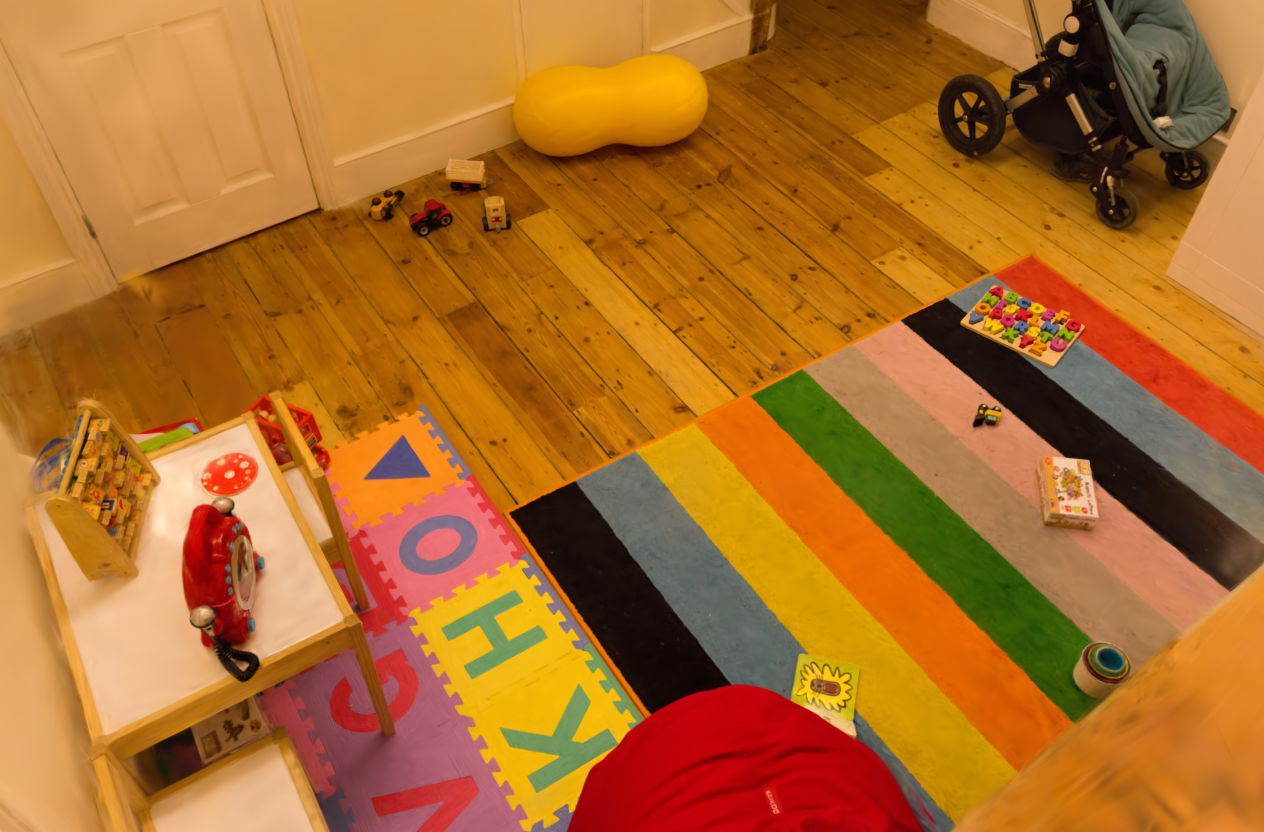}
    \end{subcaptionblock} \\

    \begin{subcaptionblock}[C]{.02\linewidth}
    \rotatebox[origin=c]{90}{\footnotesize{\revised{Ground-Truth}{Ground Truth}}}
    \end{subcaptionblock}
    \begin{subcaptionblock}[C]{.1575\linewidth}
    \pdfpxdimen=\dimexpr 1 in/72\relax
    \includegraphics[clip,width=\linewidth, viewport=0px 100px 1559px 967px]{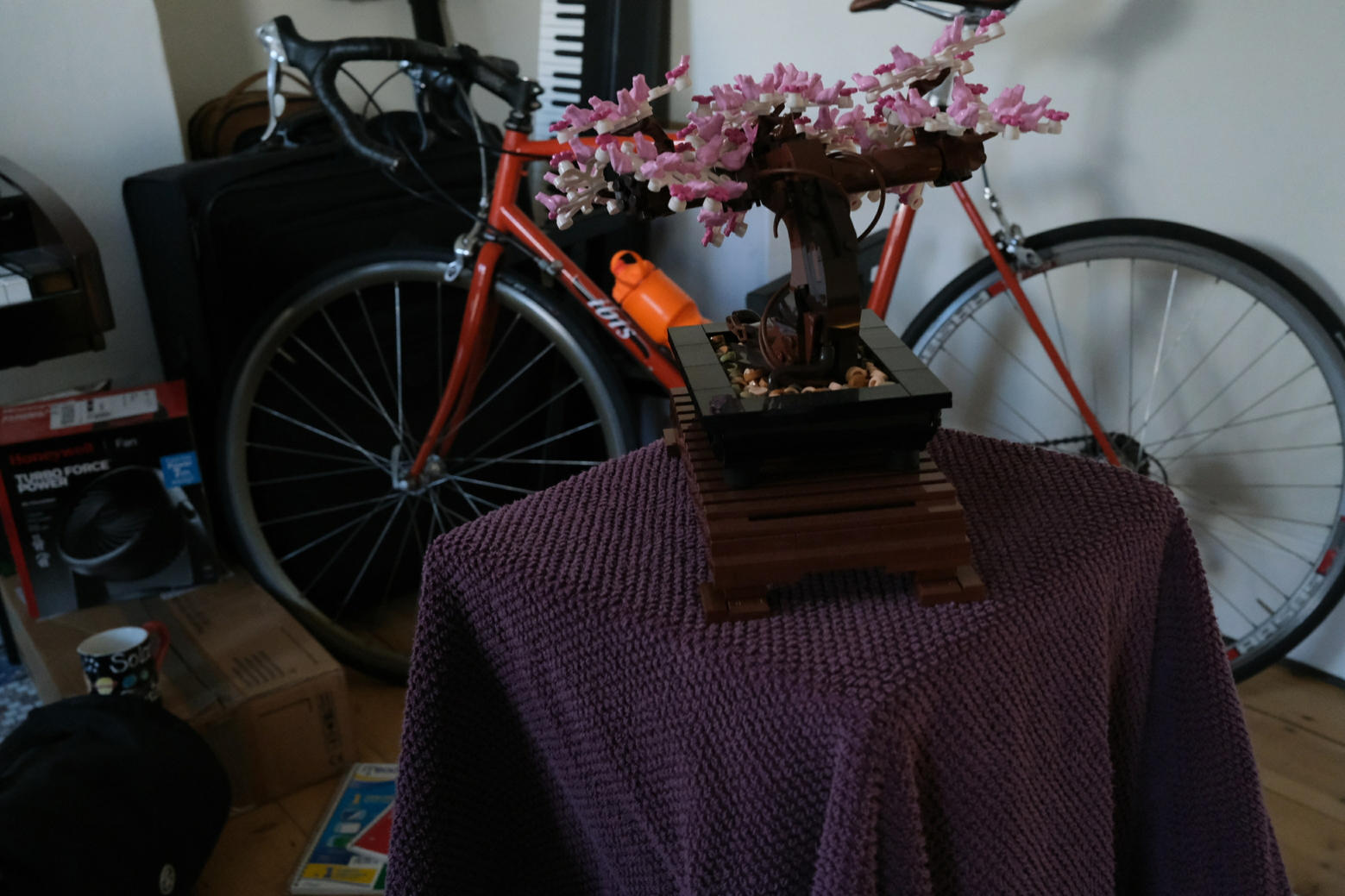}
    \end{subcaptionblock}
    \begin{subcaptionblock}[C]{.1575\linewidth}
    \pdfpxdimen=\dimexpr 1 in/72\relax
    \includegraphics[clip,width=\linewidth, viewport=0px 100px 1559px 967px]{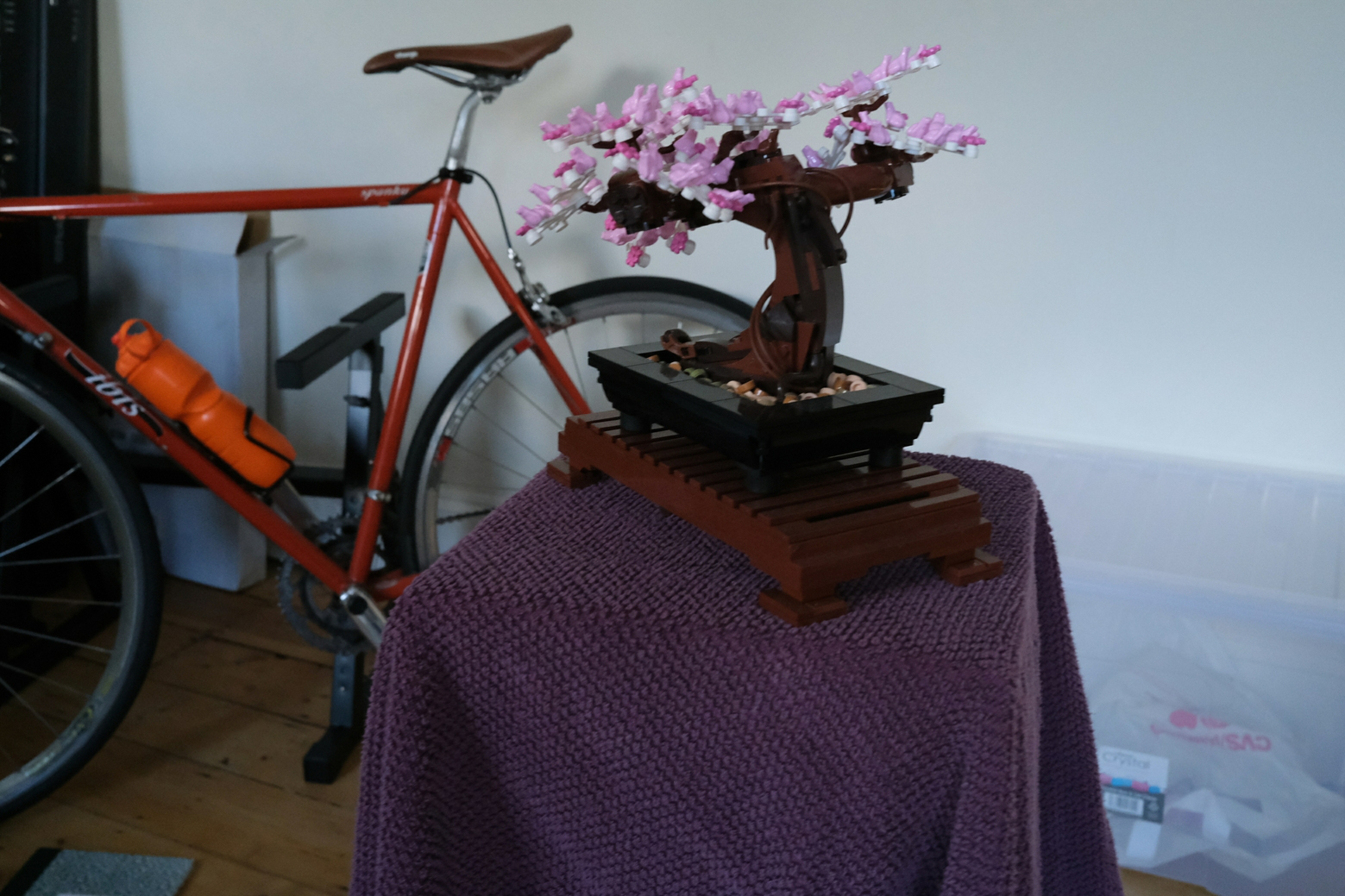}
    \end{subcaptionblock}
    \begin{subcaptionblock}[C]{.1575\linewidth}
    \includegraphics[width=\linewidth]{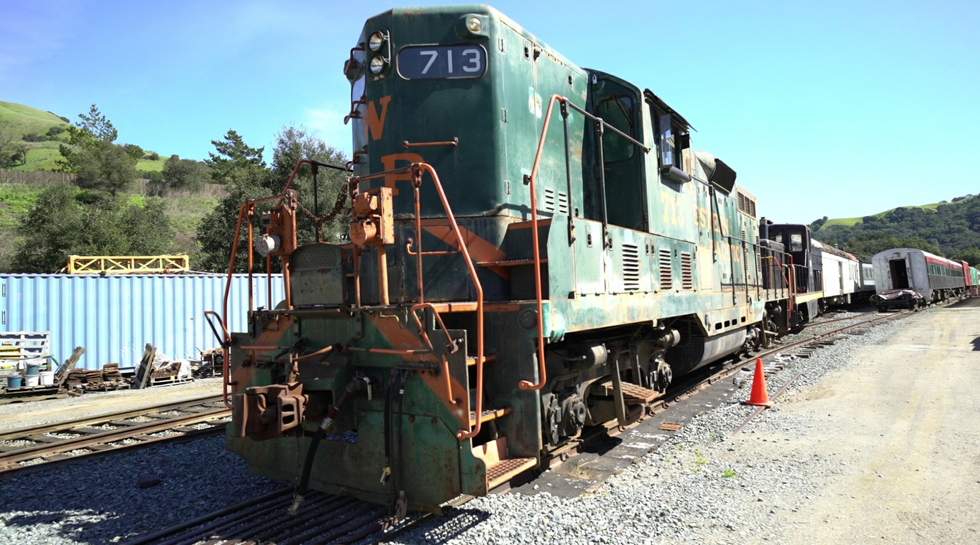}
    \end{subcaptionblock}
    \begin{subcaptionblock}[C]{.1575\linewidth}
    \pdfpxdimen=\dimexpr 1 in/72\relax
    \includegraphics[width=\linewidth]{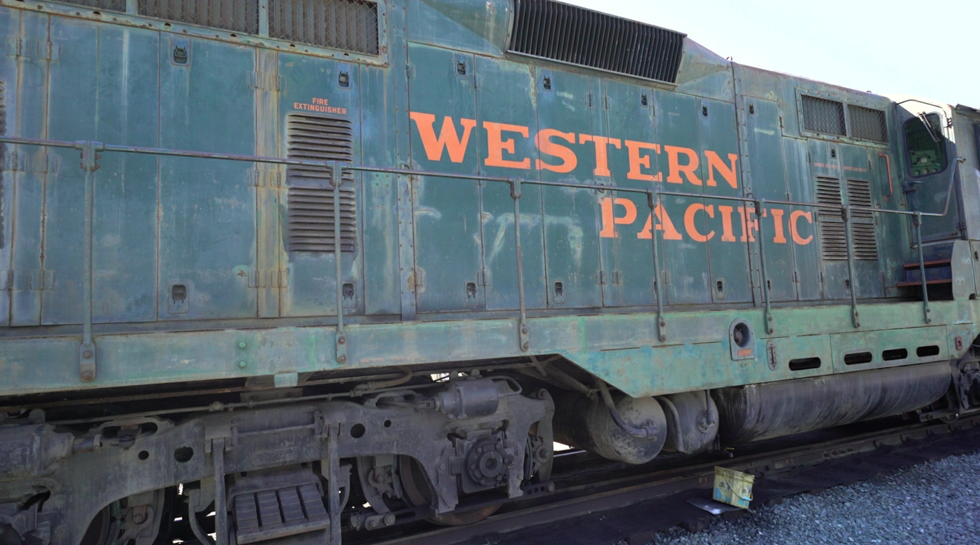}
    \end{subcaptionblock}
    \begin{subcaptionblock}[C]{.1575\linewidth}
    \includegraphics[clip,width=\linewidth, viewport=0px 50px 1264px 753px]{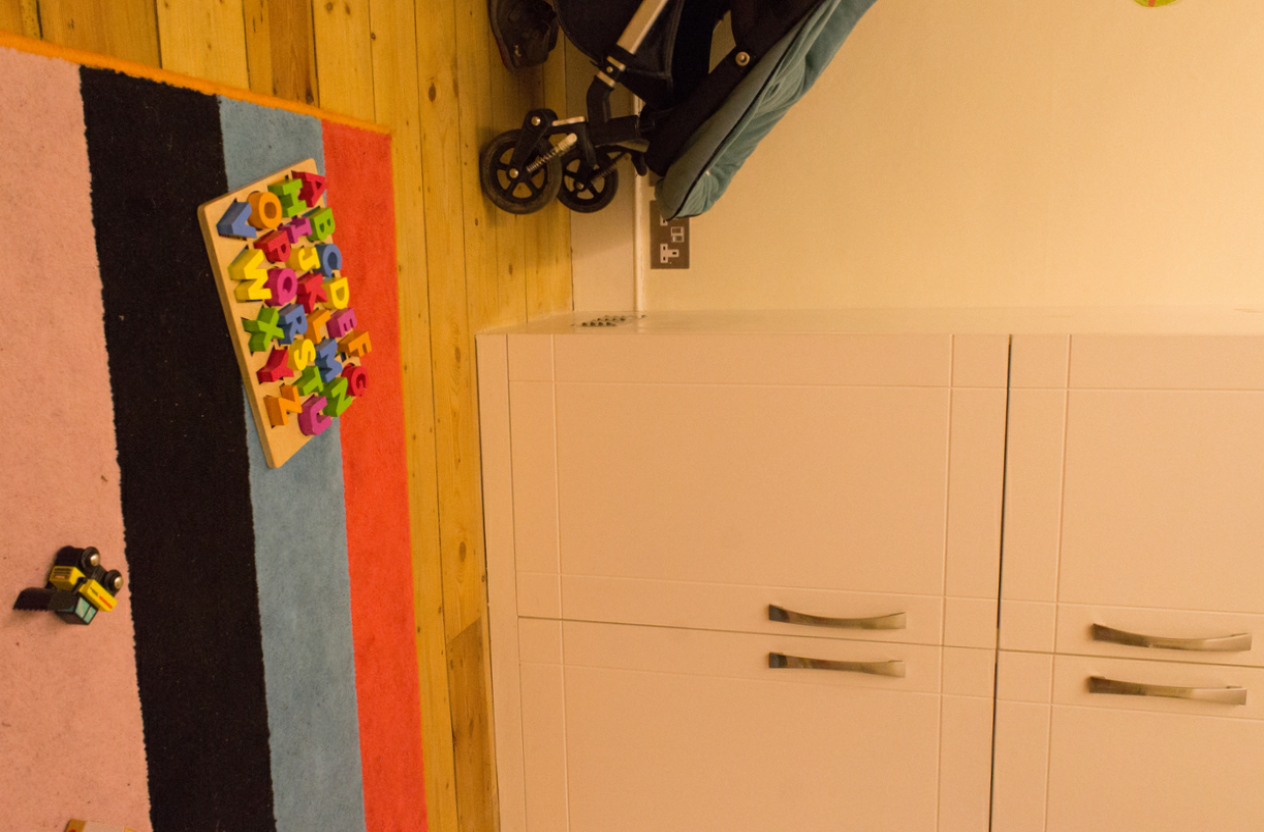}
    \end{subcaptionblock}
    \begin{subcaptionblock}[C]{.1575\linewidth}
    \includegraphics[clip,width=\linewidth, viewport=0px 50px 1264px 753px]{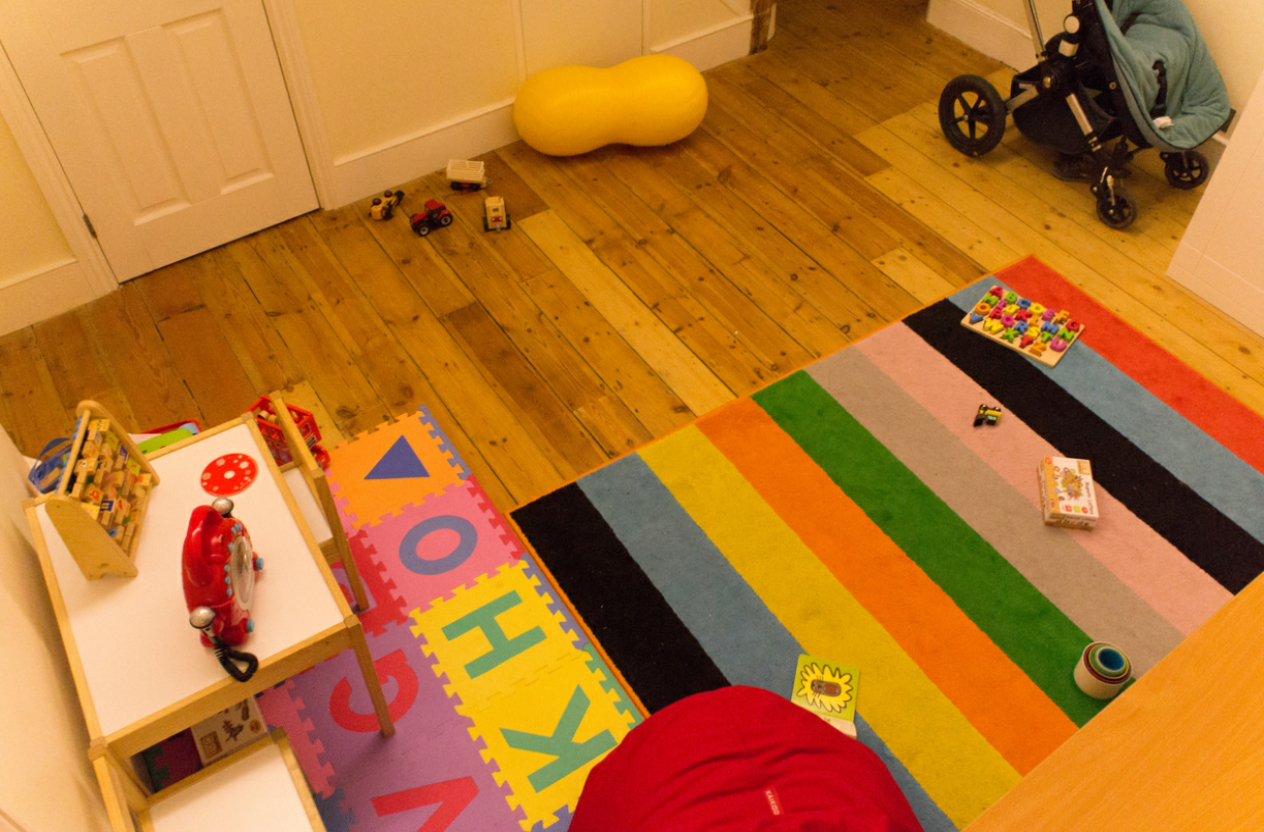}
    \end{subcaptionblock} \\

    \begin{subcaptionblock}[C]{.02\linewidth}
    \rotatebox[origin=c]{90}{\footnotesize{Ours}}
    \end{subcaptionblock}
    \begin{subcaptionblock}[C]{.1575\linewidth}
    \pdfpxdimen=\dimexpr 1 in/72\relax
    \includegraphics[clip,width=\linewidth, viewport=0px 100px 1559px 967px]{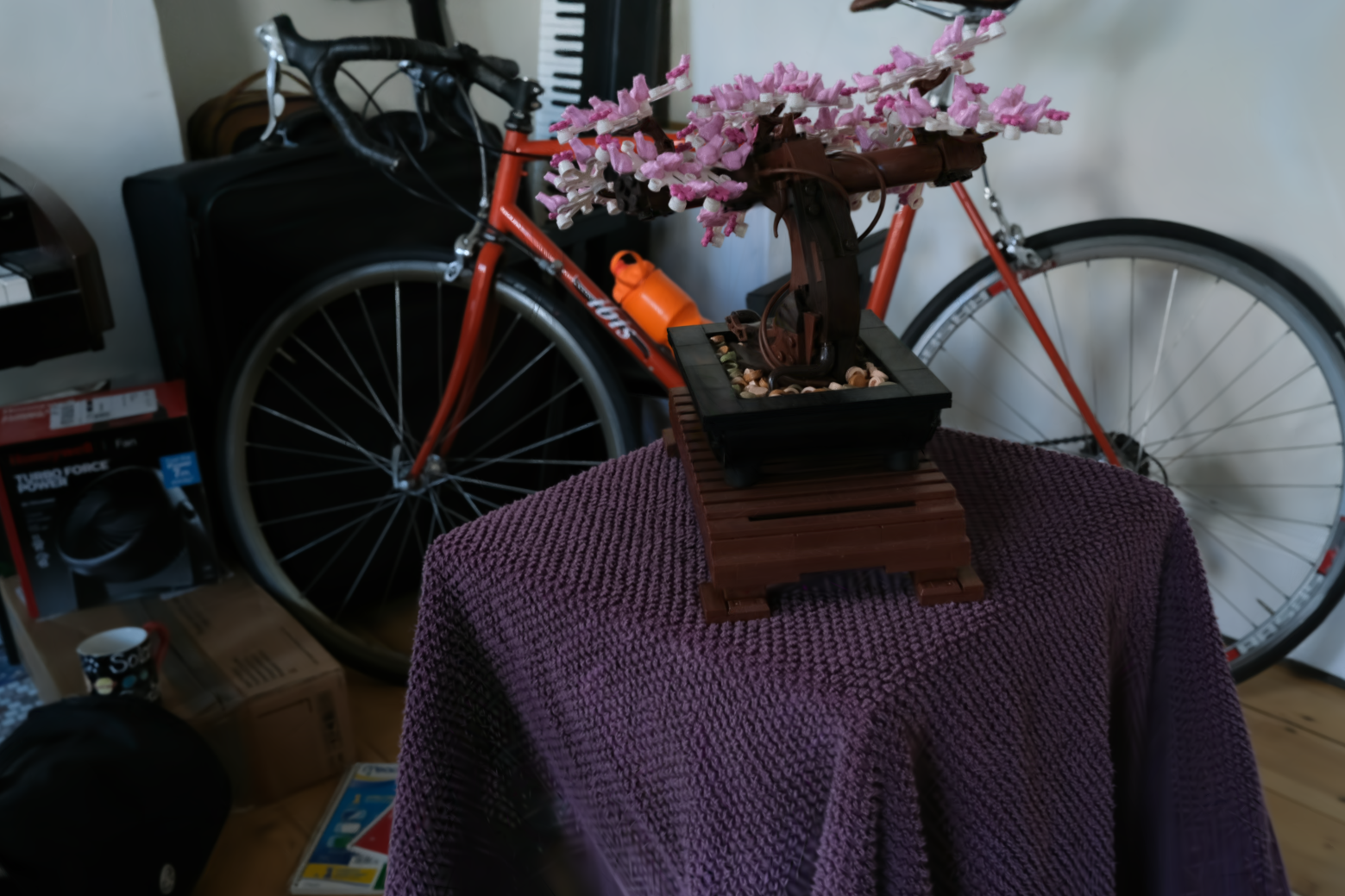}
    \caption*{\small{Ours PSNR\textsuperscript{$\uparrow$}}}
    \end{subcaptionblock}
    \begin{subcaptionblock}[C]{.1575\linewidth}
    \pdfpxdimen=\dimexpr 1 in/72\relax
    \includegraphics[clip,width=\linewidth, viewport=0px 100px 1559px 967px]{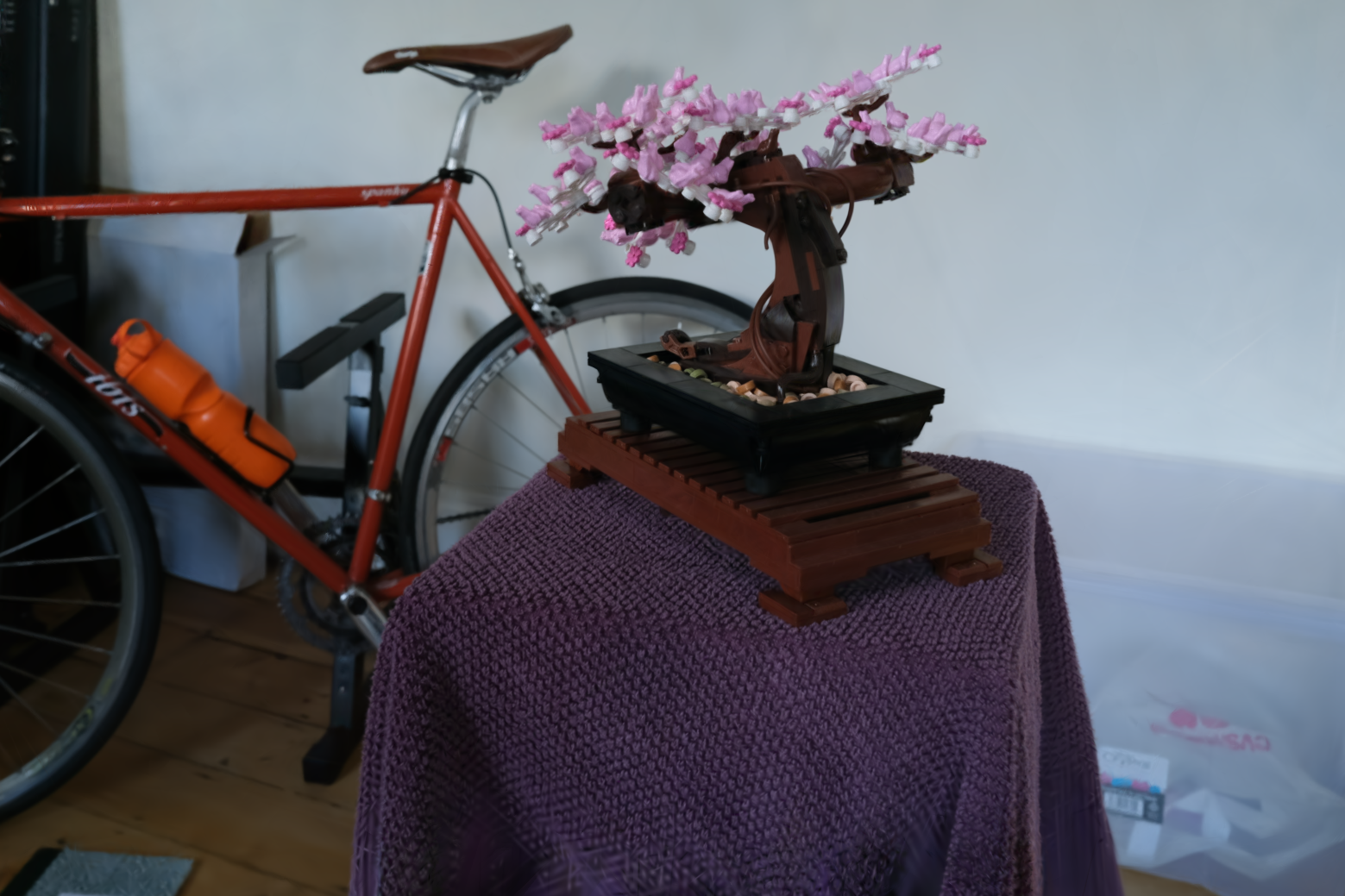}
    \caption*{\small{\gs PSNR\textsuperscript{$\uparrow$}}}
    \end{subcaptionblock}
    \begin{subcaptionblock}[C]{.1575\linewidth}
    \includegraphics[width=\linewidth]{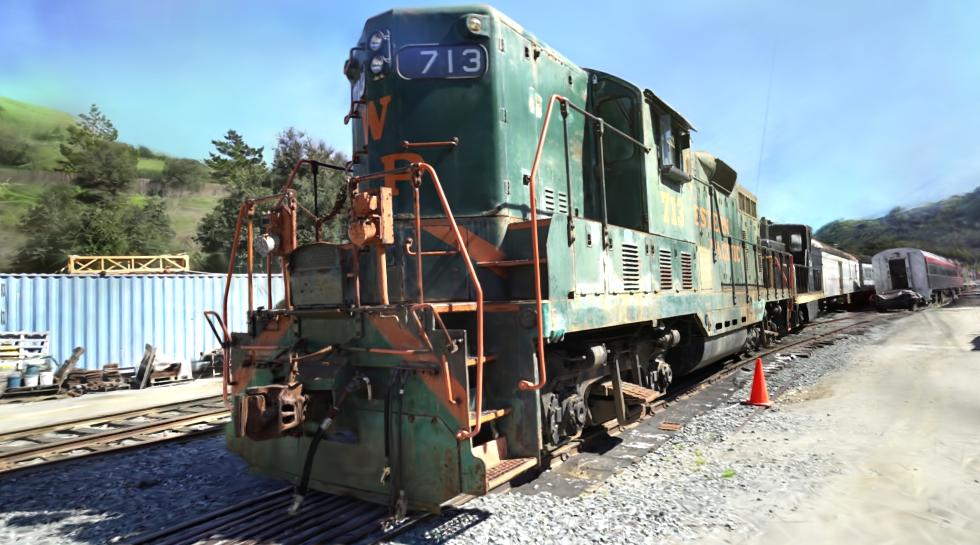}
    \caption*{\small{Ours PSNR\textsuperscript{$\uparrow$}}}
    \end{subcaptionblock}
    \begin{subcaptionblock}[C]{.1575\linewidth}
    \pdfpxdimen=\dimexpr 1 in/72\relax
    \includegraphics[width=\linewidth]{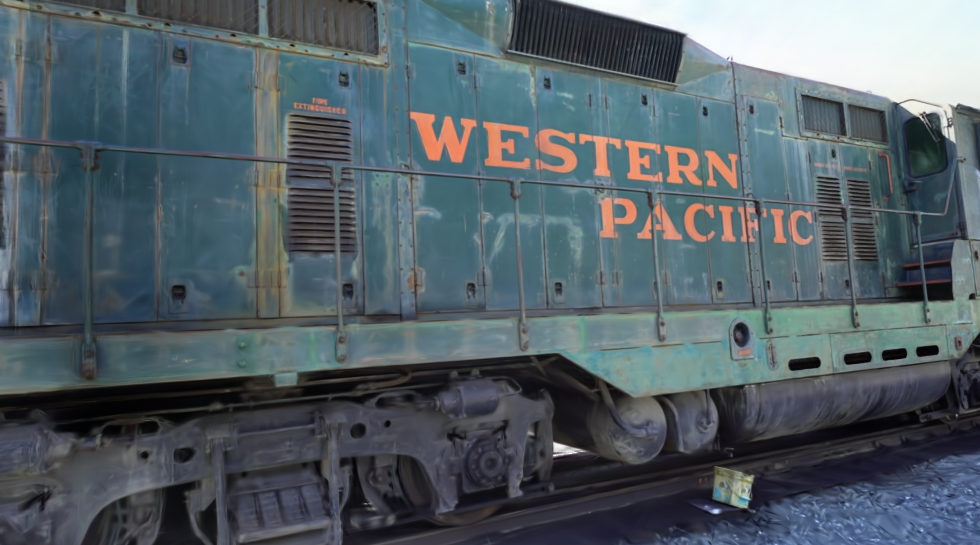}
    \caption*{\small{\gs PSNR\textsuperscript{$\uparrow$}}}
    \end{subcaptionblock}
    \begin{subcaptionblock}[C]{.1575\linewidth}
    \includegraphics[clip,width=\linewidth, viewport=0px 50px 1264px 753px]{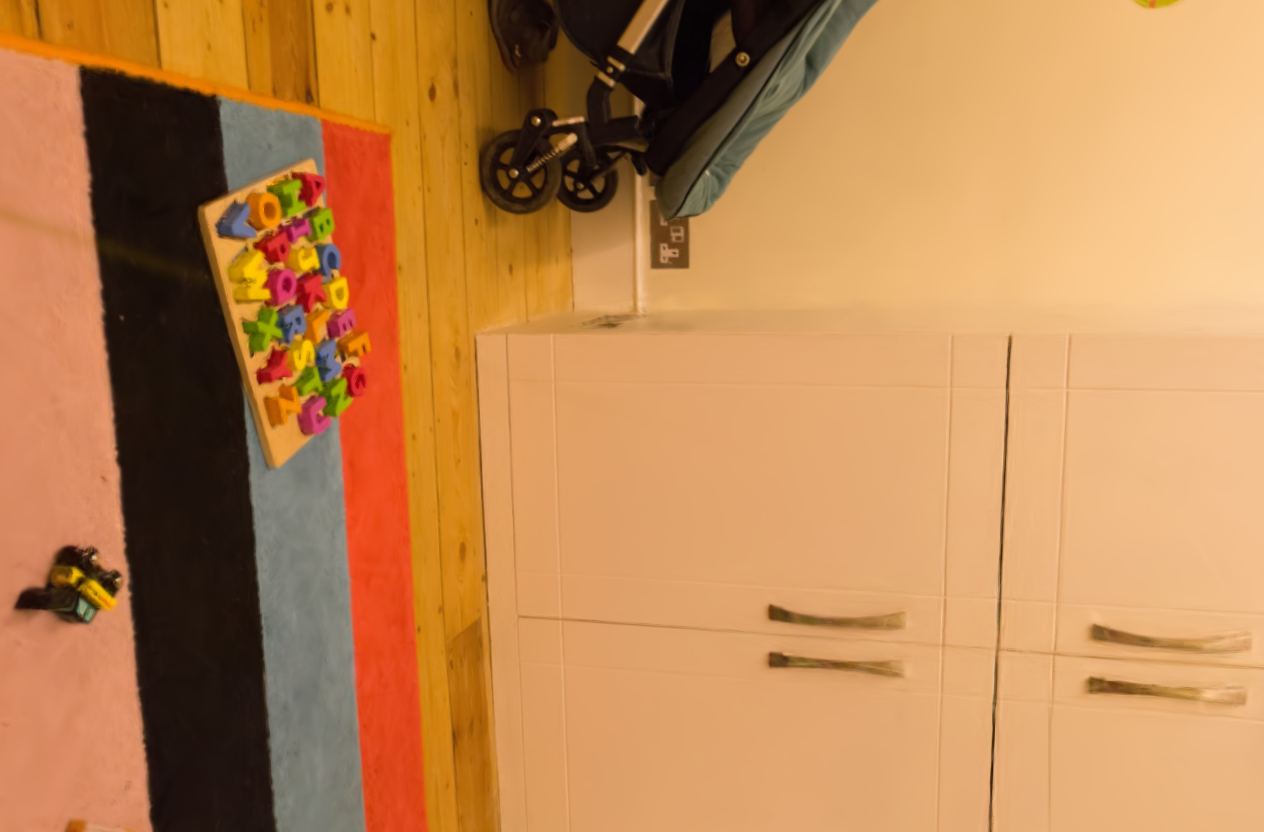}
    \caption*{\small{Ours PSNR\textsuperscript{$\uparrow$}}}
    \end{subcaptionblock}
    \begin{subcaptionblock}[C]{.1575\linewidth}
    \includegraphics[clip,width=\linewidth, viewport=0px 50px 1264px 753px]{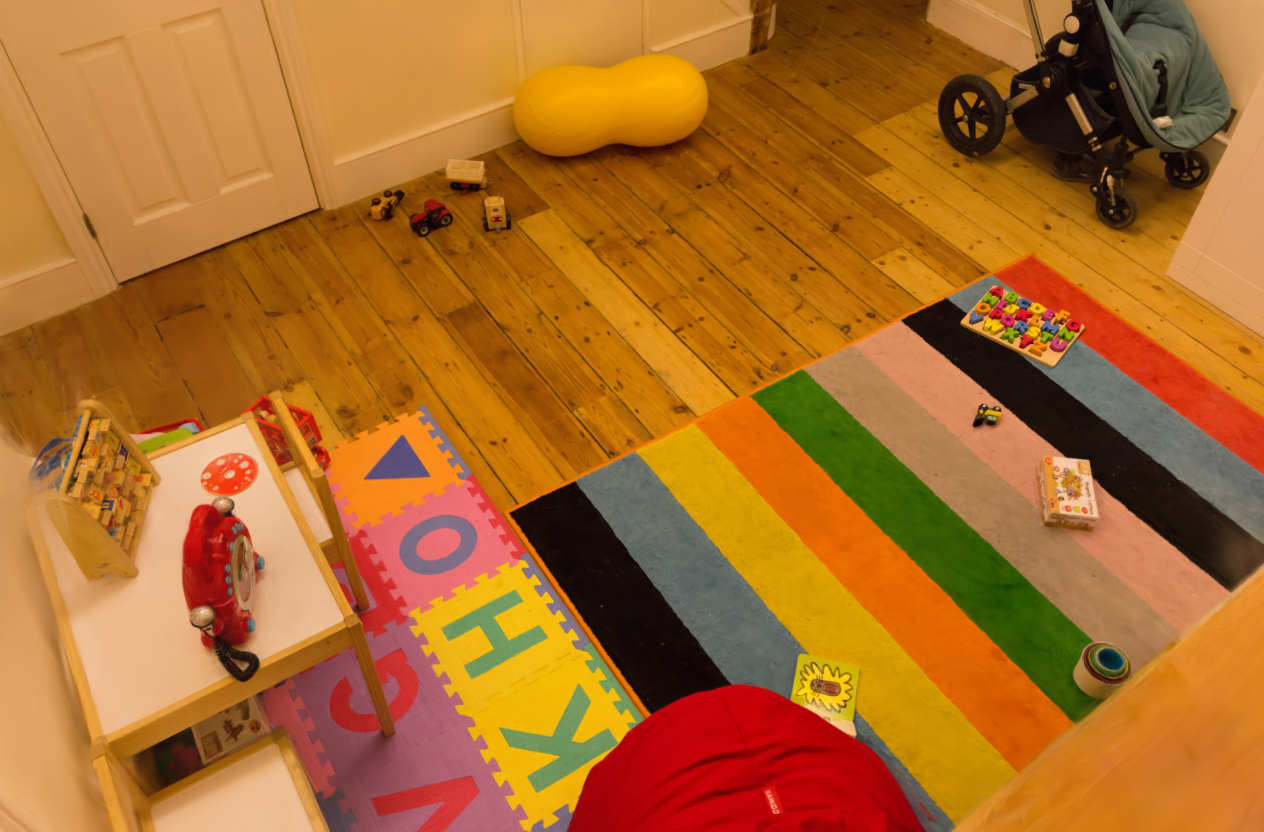}
    \caption*{\small{\gs PSNR\textsuperscript{$\uparrow$}}}
    \end{subcaptionblock} \\
    \caption{Image comparisons of our method and \gs. In most configurations, our rendered images are virtually indistinguishable from \gs.
    For each scene, we show a result where our method performs better on the left, and a result where \gs performs better on the right.
    } 
    \label{fig:img_comp}
\end{figure*}

To complement our quantitative evaluation, we provide image comparisons in \figref{img_comp} and conduct a user study to verify the effectiveness of our approach and our proposed popping detection method.

\subsubsection{User Study}
18 participants were presented with pairs of videos from our approach and \gs, following the same camera path.
The captured scenes exhibit rotation, translation, as well as a combination of the two.
We instructed the participants to rate the videos concerning view-consistency and popping artifacts.
The participants then indicated whether either of the techniques performed better or equal, which we translated into scores $s\in(-1, 0, +1)$.
On average, the results showed a clear preference for our approach ($s_{mean}=0.42$), which is statistically significant according to Wilcoxon Signed Rank tests ($Z=2276.5$, $p<.0001$).
Details about the study can be found in \appref{userstudy}.

\subsection{Performance and Ablation}
\new{
In the following, we provide a detailed performance analysis for different configurations of our method.
For our timings, we take all available COLMAP poses and interpolate a camera path between them (30 frames per pose), ensuring a variety of plausible viewpoints.
All timings were measured for Full HD rendering and averaged over 4 runs, where we used an NVIDIA RTX 4090 with CUDA 11.8.
}

\paragraph{Performance for different configurations.}
\revised{We provide a performance comparison between \gs and our renderer with different configurations in \tabref{performance_ablation}.}{In \tabref{performance_ablation} we show a performance comparison between \gs and our renderer with different configurations on an NVIDIA RTX 4090 with CUDA 11.8.} 
\revised{On average, the}{The} \emph{Render} stage takes considerably longer for our hierarchical renderer (A-E) due to additional per-ray sorting. 
Not computing the per-tile depth (B) only marginally speeds up the \emph{Duplicate} stage.
Without our load balancing scheme (C), \emph{Duplicate} takes 5$\times$ longer, as it is mostly dominated by very large Gaussians. 
Disabling tile-based culling (D) slightly accelerates \emph{Preprocess} but leads to many more entries in the global sorting data structure, which increases \emph{Sort} and \emph{Render} times. 
Disabling hierarchical culling inside the render kernel (E) leads to a drastic increase in \emph{Render} time as all Gaussians move through the entire pipeline.
Our final approach (A) with all optimizations achieves competitive runtimes on all evaluated scenes. 
Both methods see a drastic performance increase with Opacity Decay due to the significantly lower number of Gaussians---however, while our approach stays view-consistent, \gs shows even more popping artifacts.

\begin{table}
    \centering
    \footnotesize
    \caption{Performance timings for different configurations of our method and \gs. \revised{}{Times in ms for Full HD resolution, averaged over 4 runs for all scenes with an interpolated camera path between all available COLMAP poses (30 frames per pose).} The number of Gaussians is roughly the same for all methods (scene average \revised{${\sim}$2.98M}{$\sim2.98 \cdot10^6$}). Applying Opacity Decay during training leads to $\sim50\%$ fewer Gaussians (scene average \revised{${\sim}$1.54M}{$\sim1.54 \cdot 10^6$}).}
    \label{tab:performance_ablation}
    \setlength{\tabcolsep}{3pt}
    \begin{tabular}{lccccccc}
    \toprule
    Timings in ms & \emph{Preprocess} & \emph{Duplicate} & \emph{Sort} & \emph{Render} && \emph{Total} \\
    \midrule
            &\multicolumn{6}{c}{{Without Opacity Decay}} \\\cmidrule{2-7}
    \gs & \cellcolor{blue!45} 0.451 & 0.567 & 1.645 & \cellcolor{blue!45} 2.134 && \cellcolor{blue!45} 4.797 \\
    (A) Ours & \cellcolor{blue!15} 0.649 & \cellcolor{blue!30} 0.437 & \cellcolor{blue!45} 0.301 & \cellcolor{blue!15} 3.599 && \cellcolor{blue!15} 4.986 \\
    (B) Ours w/o per-tile depth & 0.658 & \cellcolor{blue!45} 0.283 & \cellcolor{blue!45} 0.301 & \cellcolor{blue!15} 3.599 && \cellcolor{blue!30} 4.841 \\
    (C) Ours w/o load balancing & 0.847 & 2.059 & 0.415 & \cellcolor{blue!30} 3.505 && 6.827 \\
    (D) Ours w/o tile-based culling & \cellcolor{blue!30} 0.610 & 0.479 & 1.180 & 5.346 && 7.614 \\
    (E) Ours w/o hier. culling & 0.649 & \cellcolor{blue!30} 0.437 & \cellcolor{blue!45} 0.301 & 5.967 && 7.364 \\
    \midrule
            &\multicolumn{6}{c}{{With Opacity Decay}} \\\cmidrule{2-7}
    \gs & \cellcolor{blue!45} 0.215 & 0.378 & 0.626 & \cellcolor{blue!45} 1.059 && \cellcolor{blue!45} 2.276 \\
    Ours & 0.366 & \cellcolor{blue!45} 0.223 & \cellcolor{blue!45} 0.161 & 2.227 && 2.976 \\
    \bottomrule
    \end{tabular}
\end{table}

\new{
\paragraph{Scene Comparison.}
Individual scenes with a similar number of Gaussians can exhibit sharp differences in runtime behavior.
In \tabref{performance_ablation_detail} and \tabref{metrics_ablation_detail}, we show detailed timings and metrics for two exemplary scenes - Bonsai and Train - which display the largest inter-method differences in performance, despite their comparable number of Gaussians $N$. 
Even though the Train scene contains slightly fewer Gaussians than Bonsai, the average number of visible (inside the view-frustum) Gaussians $N_V$, as well as their average screen-space size (indicated by avg./std. corresponding image tiles $N_t$), is considerably larger.
}

As larger Gaussian splats provide more opportunities for culling, our tile-based culling results in a larger reduction of avg. $N_t$ for Train than Bonsai (${\sim}3.5\stimes$ vs. ${\sim}2.5\stimes$).
The resulting lower number of sort entries allows Train to amortize the slower \emph{Render} stage with a much faster \emph{Sort}, while Bonsai does not experience the same gains.

\begin{table}
    \centering
    \footnotesize
    \caption{\new{Performance timings for different configurations of our method and \gs for the exemplary scenes Bonsai \& Train, which show contrary runtime behaviors.
    Times in ms for Full HD resolution.}}
    \label{tab:performance_ablation_detail}
    \setlength{\tabcolsep}{3pt}
    \begin{tabular}{lccccccc}
    \toprule
    Timings in ms & \emph{Preprocess} & \emph{Duplicate} & \emph{Sort} & \emph{Render} && \emph{Total} \\
    \midrule
    &\multicolumn{6}{c}{{Bonsai, ${\sim}$1.25M Gaussians}} \\\cmidrule{2-7}
    \gs & \cellcolor{blue!45} 0.224 & 0.384 & 0.700 & \cellcolor{blue!45} 1.266 && \cellcolor{blue!45} 2.574 \\
    (A) Ours & \cellcolor{blue!15} 0.295 & \cellcolor{blue!30} 0.321 & \cellcolor{blue!45} 0.173 & 2.610 && \cellcolor{blue!15} 3.399 \\
    (C) Ours w/o load balancing & 0.467 & 1.920 & 0.272 & \cellcolor{blue!30} 2.592 && 5.251 \\
    (D) Ours w/o tile-based culling & \cellcolor{blue!30} 0.282 & 0.331 & 0.554 & 3.680 && 4.846 \\
    \midrule
    &\multicolumn{6}{c}{{Train, ${\sim}$1.05M Gaussians}} \\\cmidrule{2-7}
    \gs & \cellcolor{blue!45} 0.288 & 0.811 & 2.451 & \cellcolor{blue!45} 1.998 && \cellcolor{blue!15} 5.548 \\
    (A) Ours & \cellcolor{blue!15} 0.409 & \cellcolor{blue!30} 0.495 & \cellcolor{blue!45} 0.270 & \cellcolor{blue!15} 3.052 && \cellcolor{blue!30} 4.225 \\
    (C) Ours w/o load balancing & 0.647 & 2.336 & 0.333 & \cellcolor{blue!30} 2.899 && 6.215 \\
    (D) Ours w/o tile-based culling & \cellcolor{blue!30} 0.323 & 0.542 & 1.550 & 5.054 && 7.469 \\
    \bottomrule
    \end{tabular}
\end{table}

\begin{table}
    \centering
    \footnotesize
    \caption{\new{Metrics of our method and \gs for exemplary scenes Bonsai \& Train, highlighting the effect of our tile-based culling. 
    Columns include total vs. visible (in view-frustum) number of Gaussians ($N$ vs. $N_V$), as well as standard deviation and average number of $16\stimes16$ tiles covered by each visible Gaussian ($N_t$).
    We additionally include an approximate number of sort entries as $N_V \cdot \text{avg}(N_t)$.}}
    \label{tab:metrics_ablation_detail}
    \setlength{\tabcolsep}{3pt}
    \begin{tabular}{llrrrrr}
    \toprule
    Scene & Method & $N$ & $N_V$ & $\text{avg}(N_t)$ & $\text{std}(N_t)$ & Sort Entries \\
    \midrule
    \multirow[c]{2}{*}{Bonsai}  & Ours & 1.26M & 0.41M & \cellcolor{blue!45}4.198  & \cellcolor{blue!45}15.282 & \cellcolor{blue!45}1.72M \\
                                & \gs  & 1.24M & 0.40M & 10.801 & 52.236 & 4.36M \\
    \cmidrule{2-7}
    \multirow[c]{2}{*}{Train}   & Ours & 1.05M & 0.57M & \cellcolor{blue!45}5.004  & \cellcolor{blue!45}20.127 & \cellcolor{blue!45}2.85M\\
                                & \gs  & 1.08M & 0.59M & 17.282 & 89.891 & 10.2M\\
    \bottomrule
    \end{tabular}
\end{table}

\paragraph{Backward Pass Performance.}
The relative performance of our backward \emph{Render} pass compared to \gs is only $1.1\times$ compared to the $1.5\times$ we see for the forward \emph{Render} stage.
This is mostly due to the backward \emph{Render} executing a large number of atomics, which are equal between both approaches.
Although the backward pass skips \emph{Duplicate} and \emph{Sort}---which are faster in our renderer---the final change in training time is only about 3\%.
The backward \emph{Render} pass is only a single step in the entire training pipeline and thus, the overall time loss is close to negligible.
Again, if we turn on Opacity Decay, training becomes proportionally faster.

\section{Conclusion, Limitations, and Future Work}
In this paper, we took a closer look at the way 3D Gaussian Splatting orders splats during blending.
A detailed analysis of the splat's depth computation revealed the reason for popping artifacts of \gs: 
the computed depth is highly inconsistent under rotation.
A per-ray depth computation which considers the highest contribution along the ray as \emph{optimal} blending depth, removes all popping artifacts but is $100\times$ more costly.
With our hierarchical renderer, which includes multiple culling and resorting stages, we are only $1.0\revised{4}{3}\times$ slower than \gs on average.
While it is difficult to identify popping in standard quality metrics, we provided a view-consistency metric based on optical flow and \FLIP, which shows that our approach significantly reduces popping.
We could also confirm this fact in a user study \new{and provided an additional metric confirming increased view-consistency and more accurate depth estimates for our method}.
Furthermore, our approach remains view-consistent even when constructing the scene with half the Gaussians; for which \gs shows a significant increase in popping artifacts.
As such, our approach can reduce memory by $2\times$ and render times by $1.6\times$ compared to \gs in this configuration, while reducing popping artifacts and achieving virtually indistinguishable quality.
 
While our approach typically removes all artifacts in our tests, resorting does not guarantee the right blend order, and thus could still lead to popping or flickering for very complex geometric relationships.
Furthermore, our approach still ignores overlaps between Gaussians along the view ray.
A fully correct volume rendering of Gaussians may not only remove artifacts completely but could lead to better scene reconstructions---a direction certainly worth exploring in the future.
Both our renderer and our optimizations for \gs are publicly available at {\color{blue}\url{https://github.com/r4dl/StopThePop}}.

\bibliographystyle{ACM-Reference-Format}
\bibliography{bib}

\appendix
\section{Deriving Depth for 3D Gaussians along a Ray}
\label{app:depth}

In order to get an accurate depth estimate for our sort order of 3D Gaussians along a view ray $\vec{r}(t)=\vec{o}+t\vec{d}$, we \revised{compute the}{perform a line search to find the} $t_{opt}$ which maximizes the Gaussian's contribution along the ray\revised{, \ie $\argmax_t G(\vec{r}(t))$}{: $t_{opt} = \argmax_t G(\vec{r}(t))$}. This optimum can be found through the following derivation:

\begin{align}
\frac{dG(\vec{r}(t))}{dt} & = -\frac{1}{2}G(\vec{r}(t))\cdot\left( (\vec{r}(t)-\vec{\mu})\Sigma^{-1}\vec{d} + \vec{d}^T\Sigma^{-1}(\vec{r}(t)-\vec{\mu})\right) \nonumber \\
& = -\frac{1}{2}G(\vec{r}(t))\cdot\left(2\cdot \vec{d}^T\Sigma^{-1}(\vec{r}(t)-\vec{\mu})\right) \nonumber \\
& = -G(\vec{o}+t\vec{d})\cdot\left(\vec{d}^T\Sigma^{-1}(\vec{o}+t\vec{d}-\vec{\mu})\right) \revised{\overset{!}{=}}{=}\ 0 \nonumber \\
  & \revised{\Rightarrow}{=}\ \vec{d}^T\Sigma^{-1}(\vec{o}+t\vec{d}-\vec{\mu}) = 0 \nonumber \\
 & \revised{\Rightarrow}{=}\ \vec{d}^T\Sigma^{-1}\vec{d}\cdot t + \vec{d}^T\Sigma^{-1}(\vec{o}-\vec{\mu}) = 0\nonumber  \\
t_{opt} & =\frac{\vec{d}^T\Sigma^{-1}(\vec{\mu}-\vec{o})}{\vec{d}^T\Sigma^{-1}\vec{d}}. \label{eq:dopt_suppl}
\end{align}

The simplification from the first to the second line relies on the fact that $\Sigma^{-1}$ is symmetric and thus both expressions are identical. $\Sigma^{-1}$ can be efficiently computed:

\[
\Sigma^{-1} = \left(RSSR^T\right)^{-1} = RS^{-1}S^{-1}R^T=R\begin{pmatrix}
s_x^{-2} & 0 & 0\\
0 & s_y^{-2} & 0\\
0 & 0 & s_z^{-2}
\end{pmatrix} R^T.
\]

\section{Additional Implementation Details}
\label{app:implementation}
This section contains a more thorough description of our implementation and various optimization strategies to make our hierarchical rasterizer viable for real-time rendering.

\subsection{Tile-based Culling}
\label{app:culling}

In Algorithm \ref{alg:max_2d_gaussian}, we describe how to find the \revised{maximally}{maximum} contributing point $\vec{\hat{x}}$ of a 2D Gaussian $G_2$ parameterized by $\revised{\vec{\mu_2}, \Sigma^{-1}_2}{\Sigma^{-1}_2, \vec{\mu_2}}$ inside an axis-aligned tile \new{$X$}. 
If $\vec{\mu_2}$ lies inside \revised{$X$}{the tile}, then it is consequently also the maximum. 
Otherwise, the maximum has to lie on one of the two edges that are reachable from $\vec{\mu_2}$. 
Those are the two edges that originate from the tile corner point $\vec{\hat{p}}$ closest to $\vec{\mu_2}$. 
We can then find the optimum by performing \revised{the same computation as in \eqnref{dopt_suppl}, but in 2D}{a line search along those two edges}.
By checking if $\vec{\mu_2}_x, \vec{\mu_2}_y$ are in range of the tile in $x,y$ direction, as well as clamping the values of $t_x,t_y$ to $[0,1]$, we ensure that the final point will lie on one of these two edges. 
The fact that the $y$ coordinate of $\vec{d}_x$ and the $x$ coordinate of $\vec{d}_y$ are zero, allows for further simplifications in the final implementation.

\begin{algorithm}
\caption{Finding maximum of 2D Gaussian inside AABB \\ \revised{$\vec{\mu_2}, \Sigma^{-1}_2$: mean and inverse covariance matrix}{$\Sigma^{-1}_2, \vec{\mu_2}$: inverse covariance matrix and mean} of 2D Gaussian $G_2$ \\
$x_{\min}, x_{\max}, y_{\min}, y_{\max}$: AABB dimensions}\label{alg:max_2d_gaussian}
\KwData{$X=\{\forall \vec{x} \in \mathbb{R}^2| x_{\min} \leq \vec{x_x} \leq x_{\max} \land y_{\min} \leq \vec{x_y} \leq y_{\max}\}$}
\KwResult{$\vec{\hat{x}}=\argmin_{\vec{x} \in X} (\vec{x}-\vec{\mu_2})^T\Sigma_2^{-1}(\vec{x}-\vec{\mu_2})$}
\eIf{$\vec{\mu_2} \in X$}{
    $\vec{\hat{x}} \gets \vec{\mu_2}$ \;
}{
    $\vec{\hat{p}} \gets$ Corner closest to $\vec{\mu_2}$ \;
    $\vec{d}_x, \vec{d}_y \gets$ vectors to next AABB corners in $x,y$ direction \;
    $t_x,t_y \gets 0$ \;
    \If{$\vec{\mu_2}_x < x_{\min} \lor \vec{\mu_2}_x > x_{\max}$}{
        $t_y \gets \min \left(1, \max \left(0, \frac{\vec{d_y^T}\Sigma^{-1}_2(\vec{\mu_2}-\vec{\hat{p}})}{\vec{d_y^T}\Sigma^{-1}_2\vec{d}_y} \right) \right)$ \;
    }
    
    \If{$\vec{\mu_2}_y < y_{\min} \lor \vec{\mu_2}_y > y_{\max}$}{
        $t_x \gets \min \left(1, \max \left(0, \frac{\vec{d_x^T}\Sigma^{-1}_2(\vec{\mu_2}-\vec{\hat{p}})}{\vec{d_x^T}\Sigma^{-1}_2\vec{d}_x}\right) \right)$ \;
    }
    $\vec{\hat{x}} \gets \vec{\hat{p}} + t_x \vec{d}_x + t_y \vec{d}_y$ \;
}
\end{algorithm}

\subsection{Tighter Bounding of 2D Gaussians}

For computing the bounding rectangle of touched tiles on screen, \citet{kerbl3Dgaussians} first bound each 2D Gaussian with a circle of radius $r=3 \cdot \lambda_{\max}$, where $\lambda_{\max}$ denotes the largest eigenvalue of the 2D covariance matrix $\Sigma_2$. They use a constant factor $t_O=3$ as a bound for a Gaussian, effectively clipping it at 0.3\% of its peak value.
We instead calculate an exact bound by considering the Gaussian's actual opacity value $\alpha$ and compute $t_O=\sqrt{2\log(\frac{\alpha}{\epsilon_O})}$, which is itself upper bounded by ${t_O}_{\max} \approx 3.3290$ (since $\alpha \in [0,1]$). 
Therefore, we conclude that the bound of $t_O=3$ by \citet{kerbl3Dgaussians} was actually chosen too small for the  opacity threshold $\epsilon_O=\frac{1}{255}$ used in the renderer. Additionally, our calculated bound allows us to fit a tighter circular bound around Gaussians with $\alpha < 1$.

\subsection{Global Sort}

Using a giant global sort for all combined (tile/depth) keys seems wasteful.
Sorting would be more efficient if the entries of each tile would be sorted individually, using a global partitioned sort.
However, this requires all the entries of a tile to be continuous in memory, with each tile knowing the range of its respective entries.
We can create such a setup by counting the number of entries per tile during the \emph{Preprocess} stage with an atomic counter per tile and computing tile ranges with a prefix sum.
In the \emph{Duplication} stage, another atomic counter per tile can be used to retrieve offsets for each entry inside this range.
While this reduces sorting costs to less than half in our experiments, the allocation using atomic operations adds an overhead that is about equal to the time saved in sorting.
Thus, we opted to keep the original sorting approach.

\subsection{Per-stage details}

\paragraph{Preprocess and Duplication} Similarly to \gs, we also prepare common values for each Gaussian during \emph{Preprocess}:
We compute and store $G_2$ for every Gaussian, evaluate Spherical Harmonics relying on the direction from the camera to the Gaussians center as view direction, establish $\Sigma^{-1}$ relying on the specifics of $R$ and  $S$, and precompute $\Sigma^{-1}(\vec \mu - \vec o)$ for the current camera position $\vec{o}$, packing the 6 unique coefficients of $\Sigma^{-1}$ with the precomputed vector for efficient loading.

We found that activating \enquote{fast math} in combination with re-scheduling in \emph{Preprocess} and \emph{Duplication} may lead to slightly different ordering of floating point instructions.
Thus, there may be slight differences in the number of tiles contributed by each Gaussian.
As we already store the number of tiles contributed by every Gaussian for memory allocation, we rely on the following simple solution: during \emph{Preprocess} we use a slightly lower threshold for culling, providing a slightly more conservative bound.
During \emph{Duplication}\new{,} we recheck whether the right number of tile contributions have been written.
If this is not the case, we simply add a dummy entry that sets a higher tile id and depth to $\infty$.
For training, we suggest to disable \enquote{fast math}, ensuring that gradient computations are as stable as possible.
However, for rendering using \enquote{fast math} may be beneficial to squeeze even more performance.

For load balancing in both \emph{Preprocess} and \emph{Duplication}, we rely on the \emph{ballot} instruction to determine which threads still require computations. 
We use \emph{shuffle} operations to broadcast already loaded register values, so each thread can perform culling and depth evaluation without additional memory loads. 
We assign successive potential tiles to each thread according to their thread rank in the warp.
For every iteration of the inner loop we again \emph{ballot} to determine which threads in the warp still want to write to a tile, $\ie$ did not cull away their tile.
We can then mask all ballot bits of lower ranked threads, compute their sum via \emph{popc} and determine the write location for each thread.

\paragraph{Render} Our hierarchical rasterizer is constructed from many steps, which are interleaved in their operation.
Due to the setup, there are special optimizations we can perform based on the current state of the pipeline:
The pipeline starts out with an initialize phase for each level, establishing a minimal fill level for each where no merge sort is performed.
In this phase, blending is not taking place either.
During the main operation, we ensure that we maintain a minimal fill level for each queue.
Finally, the pipeline is drained where the number of elements in each queue will eventually drop to zero.
Furthermore, we know that certain parts of the pipeline will always be executed a specific number of times.
The combination of these facts allows for a significant amount of specialization and loop unrolling.
However, we found that excessive code specialization and unrolling leads to a significant amount of stalls due to instruction fetches. Thus, relying on less specialized code is overall beneficial although up to 15\% more instructions are required for the increased control logics.

For Batcher Merge Sort, we use a \revised{trivial}{trival} implementation adapted from the NVIDIA CUDA examples\footnote{\url{https://github.com/NVIDIA/cuda-samples}}.
For Merge Sort, we use a custom implementation that is adapted for our use case:
each thread holds the to-be-inserted elements in registers and runs a binary search through the existing array to find where the new element should be placed with respect to the existing data.
In combination with the thread's rank, this yields the position in the final sorted array.
Still, we need to update the position of the existing data.
To this end, we switch the roles and memory locations of both data arrays and perform the exact same binary search, only switching strict comparison to non-strict comparison.
Also note that we are operating on a small fixed size array, enabling loop unrolling and leading to very few memory accesses.
For local presorting of four elements, we simply run three circular shuffles, revealing all elements among all threads to directly yield the right order via simple counting of smaller elements.
In our tests this was faster than any other method.

As we reevaluate $t_{opt}$ many times for many different ray directions, constructing and normalizing view rays can become a bottleneck.
Precomputing all view directions a single thread will need throughout the hierarchy (two for the \revised{$4\stimes4$ tile-queue}{tail queues}, one for the \revised{$2\stimes2$ tile-queue}{mid queue} and one for the \revised{per-pixel queue}{head queue}) would result in significant register pressure.
Fortunately, the same directions are needed by different threads and we can store the directions in shared memory and fetch them on demand, leading to significant performance improvements.

Obviously, we need to take some care to ensure that threads do not diverge, especially, we can only retire queues if all threads in the associated tile are done.
Also note that the loaded batches remain in registers for a potentially long time --- a 16 batch loaded by a half warp \revised{remains}{remain} in registers while four 4-thread batches are loaded and potentially up to 16 elements are blended.
However, when the 32-wide batch is loaded, no smaller batches are kept alive, somewhat reducing register pressure.

\section{Popping Detection Metric}
\label{app:popping}
\revised{For our popping detection metric, we use the RAFT~\cite{teed2020raft} model pre-trained on SINTEL~\cite{butler2012naturalistic}, which is publicly available.
We also compute the optical flow separately for each method for a fair comparison.
We follow~\citet{Nguyen2022Snerf} with timesteps $t \in \{1,7\}$ to measure short-range and long-range view-consistency, respectively.
We provide an additional ablation study for different $t \in \{3,5,9\}$ in~\tabref{different_timesteps_flip}, with three camera paths for the Garden scene of Mip-NeRF 360~\cite{barron2022mipnerf360}.
As can be seen, the consistency error grows almost linearly with increasing $t$.
Further, our method outperforms \gs for all timesteps.
}{To mitigate errors due to occlusions, we use the occlusion detection method from~\citet{ruder2016artistic}. As we compute the optical flow for each model separately, we combine the derived occlusion masks for a fair comparison. In addition, we do not consider the outermost $20$ pixels for more stable predictions. To further enhance our metric, we subtract the per-pixel minimum $\FLIP_t$ contribution before computing the mean over all pixels. It is easy to see that this does not change the inter-method difference. Finally, we use the RAFT~\cite{teed2020raft} model pre-trained on SINTEL~\cite{butler2012naturalistic}, which is publicly available.}

\begin{table}[ht!]
    \centering
    \caption{\new{$\flipview{t}$ comparison for $t \in \{1,3,5,7,9\}$ for three camera paths for the Garden scene of Mip-NeRF 360~\cite{barron2022mipnerf360}.
    As can be seen, our method outperforms \gs for each $t$, and $\flipview{t}$ scales almost linearly with increasing $t$.} }
    \label{tab:different_timesteps_flip}
    \begin{tabular}{@{}lccccc@{}}
    \toprule
    Method & $\flipview{1}$ & $\flipview{3}$ & $\flipview{5}$ & $\flipview{7}$ & $\flipview{9}$ \\
    \midrule
        \gs & 0.0080 & 0.0109 & 0.0134 & 0.0157 & 0.0180 \\
        Ours & \cellcolor{blue!45} 0.0075 & \cellcolor{blue!45} 0.0080 & \cellcolor{blue!45} 0.0082 & \cellcolor{blue!45} 0.0085 & \cellcolor{blue!45} 0.0087 \\
    \bottomrule
    \end{tabular}
\end{table}

\begin{figure}[ht!]
    \centering
     \begin{minipage}[t]{0.06\linewidth}
        \centering
        \text{}
    \end{minipage}
     \begin{minipage}[t]{0.46\linewidth}
        \centering
        \text{$\flipview{1}$}
    \end{minipage}
     \begin{minipage}[t]{0.46\linewidth}
        \centering
        \text{$\flipview{7}$}
    \end{minipage} \\
    \includegraphics[width=\linewidth]{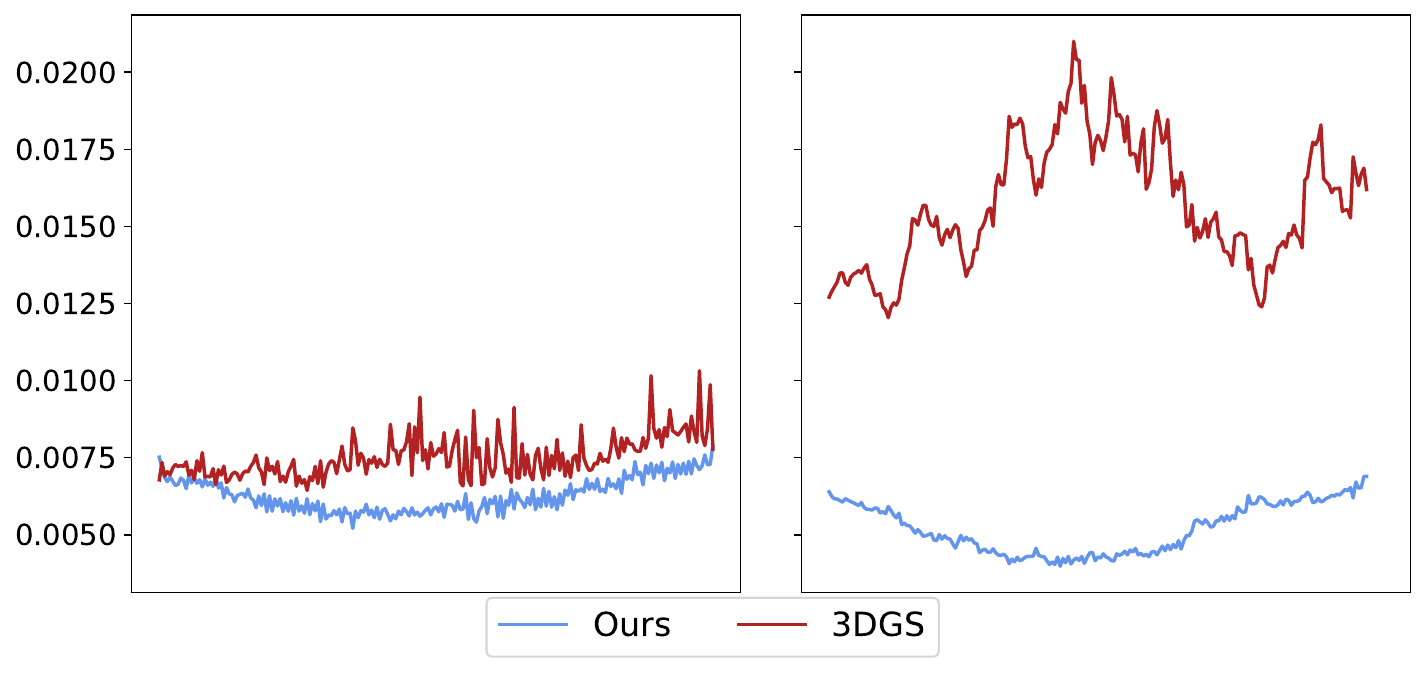}
    \caption{\new{Per-frame $\flipview{t}$ scores for $t \in \{1,7\}$ for a complete video sequence from the Garden scene.
    Popping in \gs causes significant peaks, as can be seen in the results for $\flipview{1}$.}}
    \label{fig:plot_perframe}
\end{figure}

\new{
\paragraph{Per-Frame Results.}
To gain more insight into our proposed popping detection metric, we additionally provide per-frame plots for a video of the Garden scene in \figref{plot_perframe}.
As can clearly be seen, there are significant peaks in $\flipview{1}$ for \gs, caused by popping.
Our method, on the other hand, does not suffer from such issues.
When analyzing the plot for $\flipview{7}$, \gs obtains significantly higher error rates --- using $t = 7$ accumulates artifacts over several iterations, therefore more clearly indicating popping when averaged over the complete video sequence.
}

\begin{figure}[ht!]
    \centering
     \begin{minipage}[t]{0.38\linewidth}
        \centering
        \small
        \text{Ground Truth Images}
    \end{minipage}
     \begin{minipage}[t]{0.15\linewidth}
        \centering
        \small
        \text{}
    \end{minipage}
     \begin{minipage}[t]{0.31\linewidth}
        \centering
        \small
        \text{Contrast $\uparrow$}
    \end{minipage}
     \begin{minipage}[t]{0.12\linewidth}
        \centering
        \small
        \text{\FLIP}
    \end{minipage} \\[-1.1ex]
         \begin{minipage}[t]{0.38\linewidth}
        \centering
        \small
        \text{}
    \end{minipage}
     \begin{minipage}[t]{0.17\linewidth}
        \centering
        \small
        \text{}
    \end{minipage}
     \begin{minipage}[t]{0.14\linewidth}
        \centering
        \scriptsize
        \text{Original}
    \end{minipage}
     \begin{minipage}[t]{0.14\linewidth}
        \centering
        \scriptsize
        \text{Rotated}
    \end{minipage}
     \begin{minipage}[t]{0.13\linewidth}
        \centering
        \small
        \text{}
    \end{minipage} \\
    \includegraphics[width=\linewidth]{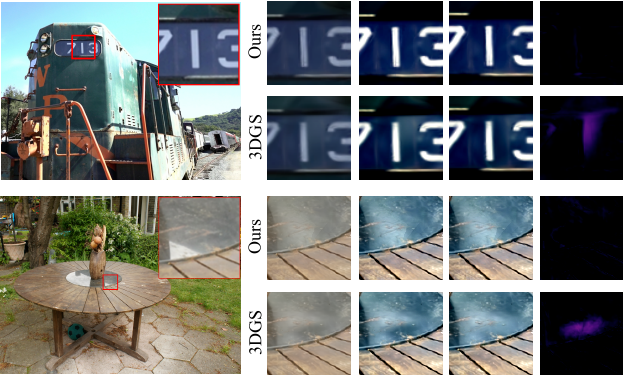}
    \caption{\new{\gs can fake view-dependent effects with popping.
    We slightly rotate test set views, and \gs's results are significantly less consistent compared to our results. 
    We increase contrast for zoomed-in views and include a \FLIP view for a better comparison.}}
    \label{fig:supp:gscheating}
\end{figure}

\new{
\paragraph{3DGS Cheating.}
To support our claim that \gs indeed cheats with popping to produce view-dependent effects, we provide additional images in \figref{supp:gscheating}.
We choose a ground-truth view from Train and Garden and sample a random rotation from \([-0.5^\circ, 0.5^\circ]^3\), which we apply to the ground-truth camera rotation.
Subsequently, we compare the rendering from the ground-truth camera pose and the rendering from the slightly rotated pose for \gs, as well as our method.
}

\new{
As can be seen, our approach produces more consistent results under view rotation.
Due to \gs's popping, the appearance changes significantly around test set views, which results in better image metrics in some configurations.
In \figref{supp:gscheating}, we increase contrast for the zoomed-in views and provide \FLIP comparisons to more clearly illustrate view inconsistencies.
}

\section{User Study}
\label{app:userstudy}
For our user study we recruited 18 participants from a local university, age 26 to 34, all normal or corrected vision, no color blindness.
All participants indicated that they are familiar with computer graphics (3-5 on a 5-point Likert scale).

We pre-recorded camera paths for all 13 scenes, looking at the main object present in the scene.
For \gs and ours, we used the version specifically trained for these approaches without Opacity Decay.
The paths all exhibit translation and rotation.
The recorded video clips were between $8$ and $19$ seconds long.

After a pre-questionnaire, we instructed the participants that they will be presented with video pairs and they should specifically look for consistency in the rendering and then rate whether either of the video clips was {more consistent} than the other.
If they did not consider any clip more consistent, they were allowed to rate them as equal.
We mapped those answers onto \revised{scores $s$:
\begin{equation}
    s = \begin{cases}
        -1 & \text{\gs is more consistent}, \\
        \phantom{-}0 & \text{both are equal}, \\
        \phantom{-}1 & \text{ours is more consistent}. \\
    \end{cases}\nonumber
\end{equation}
}{three values: ($-1$) \gs is more consistent, ($0$) they are equal, ($+1$) ours is more consistent.}
We presented both videos side-by-side and played them in a loop.
We did not restrict the answer times, allowing participants to watch the clips \revised{repeatedly}{multiple times}.
We randomized the order of videos (left, right) as well as the order of scenes.

\begin{figure}[ht!]
    \centering
    \includegraphics[width=\linewidth]{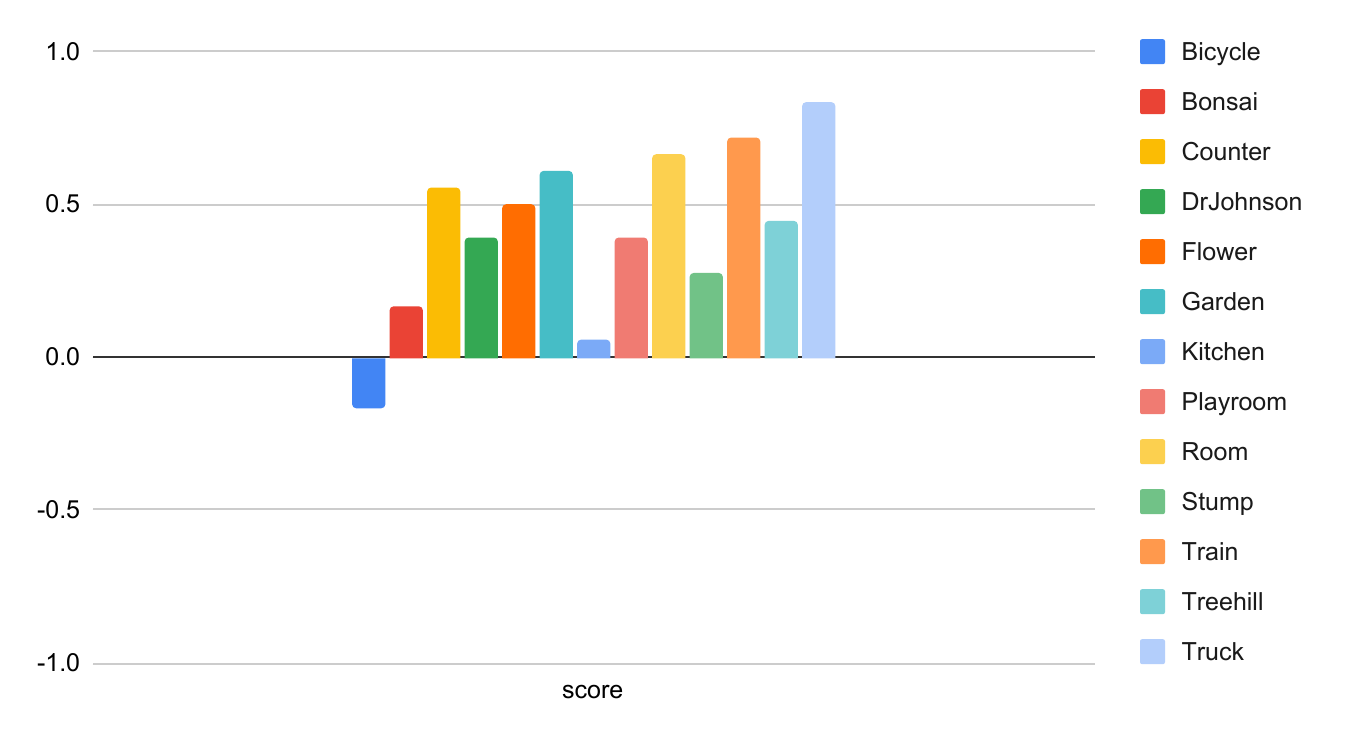}
    \caption{Average per-scene user study score. 
    A positive score indicates a preference for our method, whereas a negative score indicates a preference for \gs.
    \new{Our method clearly outperforms \gs.}}
    \label{fig:study}
\end{figure}

Overall, participants considered our method more consistent in $54.3\%$ of the cases, voted for equal in $33.3\%$ and preferred \gs in $12.4\%$, leading to an average preference score of $s_{mean}=0.42$.
The result is statistically significant according to Wilcoxon Signed Rank tests ($Z=2276.5$, $p<.0001$)~\cite{wilcoxon2008}.
As can be seen in \figref{study}, we observe inter-scene differences.
For scenes with mostly small Gaussians, like in Bonsai or Kitchen, we expected less difference in the voted scores, as there is also less popping.
In contrast, for scenes with large Gaussians, where popping occurs more often, like Room, Train or Truck, it is not surprising that our method is preferred by a large margin.
We were not able to assess why participants slightly preferred \gs for Bicycle.

\section{Detailed Performance Timings}
\label{app:performance}
\new{
In this section, we provide additional performance ablation studies.
We follow the evaluation setup from the main material, interpolating between all available COLMAP poses (30 frames per pose), and rendering in Full HD on an NVIDIA RTX 4090 with CUDA 11.8.}

\paragraph{Per-Scene Performance Timings.}
In \tabref{performance_total_per_scene}, we show per-scene performance timings for the total render time in ms. For the Mip-NeRF 360~\cite{barron2022mipnerf360} Indoor and Outdoor scenes, our method is slightly slower than \gs. For the Tanks \& Temples~\cite{Knapitsch2017} and Deep Blending~\cite{DeepBlending2018} datasets, we achieve higher performance than \gs for most scenes.
Analyzing the performance in more detail, we could verify that our method outperforms \gs when Gaussian are larger and/or more anisotropic, as our culling and load balancing can speed up rendering.
If Gaussians are small and uniformly sized, the main load stems from the final stages of the render kernel, where sorting of course creates an overhead compared to \gs.

\begin{table}[ht!]
    \centering
    \footnotesize
    \setlength{\tabcolsep}{2.5pt}
    \caption{Total performance timings for different configurations of our method and \gs, with the respective number of Gaussians per scene for comparison. 
    \revised{Although scenes may exhibit a similar number of Gaussians, performance timings vary significantly.}{Times in ms for Full HD resolution, averaged over 4 runs with an interpolated camera path between all available COLMAP poses (30 frames per pose).}}
    \label{tab:performance_total_per_scene}
    \begin{tabular}{lrrrrr}
    \toprule
    Scene & Bicycle & Flowers & Garden & Stump & Treehill \\
    \#Gaussians & 5.95M & 3.60M & 5.49M & 4.84M & 3.85M \\
    \midrule

    (A) Ours & \cellcolor{blue!30} 6.829 & \cellcolor{blue!15} 4.921 & \cellcolor{blue!15} 7.247 & \cellcolor{blue!15} 4.693 & \cellcolor{blue!15} 5.012 \\
    (B) Ours w/o per-tile depth & \cellcolor{blue!45} 6.730 & \cellcolor{blue!30} 4.693 & \cellcolor{blue!30} 7.160 & \cellcolor{blue!30} 4.509 & \cellcolor{blue!30} 4.879 \\
    (C) Ours w/o load balancing & 8.482 & 6.732 & 9.167 & 6.496 & 6.919 \\
    (D) Ours w/o tile-based culling & 10.066 & 7.338 & 9.796 & 6.584 & 7.884 \\
    (E) Ours w/o hier. culling & 11.087 & 7.589 & 11.788 & 7.178 & 7.773 \\
    \gs & \cellcolor{blue!15} 7.438 & \cellcolor{blue!45} 4.002 & \cellcolor{blue!45} 6.034 & \cellcolor{blue!45} 3.708 & \cellcolor{blue!45} 4.492 \\
    \bottomrule
    & \\
    & \\[-1.5ex]
    \toprule
    Scene && Bonsai & Counter & Kitchen & Room \\
    \#Gaussians && 1.25M & 1.20M & 1.81M & 1.55M \\
    \midrule
    (A) Ours && \cellcolor{blue!15} 3.399 & \cellcolor{blue!15} 4.390 & \cellcolor{blue!15} 5.695 & \cellcolor{blue!30} 3.990 \\
    (B) Ours w/o per-tile depth && \cellcolor{blue!30} 3.285 & \cellcolor{blue!30} 4.250 & \cellcolor{blue!30} 5.587 & \cellcolor{blue!45} 3.844 \\
    (C) Ours w/o load balancing && 5.251 & 6.217 & 7.558 & 5.843 \\
    (D) Ours w/o tile-based culling && 4.846 & 6.977 & 8.214 & 6.155 \\
    (E) Ours w/o hier. culling && 4.608 & 6.142 & 8.916 & 5.450 \\
    \gs && \cellcolor{blue!45} 2.574 & \cellcolor{blue!45} 4.043 & \cellcolor{blue!45} 4.783 & \cellcolor{blue!15} 4.180 \\
    \bottomrule
    & \\
    & \\[-1.5ex]
    \toprule
    Dataset && \multicolumn{2}{c}{Deep Blending} & \multicolumn{2}{c}{Tanks \& Temples} \\
    \cmidrule(lr){3-4}\cmidrule(lr){5-6}
    Scene && DrJohnson & Playroom & Train & Truck \\
    \#Gaussians && 3.28M & 2.33M & 1.05M & 2.56M \\
    \midrule
    (A) Ours && \cellcolor{blue!30} 4.763 & \cellcolor{blue!15} 4.549 & \cellcolor{blue!30} 4.225 & \cellcolor{blue!30} 5.100 \\
    (B) Ours w/o per-tile depth && \cellcolor{blue!45} 4.612 & \cellcolor{blue!30} 4.373 & \cellcolor{blue!45} 4.099 & \cellcolor{blue!45} 4.898 \\
    (C) Ours w/o load balancing && 6.648 & 6.275 & 6.215 & 6.942 \\
    (D) Ours w/o tile-based culling && 7.998 & 7.295 & 7.469 & 8.363 \\
    (E) Ours w/o hier. culling && 6.418 & 5.999 & 5.675 & 7.113 \\
    \gs && \cellcolor{blue!15} 5.752 & \cellcolor{blue!45} 4.303 & \cellcolor{blue!15} 5.548 & \cellcolor{blue!15} 5.506 \\
    \bottomrule
    \end{tabular}
\end{table}

\new{
\paragraph{Relative Performance Timings.}
In \tabref{relative_performance}, we report per-stage performance timings of our method relative to \gs for each scene.
\emph{Preprocess} is generally slower due to the additional workload of tile-based culling and computation of $\Sigma^{-1}$.
Due to our load balancing strategy, our \emph{Duplicate} stage is faster for every tested scene, except for Flowers --- here, our load balancing scheme is not able to amortize the additional workload of per-tile depth evaluations and tile-based culling.
\emph{Sort} is accelerated drastically, as the modifications for the previous stages result in fewer 2D splats to sort.
The \emph{Render} stage is naturally slower due to the overhead of our hierarchical rasterizer.}

\begin{table}[ht!]
    \centering
    \caption{\new{Relative per-scene performance timings of our method with respect to \gs for each stage.
    Shades of blue indicate scenes where our method performed favorably, whereas shades of red indicate the opposite.
    We also report average percentages in the final row (not average runtime).
    }}
    \label{tab:relative_performance}
    \footnotesize
    \setlength{\tabcolsep}{3.5pt}
\begin{tabular}{@{}llcrrrrr@{}}
\toprule
Dataset & Scene & \#Gaussians &\emph{Preprocess} & \emph{Duplicate} & \emph{Sort} & \emph{Render} & \emph{Total} \\
\midrule
\multirow{5}{3.5pt}{M360 Outdoor} & Bicycle & 5.95M &1.38 & \cellcolor{blue!15} 0.65 & 0.19 & \cellcolor{blue!30} 1.40 & \cellcolor{blue!15} 0.92 \\
 & Flowers& 3.60M & 1.53 & \cellcolor{red!45} 1.21 & \cellcolor{red!15} 0.25 & 1.71 & \cellcolor{red!15} 1.23 \\
 & Garden& 5.49M & \cellcolor{blue!30} 1.27 & 0.78 & \cellcolor{red!45} 0.29 & \cellcolor{red!30} 2.04 & 1.20 \\
 & Stump& 4.84M & 1.52 & \cellcolor{red!15} 0.90 & \cellcolor{red!30} 0.26 & 1.82 & \cellcolor{red!30} 1.27 \\
 & Treehill& 3.85M & 1.45 & 0.88 & 0.22 & 1.70 & 1.12 \\\midrule
\multirow{4}{3.5pt}{M360 Indoor} & Bonsai& 1.25M & \cellcolor{blue!15} 1.32 & 0.84 & 0.25 & \cellcolor{red!45} 2.06 & \cellcolor{red!45} 1.32 \\
 & Counter& 1.20M & 1.46 & 0.75 & 0.17 & 1.80 & 1.09 \\
 & Kitchen& 1.81M & \cellcolor{blue!45} 1.25 & 0.66 & 0.25 & \cellcolor{red!15} 2.02 & 1.19 \\
 & Room& 1.55M & \cellcolor{red!15} 1.60 & 0.70 & \cellcolor{blue!15} 0.13 & 1.53 & 0.95 \\\midrule
\multirow{2}{3.5pt}{DB} & DrJohnson& 3.28M & \cellcolor{red!30} 1.60 & \cellcolor{blue!30} 0.62 & \cellcolor{blue!30} 0.12 & \cellcolor{blue!45} 1.36 & \cellcolor{blue!30} 0.83 \\
 & Playroom& 2.33M & \cellcolor{red!45} 1.70 & \cellcolor{red!30} 0.93 & 0.14 & 1.63 & 1.06 \\\midrule
\multirow{2}{3.5pt}{T\&T} & Train& 1.05M & 1.42 & \cellcolor{blue!45} 0.61 & \cellcolor{blue!45} 0.11 & \cellcolor{blue!15} 1.53 & \cellcolor{blue!45} 0.76 \\
 & Truck& 2.56M & 1.57 & 0.79 & 0.13 & 1.69 & 0.93 \\\midrule
\multicolumn{2}{c}{Average} & 2.98M & 1.47 & 0.79 &  0.19 & 1.71 &  1.07 \\
\bottomrule
\end{tabular}
\end{table}

\new{
As previously discussed, Bonsai and Train exhibit the biggest inter-method differences despite the similar number of Gaussians due to differently sized splats in screen space.
Bicycle is the only scene of Mip-NeRF 360 Outdoor where our approach outperforms \gs.
Although Garden has a similarly high number of Gaussians, Bicycle allows for more excessive culling and, therefore, has a significantly lower avg. $N_t$ for our method ($2.7$ vs. $2.41$) compared to \gs ($6.24$ vs. $4.45$).
}

\new{
\paragraph{Hyperparameter Performance Ablation.}
In \tabref{ablation_queue_sizes}, we compare timings for the \emph{Render} stage for different size combinations of the $2\stimes2$ tile-queue and per-pixel queue.
As expected, our final chosen values (8 and 4) perform best, since they have the lowest memory and compute requirements.
Runtime increases significantly for very large per-pixel queue sizes, as register pressure increases and spills into local memory.
In \tabref{ablation_queue_sizes}, we also show performance timings of the \emph{Preprocess} and \emph{Duplicate} stages for different load balancing tile thresholds ($\infty=$ no load balancing).
Gaussians whose 2D splat covers more tiles than this threshold will be computed cooperatively by all threads of a warp.
Our chosen value of 32 performs well for both stages. \emph{Preprocess}, which only performs tile-based culling, does not show large differences and is even slightly faster for small thresholds.
\emph{Duplicate} shows significant improvements, as it performs many more computations per tile, \ie tile-based culling, per-tile depth calculation, and key/value writes.
Choosing a value of 32 also fits GPU warp sizes, and allows for better warp utilization, as well as fast concurrent memory writes.
}

\section{Per-Scene Quality Metrics}
\label{app:perscenequality}
We provide per-scene results for Mip-NeRF 360~\cite{barron2022mipnerf360}, Tanks and Temples~\cite{Knapitsch2017} and Deep Blending~\cite{DeepBlending2018} in Tabs.~\ref{tab:per_scene_tatdb_psnrssim} and \ref{tab:per_scene_360_psnr}.
Results with dagger ($\dagger$) were reproduced from~\citet{kerbl3Dgaussians}\new{: this includes results for \mipnerf~\cite{barron2022mipnerf360}, Instant-NGP~\cite{mueller2022instant} and Plenoxels~\cite{fridovich2022plenoxels}}.
We evaluate our final hierarchical rasterizer (\enquote{Ours}), as well as the fixed-size head sorting method for two different resorting window sizes (\enquote{Head 8} and \enquote{Head 16}), with and without per-tile depth (\enquote{\woptd}).

\begin{table}[h!]
    \centering
    \small
    \setlength{\tabcolsep}{4pt}
    \caption{\new{Ablation of our method for different sizes of the $2\stimes2$ tile-queue and the per-pixel queue in addition to different load balancing thresholds, with the same underlying model (no retraining).} 
    }
    \label{tab:ablation_queue_sizes}
    \begin{tabular}{lrrrrrrr}
    \toprule
    $2\stimes2$ tile-queue  &&    8 &    8 &   12 &   12 &  20 &   20  \\
    per-pixel queue         &&    4 &    8 &    4 &    8 &   8 &   16  \\
    \midrule
    \emph{Render} && \cellcolor{blue!45} 3.646 & \cellcolor{blue!30} 4.000 & \cellcolor{blue!15} 4.052 & 4.364 & 4.470 & 8.549 \\
    \bottomrule
    & \\[-1.5ex]
    & \\[-1.5ex]
    \toprule
    Threshold          &       1 &      4 &      16 &      32 &      64 &      256 &   $\infty$  \\
    \midrule
    \emph{Preprocess}  &   \cellcolor{blue!30} 0.644 & \cellcolor{blue!45} 0.643 & \cellcolor{blue!15} 0.648 &  \cellcolor{blue!15} 0.648 &   0.649 &   0.652 &  0.847  \\
    \emph{Duplicate}   &   0.607 & 0.541 & \cellcolor{blue!30}   0.437 &  \cellcolor{blue!45} 0.435 &  \cellcolor{blue!15}  0.447 &   0.515 &  2.059  \\
    \bottomrule
    \end{tabular}
\end{table}

\begin{table}[h!]
    \centering
    \scriptsize
    \setlength{\tabcolsep}{1.8pt}
    \caption{\revised{Combined per-scene scores for PSNR, SSIM, LPIPS and \FLIP for Tanks \& Temples and Deep Blending.}{}}
    \label{tab:per_scene_tatdb_psnrssim}
    \begin{tabular}{lrrrrrrrr}
    \toprule
    Metric & \multicolumn{4}{c}{PSNR} & \multicolumn{4}{c}{SSIM} \\
    \cmidrule(lr){2-5}\cmidrule(lr){6-9}
    Dataset & \multicolumn{2}{c}{Tanks \& Temples} & \multicolumn{2}{c}{Deep Blending} & \multicolumn{2}{c}{Tanks \& Temples} & \multicolumn{2}{c}{Deep Blending} \\
\cmidrule(lr){2-3}\cmidrule(lr){4-5}\cmidrule(lr){6-7}\cmidrule(lr){8-9}
    Scene & Truck & Train & DrJ. & Playroom & Truck & Train & DrJ. & Playroom\\
    \midrule
    \revised{Mip-NeRF 360\textsuperscript{$\dagger$}}{\citet{barron2022mipnerf360}\textsuperscript{$\dagger$}}& 24.91 & 19.52 & 29.14 & 29.66  & 0.857 & 0.660 & 0.901 & 0.900 \\
    \revised{Instant-NGP (base)\textsuperscript{$\dagger$}}{\citet{mueller2022instant}\textsuperscript{$\dagger$}}& 23.26 & 20.17 & 27.75 & 19.48 & 0.779 & 0.666 & 0.839 & 0.754 \\
    \revised{Instant-NGP (big)\textsuperscript{$\dagger$}}{\citet{mueller2022instant}\textsuperscript{$\dagger$}}& 23.38 & 20.46 & 28.26 & 21.67 & 0.800 & 0.689 & 0.854 & 0.780 \\
    \revised{Plenoxels}{\citet{fridovich2022plenoxels}\textsuperscript{$\dagger$}}& 23.23 & 18.94 & 23.16 & 23.02 & 0.774 & 0.663 & 0.787 & 0.802 \\
    \gs & \cellcolor{blue!45} 25.39 & \cellcolor{blue!45} 22.04 & 29.06 & 29.86 & 0.878 & \cellcolor{blue!45} 0.813 & 0.898 & 0.901 \\
\midrule
    Head 8 \woptd & 24.79 & 21.52 & 29.40 & 30.29 & 0.877 & 0.809 & 0.902 & 0.905 \\
    Head 8 & 24.81 & 21.41 & \cellcolor{blue!45} 29.51 & \cellcolor{blue!15} 30.31 & 0.878 & \cellcolor{blue!15} 0.810 & \cellcolor{blue!45} 0.904 & \cellcolor{blue!15} 0.905 \\
    Head 16 \woptd & 24.84 & \cellcolor{blue!15} 21.60 & 29.40 & \cellcolor{blue!45} 30.36 & \cellcolor{blue!30} 0.878 & \cellcolor{blue!30} 0.810 & \cellcolor{blue!30} 0.904 & 0.905 \\
    Head 16 & 24.81 & 21.36 & \cellcolor{blue!15} 29.44 & 30.31 & 0.877 & 0.809 & 0.903 & \cellcolor{blue!30} 0.906 \\
    Ours \woptd & 24.93 & 21.53 & \cellcolor{blue!30} 29.44 & \cellcolor{blue!30} 30.31 & \cellcolor{blue!15} 0.878 & 0.810 & 0.903 & 0.905 \\
    Ours & \cellcolor{blue!15} 24.93 & 21.48 & 29.42 & 30.31 & \cellcolor{blue!45} 0.878 & 0.808 & \cellcolor{blue!15} 0.903 & 0.905 \\
\midrule
    \gs (Opacity Decay) & \cellcolor{blue!30} 25.31 & \cellcolor{blue!30} 21.73 & 28.18 & 29.69  & 0.874 & 0.804 & 0.888 & 0.899 \\
    Ours (Opacity Decay)& 24.90 & 21.46 & 29.38 & 30.30 & 0.875 & 0.804 & 0.903 & \cellcolor{blue!45} 0.907 \\
    \bottomrule
    & \\
    & \\[-1.5ex]
    \toprule
    & \multicolumn{4}{c}{LPIPS} & \multicolumn{4}{c}{\FLIP} \\
    \cmidrule(lr){2-5}\cmidrule(lr){6-9}
     Dataset & \multicolumn{2}{c}{Tanks \& Temples} & \multicolumn{2}{c}{Deep Blending} & \multicolumn{2}{c}{Tanks \& Temples} & \multicolumn{2}{c}{Deep Blending} \\
 \cmidrule(lr){2-3}\cmidrule(lr){4-5}\cmidrule(lr){6-7}\cmidrule(lr){8-9}
     Scene & Truck & Train & DrJ. & Playroom & Truck & Train & DrJ. & Playroom\\
    \midrule
    \revised{Mip-NeRF 360\textsuperscript{$\dagger$}}{\citet{barron2022mipnerf360}\textsuperscript{$\dagger$}}& 0.159 & 0.354 & 0.237 & 0.252& 0.162 & 0.302 & 0.117 & 0.158 \\
    \revised{Instant-NGP (base)\textsuperscript{$\dagger$}}{\citet{mueller2022instant}\textsuperscript{$\dagger$}}& 0.274 & 0.386 & 0.381 & 0.465 & 0.194 & 0.297 & 0.141 & 0.375 \\
    \revised{Instant-NGP (big)\textsuperscript{$\dagger$}}{\citet{mueller2022instant}\textsuperscript{$\dagger$}}& 0.249 & 0.360 & 0.352 & 0.428  & 0.190 & 0.291 & 0.133 & 0.311 \\
    \revised{Plenoxels}{\citet{fridovich2022plenoxels}\textsuperscript{$\dagger$}}& 0.308 & 0.379 & 0.433 & 0.418 & 0.196 & 0.328 & 0.222 & 0.266 \\
    \gs  & 0.148 & 0.208 & 0.247 & 0.246& \cellcolor{blue!45} 0.148 & \cellcolor{blue!45} 0.250 & 0.119 & 0.143 \\
\midrule
    Head 8 \woptd & 0.143 & 0.204 & 0.236 & 0.237 & 0.165 & 0.265 & 0.116 & 0.140 \\
    Head 8 & 0.142 & \cellcolor{blue!30} 0.203 & \cellcolor{blue!15} 0.234 & \cellcolor{blue!15} 0.235& 0.166 & 0.266 & \cellcolor{blue!45} 0.114 & 0.139 \\
    Head 16 \woptd & \cellcolor{blue!30} 0.142 & \cellcolor{blue!15} 0.203 & 0.234 & 0.236 & 0.166 & \cellcolor{blue!15} 0.262 & 0.115 & \cellcolor{blue!15} 0.138 \\
    Head 16 & 0.142 & \cellcolor{blue!45} 0.203 & \cellcolor{blue!45} 0.234 & 0.235 & 0.164 & 0.267 & 0.116 & 0.139 \\
    Ours \woptd & \cellcolor{blue!45} 0.142 & 0.204 & 0.234 & \cellcolor{blue!30} 0.235 & 0.163 & 0.264 & 0.115 & 0.139 \\
    Ours & \cellcolor{blue!15} 0.142 & 0.204 & \cellcolor{blue!30} 0.234 & \cellcolor{blue!45} 0.235  & 0.164 & 0.267 & \cellcolor{blue!15} 0.115 & \cellcolor{blue!30} 0.138 \\
\midrule
    \gs (Opacity Decay) & 0.160 & 0.228 & 0.265 & 0.260 & \cellcolor{blue!30} 0.148 & \cellcolor{blue!30} 0.261 & 0.124 & 0.144 \\
    Ours (Opacity Decay)& 0.151 & 0.218 & 0.241 & 0.241 & \cellcolor{blue!15} 0.160 & 0.267 & \cellcolor{blue!30} 0.115 & \cellcolor{blue!45} 0.138 \\
    \bottomrule
    \end{tabular}
\end{table}

\begin{table}[ht!]
    \centering
    \scriptsize
    \setlength{\tabcolsep}{1.6pt}
    \caption{\revised{Combined per-scene scores for PSNR, SSIM, LPIPS \& \FLIP for the Mip-NeRF 360 dataset.}{}}
    \label{tab:per_scene_360_psnr}
    \begin{tabular}{lrrrrrrrrr}
    \toprule
    Dataset & \multicolumn{5}{c}{Mip-NeRF 360 Outdoor} &  \multicolumn{4}{c}{Mip-NeRF 360 Indoor} \\
\cmidrule(lr){2-6}\cmidrule(lr){7-10}
    Scene & Bicycle & Flowers & Garden & Stump & Treehill & Room & Counter & Kitchen & Bonsai \\
    \midrule
    & \multicolumn{9}{c}{PSNR} \\
    \cmidrule(lr){2-10}
    \revised{Mip-NeRF 360\textsuperscript{$\dagger$}}{\citet{barron2022mipnerf360}\textsuperscript{$\dagger$}} & 24.30 & \cellcolor{blue!45} 21.65 & 26.88 & 26.36 & \cellcolor{blue!45} 22.93 & \cellcolor{blue!30} 31.47 & \cellcolor{blue!45} 29.45 & \cellcolor{blue!45} 31.99 & \cellcolor{blue!45} 33.40 \\
    \revised{Instant-NGP (base)\textsuperscript{$\dagger$}}{\citet{mueller2022instant}\textsuperscript{$\dagger$}}& 22.19 & 20.35 & 24.60 & 23.63 & 22.36 & 29.27 & 26.44 & 28.55 & 30.34 \\
    \revised{Instant-NGP (big)\textsuperscript{$\dagger$}}{\citet{mueller2022instant}\textsuperscript{$\dagger$}}& 22.17 & 20.65 & 25.07 & 23.47 & 22.37 & 29.69 & 26.69 & 29.48 & 30.69 \\
    \revised{Plenoxels}{\citet{fridovich2022plenoxels}\textsuperscript{$\dagger$}} & 21.90 & 20.10 & 23.50 & 20.68 & 22.26 & 27.57 & 23.64 & 23.43 & 24.71 \\
    \gs & 25.18 & 21.48 & \cellcolor{blue!45} 27.24 & 26.62 & 22.45 & \cellcolor{blue!45} 31.49 & \cellcolor{blue!30} 28.98 & \cellcolor{blue!30} 31.35 & \cellcolor{blue!30} 32.10 \\
    \midrule
    Head 8 \woptd & 25.18 & 21.49 & 27.14 & 26.64 & 22.41 & 30.77 & 28.83 & 31.06 & 31.85 \\
    Head 8 & 25.19 & \cellcolor{blue!15} 21.50 & \cellcolor{blue!30} 27.20 & 26.62 & \cellcolor{blue!30} 22.52 & 30.88 & 28.78 & 31.04 & \cellcolor{blue!15} 31.98 \\
    Head 16 \woptd & 25.20 & 21.48 & \cellcolor{blue!15} 27.18 & 26.62 & 22.45 & 30.84 & \cellcolor{blue!15} 28.84 & 30.89 & 31.63 \\
    Head 16 & \cellcolor{blue!45} 25.22 & \cellcolor{blue!30} 21.55 & 27.12 & 26.59 & \cellcolor{blue!15} 22.50 & 30.81 & 28.78 & 31.06 & 31.88 \\
    Ours \woptd & \cellcolor{blue!30} 25.21 & 21.45 & 27.17 & \cellcolor{blue!30} 26.68 & 22.47 & 30.84 & 28.70 & \cellcolor{blue!15} 31.23 & 31.90 \\
    Ours & \cellcolor{blue!15} 25.20 & 21.50 & 27.16 & \cellcolor{blue!45} 26.69 & 22.43 & 30.83 & 28.59 & 31.13 & 31.93 \\\midrule
    \gs (Opacity Decay) & 24.93 & 21.30 & 27.05 & 26.57 & 22.39 & \cellcolor{blue!15} 31.03 & 28.64 & 31.07 & 31.52 \\
    Ours (Opacity Decay) & 25.00 & 21.30 & 26.95 & \cellcolor{blue!15} 26.67 & 22.39 & 30.58 & 28.33 & 30.46 & 30.76 \\
    \midrule
    & \multicolumn{9}{c}{SSIM} \\
    \cmidrule(lr){2-10}
    \revised{Mip-NeRF 360\textsuperscript{$\dagger$}}{\citet{barron2022mipnerf360}\textsuperscript{$\dagger$}} & 0.685 & 0.584 & 0.809 & 0.745 & 0.631 & 0.910 & 0.892 & 0.917 & 0.938 \\
    \revised{Instant-NGP (base)\textsuperscript{$\dagger$}}{\citet{mueller2022instant}\textsuperscript{$\dagger$}} & 0.491 & 0.450 & 0.649 & 0.574 & 0.518 & 0.855 & 0.798 & 0.818 & 0.890 \\
    \revised{Instant-NGP (big)\textsuperscript{$\dagger$}}{\citet{mueller2022instant}\textsuperscript{$\dagger$}}& 0.512 & 0.486 & 0.701 & 0.594 & 0.542 & 0.871 & 0.817 & 0.858 & 0.906 \\
    \revised{Plenoxels}{\citet{fridovich2022plenoxels}\textsuperscript{$\dagger$}} & 0.495 & 0.432 & 0.606 & 0.523 & 0.510 & 0.840 & 0.758 & 0.648 & 0.814 \\
    \gs & 0.763 & 0.603 & \cellcolor{blue!45} 0.862 & 0.772 & 0.632 & \cellcolor{blue!45} 0.917 & \cellcolor{blue!45} 0.906 & \cellcolor{blue!45} 0.925 & \cellcolor{blue!30} 0.939 \\
\midrule
    Head 8 \woptd & 0.766 & 0.602 & \cellcolor{blue!15} 0.862 & 0.773 & 0.633 & \cellcolor{blue!15} 0.917 & \cellcolor{blue!15} 0.905 & 0.925 & 0.939 \\
    Head 8 & 0.766 & \cellcolor{blue!15} 0.604 & \cellcolor{blue!30} 0.862 & 0.773 & 0.634 & 0.916 & 0.905 & 0.924 & \cellcolor{blue!45} 0.939 \\
    Head 16 \woptd & \cellcolor{blue!15} 0.767 & 0.603 & 0.861 & 0.773 & 0.633 & 0.917 & \cellcolor{blue!30} 0.905 & 0.922 & 0.939 \\
    Head 16  & \cellcolor{blue!45} 0.767 & \cellcolor{blue!45} 0.604 & 0.861 & 0.773 & \cellcolor{blue!30} 0.635 & \cellcolor{blue!30} 0.917 & 0.905 & 0.925 & 0.939 \\
    Ours \woptd & \cellcolor{blue!30} 0.767 & 0.603 & 0.862 & \cellcolor{blue!15} 0.775 & \cellcolor{blue!45} 0.635 & 0.917 & 0.904 & \cellcolor{blue!30} 0.925 & 0.939 \\
    Ours & 0.767 & \cellcolor{blue!30} 0.604 & 0.862 & \cellcolor{blue!30} 0.775 & \cellcolor{blue!15} 0.635 & 0.917 & 0.903 & \cellcolor{blue!15} 0.925 & \cellcolor{blue!15} 0.939 \\
\midrule
    \gs (Opacity Decay)  & 0.749 & 0.592 & 0.854 & 0.770 & 0.626 & 0.914 & 0.899 & 0.921 & 0.937 \\
    Ours (Opacity Decay)& 0.756 & 0.593 & 0.855 & \cellcolor{blue!45} 0.775 & 0.629 & 0.914 & 0.898 & 0.920 & 0.935 \\
    \bottomrule
    & \\
    & \\[-1.5ex]
    \toprule
    Dataset & \multicolumn{5}{c}{Mip-NeRF 360 Outdoor} &  \multicolumn{4}{c}{Mip-NeRF 360 Indoor} \\
\cmidrule(lr){2-6}\cmidrule(lr){7-10}
    Scene & Bicycle & Flowers & Garden & Stump & Treehill & Room & Counter & Kitchen & Bonsai \\
    \midrule
    & \multicolumn{9}{c}{LPIPS} \\
    \cmidrule(lr){2-10}
    \revised{Mip-NeRF 360\textsuperscript{$\dagger$}}{\citet{barron2022mipnerf360}\textsuperscript{$\dagger$}} & 0.305 & 0.346 & 0.171 & 0.261 & 0.347 & \cellcolor{blue!45} 0.213 & 0.207 & 0.128 & \cellcolor{blue!45} 0.179 \\
    \revised{Instant-NGP (base)\textsuperscript{$\dagger$}}{\citet{mueller2022instant}\textsuperscript{$\dagger$}} & 0.487 & 0.481 & 0.312 & 0.450 & 0.489 & 0.301 & 0.342 & 0.254 & 0.227 \\
    \revised{Instant-NGP (big)\textsuperscript{$\dagger$}}{\citet{mueller2022instant}\textsuperscript{$\dagger$}}& 0.446 & 0.441 & 0.257 & 0.421 & 0.450 & 0.261 & 0.306 & 0.205 & \cellcolor{blue!30} 0.193 \\
    \revised{Plenoxels}{\citet{fridovich2022plenoxels}\textsuperscript{$\dagger$}}& 0.490 & 0.506 & 0.374 & 0.468 & 0.495 & 0.344 & 0.378 & 0.404 & 0.336 \\
    \gs & 0.213 & 0.338 & 0.109 & 0.216 & 0.327 & 0.221 & 0.202 & 0.127 & 0.206 \\
\midrule
    Head 8 \woptd & 0.207 & 0.336 & 0.107 & 0.211 & 0.322 & 0.216 & \cellcolor{blue!30} 0.199 & 0.126 & 0.203 \\
    Head 8 & 0.207 & 0.335 & \cellcolor{blue!30} 0.107 & 0.211 & 0.320 & 0.217 & 0.199 & 0.126 & \cellcolor{blue!15} 0.202 \\
    Head 16 \woptd & \cellcolor{blue!15} 0.206 & 0.336 & 0.107 & 0.211 & 0.321 & \cellcolor{blue!15} 0.216 & \cellcolor{blue!45} 0.198 & 0.128 & 0.203 \\
    Head 16 & \cellcolor{blue!30} 0.206 & \cellcolor{blue!15} 0.335 & 0.107 & \cellcolor{blue!15} 0.211 & \cellcolor{blue!45} 0.319 & 0.216 & \cellcolor{blue!15} 0.199 & \cellcolor{blue!15} 0.126 & 0.202 \\
    Ours \woptd & \cellcolor{blue!45} 0.205 & \cellcolor{blue!30} 0.335 & \cellcolor{blue!45} 0.107 & \cellcolor{blue!30} 0.210 & \cellcolor{blue!30} 0.319 & \cellcolor{blue!30} 0.216 & 0.199 & \cellcolor{blue!45} 0.126 & 0.203 \\
    Ours & 0.206 & \cellcolor{blue!45} 0.335 & \cellcolor{blue!15} 0.107 & \cellcolor{blue!45} 0.210 & \cellcolor{blue!15} 0.319 & 0.216 & 0.200 & \cellcolor{blue!30} 0.126 & 0.202 \\
\midrule
    \gs (Opacity Decay) & 0.244 & 0.358 & 0.125 & 0.232 & 0.347 & 0.230 & 0.215 & 0.137 & 0.210 \\
    Ours (Opacity Decay)& 0.232 & 0.354 & 0.122 & 0.224 & 0.336 & 0.224 & 0.211 & 0.135 & 0.207 \\\midrule
    & \multicolumn{9}{c}{\FLIP} \\
    \cmidrule(lr){2-10}
    \revised{Mip-NeRF 360\textsuperscript{$\dagger$}}{\citet{barron2022mipnerf360}\textsuperscript{$\dagger$}}& 0.169 & \cellcolor{blue!45} 0.217 & 0.124 & 0.156 & 0.184 & \cellcolor{blue!30} 0.095 & \cellcolor{blue!45} 0.100 & \cellcolor{blue!45} 0.088 & \cellcolor{blue!45} 0.069 \\
    \revised{Instant-NGP (base)\textsuperscript{$\dagger$}}{\citet{mueller2022instant}\textsuperscript{$\dagger$}}& 0.203 & 0.260 & 0.155 & 0.209 & 0.189 & 0.118 & 0.144 & 0.123 & 0.093 \\
    \revised{Instant-NGP (big)\textsuperscript{$\dagger$}}{\citet{mueller2022instant}\textsuperscript{$\dagger$}}& 0.201 & 0.251 & 0.146 & 0.213 & 0.189 & 0.112 & 0.139 & 0.113 & 0.089 \\
    \revised{Plenoxels}{\citet{fridovich2022plenoxels}\textsuperscript{$\dagger$}}& 0.211 & 0.271 & 0.181 & 0.276 & 0.206 & 0.143 & 0.201 & 0.218 & 0.165 \\
    \gs & \cellcolor{blue!45} 0.158 & 0.225 & \cellcolor{blue!45} 0.118 & 0.150 & 0.186 & \cellcolor{blue!45} 0.093 & \cellcolor{blue!30} 0.105 & \cellcolor{blue!30} 0.096 & \cellcolor{blue!30} 0.082 \\
\midrule
    Head 8 \woptd & 0.160 & \cellcolor{blue!15} 0.223 & 0.120 & 0.150 & 0.184 & 0.102 & \cellcolor{blue!15} 0.107 & 0.100 & 0.086 \\
    Head 8 & 0.159 & 0.223 & 0.119 & 0.150 & \cellcolor{blue!45} 0.181 & 0.101 & 0.108 & 0.099 & \cellcolor{blue!15} 0.083 \\
    Head 16 \woptd & 0.159 & 0.224 & \cellcolor{blue!15} 0.119 & 0.151 & \cellcolor{blue!30} 0.182 & 0.101 & 0.107 & 0.103 & 0.086 \\
    Head 16 & \cellcolor{blue!15} 0.159 & \cellcolor{blue!30} 0.222 & 0.121 & 0.151 & 0.183 & 0.102 & 0.108 & 0.099 & 0.085 \\
    Ours \woptd & 0.160 & 0.225 & 0.120 & \cellcolor{blue!15} 0.149 & 0.183 & 0.101 & 0.110 & \cellcolor{blue!15} 0.099 & 0.085 \\
    Ours & \cellcolor{blue!30} 0.159 & 0.224 & \cellcolor{blue!30} 0.119 & \cellcolor{blue!30} 0.149 & 0.184 & 0.101 & 0.111 & 0.099 & 0.084 \\
\midrule
    \gs (Opacity Decay) & 0.162 & 0.228 & 0.120 & 0.151 & \cellcolor{blue!15} 0.182 & \cellcolor{blue!15} 0.096 & 0.107 & 0.099 & 0.085 \\
    Ours (Opacity Decay)& 0.162 & 0.228 & 0.122 & \cellcolor{blue!45} 0.148 & 0.182 & 0.103 & 0.112 & 0.106 & 0.090 \\
    \bottomrule
    \end{tabular}
\end{table}
\clearpage

\end{document}